\documentclass[acmsmall,screen]{acmart}

\AtBeginDocument{%
  \providecommand\BibTeX{{%
    \normalfont B\kern-0.5em{\scshape i\kern-0.25em b}\kern-0.8em\TeX}}}

\acmJournal{JACM}

\usepackage{fixltx2e}
\usepackage{amsmath, amsfonts}
\usepackage{subcaption}
\usepackage{graphicx}
\usepackage{algorithm}
\usepackage{algorithmicx}
\usepackage{amsmath,amsfonts}
\usepackage{array}
\usepackage{balance}
 \usepackage{booktabs}
\usepackage[skip=1pt,labelfont=bf]{caption}
 \usepackage{calligra}
\usepackage{color}
\usepackage{colortbl}
\usepackage{courier} 
\usepackage{csvsimple}
\usepackage{enumitem}
\usepackage{fancybox}
\usepackage{fontenc}
\usepackage{graphicx}
\usepackage{listings}
\usepackage{longtable}
\usepackage{lscape}
\usepackage{makecell}
\usepackage{marvosym}
\usepackage{moreverb}
\usepackage{multicol}
\usepackage{multirow}
\usepackage{pifont}
\usepackage{rotating}
\usepackage{setspace}

\usepackage[most]{tcolorbox}
\usepackage{threeparttable}
\usepackage{tikz}
\usepackage[normalem]{ulem}
\usepackage{url}
\usepackage{soul}
\usepackage{wasysym}
\usepackage{xspace}
\usepackage{xcolor}
\usepackage{lineno}\usepackage{pifont}
\definecolor{myred}{RGB}{200, 50, 50}

\definecolor{darkpastelred}{rgb}{0.76, 0.23, 0.13}
\definecolor{ao(english)}{rgb}{0.0, 0.5, 0.0}
\definecolor{darkpastelred}{rgb}{0.76, 0.23, 0.13}
\definecolor{ao(english)}{rgb}{0.0, 0.5, 0.0}

\definecolor{yellow}{RGB}{255,255,153}
\definecolor{grey}{RGB}{224,224,224}

\lstdefinestyle{test-smell-listing}{
  language=Java,
  basicstyle=\ttfamily\footnotesize,
  keywordstyle=\color{blue},
  commentstyle=\color{gray},
  stringstyle=\color{red},
  numbers=left,
  numberstyle=\tiny\color{gray},
  stepnumber=1,
  numbersep=5pt,
  frame=single,
  breaklines=true,
  captionpos=b
}

\definecolor{DarkOrange}{rgb}{0.8,0.3,0.0}
\definecolor{DarkCyan}{rgb}{0.0, 0.55, 0.55}
\definecolor{DarkCyel}{rgb}{1.0, 0.49, 0.0}
\definecolor{yellow-green}{rgb}{0.6, 0.8, 0.2}
\definecolor{darkpastelred}{rgb}{0.76, 0.23, 0.13}
\definecolor{ao(english)}{rgb}{0.0, 0.5, 0.0}
\definecolor{darkpastelred}{rgb}{0.76, 0.23, 0.13}
\definecolor{ao(english)}{rgb}{0.0, 0.5, 0.0}
\definecolor{yellow}{RGB}{255,255,153}
\definecolor{grey}{RGB}{224,224,224}

\usepackage{ifthen}

\newboolean{showcomments}
\setboolean{showcomments}{true} 

\newcommand{\mynote}[2]{}

\ifthenelse{\boolean{showcomments}}{
  \renewcommand{\mynote}[2]{%
    \fbox{\bfseries\sffamily\scriptsize#1}%
    {\small$\blacktriangleright$\textsf{\emph{#2}}$\blacktriangleleft$}%
  }
}{}

\newcolumntype{?}{!{\vrule width 1pt}}

\newcommand{\highlight}[1]{\begin{tcolorbox}[leftrule=0mm,rightrule=0mm,toprule=0mm,bottomrule=0mm,left=2pt,right=2pt,top=2pt,bottom=2pt]
  #1
  \end{tcolorbox}
}

\def\BibTeX{{\rm B\kern-.05em{\sc i\kern-.025em b}\kern-.08em
    T\kern-.1667em\lower.7ex\hbox{E}\kern-.125emX}}

\begin{document}
\title{Humanizing Automatically Generated Unit Test Suites with LLM-Based Refactoring}

\author{Wendkûuni C. Ouédraogo}
\email{wendkuuni.ouedraogo@uni.lu}
\affiliation{
  \institution{University of Luxembourg}
 	\country{Luxembourg}
}

\author{Yinghua Li}\authornote{Corresponding author.}
\email{yinghua.li@njust.edu.cn}
\affiliation{
  \institution{Nanjing University of Science and Technology}
 	\country{China}
}

\author{Xueqi Dang}
\email{xueqi.dang@uni.lu}
\affiliation{
  \institution{University of Luxembourg}
 	\country{Luxembourg}
}

\author{Paweł Borsukiewicz}
\email{pawel.borsukiewicz@uni.lu}
\affiliation{
  \institution{University of Luxembourg}
  \country{Luxembourg}
}

\author{Liang Xiao}
\email{xiaoliang@mail.njust.edu.cn}
\affiliation{
  \institution{Nanjing University of Science and Technology}
 	\country{China}
}

\author{Lingfeng Bao}
\email{lingfengbao@zju.edu.cn}
\affiliation{
  \institution{Zhejiang University}
 	\country{China}
}

\author{Anil Koyuncu}
\email{anil.koyuncu@cs.bilkent.edu.tr}
\affiliation{
  \institution{Bilkent University}
 	\country{Turkey}
}

\author{Jacques Klein}
\email{jacques.klein@uni.lu}
\affiliation{
  \institution{University of Luxembourg}
  \country{Luxembourg}
}

\author{David Lo}
\email{davidlo@smu.edu.sg}
\affiliation{
  \institution{Singapore Management University}
  \country{Singapore}
}

\author{Tegawend\'e F. Bissyand\'e}
\email{tegawende.bissyande@uni.lu}
\affiliation{
  \institution{University of Luxembourg}
 	\country{Luxembourg}
}

\renewcommand{\shortauthors}{W. Ouédraogo et al.}

\begin{abstract}
Search-based test generation tools such as EvoSuite can produce compilable and high-coverage unit tests at scale, yet the resulting suites are often difficult to read and maintain, limiting practical 
adoption. Large language models (LLMs) can generate more natural tests, but one-shot generation remains brittle, with compilation rates as low as 51--78\% in our study. We introduce \textsc{TestHumanizer}, a hybrid SBST+LLM approach that treats LLMs as controlled \emph{refactoring layers} over compilable SBST suites, 
improving naming, structure, and developer-oriented clarity while preserving test behavior and compilation validity.
We evaluate \textsc{TestHumanizer} on 350 classes from Defects4J (147) and SF110 (203). EvoSuite generates 15 suites per class, each refactored under three context configurations (tests-only, code-centric, summary-based) using two general-purpose LLMs, yielding 31{,}500 refactorings. \textsc{TestHumanizer} achieves 
compilation rates of 88--98\%, approaching EvoSuite's 100\% compilation baseline and clearly outperforming direct LLM generation in both compilability and coverage preservation. Structural coverage is largely preserved (typically within 1--2 percentage points, with median method coverage remaining at 100\%), and 86--95\% of refactorings remain above a composite faithful-refactoring threshold. Refactored suites exhibit improved predicted readability, reduced control-flow and cognitive complexity, and substantial mitigation of 
structural smells (e.g., Conditional Logic Test reduced from 36.97\% to 5.02\% on Defects4J and from 48.31\% to 4.37\% on SF110 in the summary-based setting). The summary-based configuration provides the 
most robust trade-off, whereas long code-centric prompts are more prone to hallucination-induced failures and API-level rejections.
A developer study on 30 classes (444 test methods) confirms significant gains in perceived readability and willingness to adopt (Wilcoxon, $p<0.01$, medium-to-large effects), with substantial inter-rater agreement ($\alpha=0.71$--$0.85$). Overall, LLMs are most effective not as standalone generators but as validation-gated 
refinement layers over robust SBST outputs, motivating agentic and retrieval-augmented refactoring pipelines and evaluation protocols that go beyond surface similarity.
\end{abstract}



\maketitle

\section{Introduction}
\label{sec:introduction}

Unit tests are essential for detecting defects early, documenting expected behavior, and supporting safe software 
evolution~\cite{beck2000extreme,shore2021art,siddiqui2021learning}. Over the last decade, automated test generation has reduced the manual effort of building unit test suites while achieving high structural coverage~\cite{fraser2011evosuite,panichella2017automated}. Among these tools, ~\cite{fraser2011evosuite} has emerged as a state-of-the-art Search-Based Software Testing (SBST) framework and a coverage-oriented baseline 
validated by large-scale studies~\cite{fraser2014large,tang2024chatgpt,ouedraogo2024large,ouedraogo2025enriching}. However, despite strong coverage, EvoSuite-generated tests are rarely adopted in practice because developers perceive them as difficult to 
read, understand, and maintain~\cite{almasi2017industrial,panichella2020revisiting,ouedraogo2024large}. 
This usability gap is largely explained by well-documented readability and maintainability issues in SBST suites---cryptic naming, implementation-driven assertions, and high densities of test smells (e.g., Assertion Roulette, General Fixture, Redundant Assertions)---whose profiles differ markedly from human-written tests 
and inflate maintenance effort~\cite{bacchelli2008effectiveness,shamshiri2015automatically,daka2015modeling,bavota2012empirical,ouedraogo2024test}. 
Optimized for coverage rather than domain intent~\cite{fraser2011evosuite}, these suites can obscure scenario semantics, hinder debugging and evolution, and are often rejected or rewritten, limiting their value as executable documentation and their industrial 
uptake~\cite{shamshiri2015automatically,almasi2017industrial,bacchelli2008effectiveness}.

Large Language Models (LLMs) offer new opportunities for improving test quality~\cite{tang2024chatgpt,wang2024software,siddiq2024using}. However, empirical evaluations reveal recurring trade-offs: although LLM-generated tests often improve readability, they may fail to compile, hallucinate APIs, or achieve lower coverage in practice~\cite{ouedraogo2024large}, and exhibit distinct smell profiles under minimal prompting~\cite{ouedraogo2024test}. Recent work has started to combine SBST and LLM strengths in hybrid pipelines: some approaches use LLMs to enrich test inputs or guide search~\cite{ouedraogo2025enriching,lemieux2023codamosa}, others improve input consistency~\cite{zhou2024llm} or apply post-processing transformations such as renaming and commenting~\cite{gay2023improving,deljouyi2024leveraging,biagiola2025improving}. Yet these approaches address only a fraction of the coverage--usability gap: input-level and search-guidance methods do not target test readability, while post-processing approaches either restrict transformations to identifier renaming or lack a systematic study of how contextual information affects suite-level refactoring quality, test smells, and developer acceptance. This leaves open the question of whether LLMs can be used as principled, context-aware refactoring layers that improve the full structure of SBST-generated suites without compromising their behavioral correctness.

\textbf{In this paper}, we investigate whether LLMs can systematically \emph{refactor} EvoSuite-generated test suites into 
suites that developers find easier to read, maintain, and reuse, without altering their behavior. We propose \emph{TestHumanizer}, an LLM-based framework that refactors EvoSuite-generated test suites to improve developer comprehension and maintainability without changing their behavior. Unlike approaches that focus mainly on renaming identifiers, TestHumanizer operates at the level of entire test suites, preserving EvoSuite-specific structural elements (packages, imports, annotations, and class declarations) while targeting increased readability, maintainability, and modularity through structural changes such as \emph{Given--When--Then} documentation, grouping of related logic, and descriptive renaming. We examine how the form and scope of contextual information influence refactoring quality, and instantiate three configurations: (i)~\emph{tests-only}, providing only the EvoSuite suite; (ii)~\emph{code-centric}, adding the full source code of the class under test; and (iii)~\emph{summary-based}, providing tests plus an automatically generated natural-language summary of the class---a compact 
alternative that mitigates the ``lost-in-the-middle'' effect inherent to long-context prompting~\cite{liu2024lost,hsieh2024found}. We evaluate \textsc{TestHumanizer} both quantitatively, using metrics that measure refactoring quality across feasibility, semantic preservation, coverage, and structural smells, and qualitatively, via a human study in which professional developers compare EvoSuite tests with TestHumanizer-refactored versions on readability, understandability, and adoption willingness.

The main contributions of this work are as follows:
\begin{enumerate}
    \item \textbf{TestHumanizer:} An LLM-based framework that refactors SBST-generated test suites (e.g., EvoSuite) to improve 
    readability and maintainability while preserving test behavior and coverage.

    \item \textbf{Contextual Refactoring Study:} A systematic comparison of three context configurations (tests-only, code-centric, summary-based) to quantify how the form and scope of contextual information affect LLM-driven refactoring quality.

    \item \textbf{Large-Scale Empirical Evaluation:} An extensive study over 350 classes (147 Defects4J, 203 SF110), generating 31{,}500 refactored suites using two LLMs, and assessing readability, understandability, coverage, and test smells.

    \item \textbf{Human Evaluation:} A developer study involving four evaluators who compare EvoSuite-generated suites with their TestHumanizer-refactored counterparts over 30 classes (444 test methods), rating perceived readability, clarity of intent, and willingness to adopt the tests in real-world scenarios.

    \item \textbf{Replication Package:} All relevant artifacts are made publicly available to ensure transparency and enable 
    replication: \url{https://anonymous.4open.science/r/TestHumanizer-06FB/}
\end{enumerate}

The rest of the paper is organized as follows. Section~\ref{sec:background} presents background and related work on 
automated test generation, test smells, and LLM-based code transformation. Section~\ref{sec:setup} describes the TestHumanizer 
framework, including our prompting strategies, context configurations, study design, datasets, and evaluation metrics. 
Section~\ref{sec:results} reports our empirical and human-study results. Section~\ref{sec:discussion} discusses implications and threats to validity. Section~\ref{sec:relatedwork} reviews prior work, and Section~\ref{sec:conclusion} concludes with directions for future research.

\section{Background and Related Work}
\label{sec:background}

\subsection{From SBST Tools to LLM-Driven Tests}
\label{subsec:analysis-llms}

Automated unit test generation has traditionally relied on SBST tools such as EvoSuite~\cite{fraser2011evosuite} and Pynguin~\cite{lukasczyk2022pynguin}, which optimize for high code 
coverage but often produce tests that lack readability and maintainability~\cite{almasi2017industrial,grano2018empirical,roy2020deeptc,daka2015modeling,panichella2020revisiting}. In contrast, LLMs generate more human-readable tests by leveraging natural language 
and code priors~\cite{tang2024chatgpt,wang2024software,siddiq2024using,
bhatia2024unit,ouedraogo2024llms,yuan2023no,ouedraogo2024large}. However, LLM-generated tests generally achieve lower coverage than SBST tools and exhibit higher rates of compilation failures, 
hallucinated APIs, and brittle assertions~\cite{tang2024chatgpt,siddiq2024using,ouedraogo2024large}. 
LLMs also face token limitations that make it difficult to handle large classes or complex dependencies, whereas SBST tools can process entire projects or JAR files in a uniform way.

To address these trade-offs, recent work has started to combine SBST, static analysis, and LLMs. CodaMOSA~\cite{lemieux2023codamosa} uses LLMs to escape EvoSuite-like coverage plateaus, while Gay 
et al.~\cite{gay2023improving} apply GPT-4 to improve readability by refining test names and comments. Zhou et al.~\cite{zhou2024llm} introduce the C3 criterion to align LLM-generated test inputs with 
source-code usage patterns, and UTGen~\cite{deljouyi2024leveraging} refines SBST-generated tests but remains sensitive to incomplete context. ASTER~\cite{pan2025aster} pushes this line further by guiding 
LLM-based test generation with lightweight static analysis and iterative repair: it achieves coverage competitive with, or superior to, EvoSuite and CodaMOSA on Java SE, Java EE, and Python projects, 
while producing substantially more ``natural'' tests that 161 professional developers often consider ready to add to regression suites with minor or no changes. These studies show that SBST and LLMs 
excel at complementary aspects---coverage and robustness on one side, readability and naturalness on the other---and motivate approaches that combine their strengths rather than using LLMs as drop-in replacements.

Beyond test-specific refactoring, LLM-based agentic frameworks have recently emerged for general-purpose code refactoring. RefAgent~\cite{oueslati2025refagent} orchestrates specialized planner, 
execution, compilation, and testing agents in a closed feedback loop, achieving substantial code smell reduction and high unit-test pass rates while autonomously adapting to evolving project context. This 
emerging line of work motivates exploring agentic extensions of suite-level test refactoring, a direction we revisit in  Section~\ref{subsec:agentic-rag}.

\subsection{Readability vs.\ Understandability}
\label{subsec:readability-vs-understandability}

Readability and understandability are related but distinct dimensions of test quality. \emph{Readability} concerns how easily developers can visually parse and skim code, influenced by layout, naming, and 
comments~\cite{oliveira2022systematic,buse2009learning}. \emph{Understandability} captures the cognitive effort required to grasp a test's intent, control flow, and assertion logic~\cite{grano2018empirical}.

Buse and Weimer~\cite{buse2009learning} introduced a learned readability model later adapted to unit 
tests~\cite{daka2015modeling}, and Scalabrino et al.~\cite{scalabrino2018comprehensive} proposed a comprehensive Java readability model combining structural and lexical features. Recent work shows that modern LLMs can approximate or surpass such models in human alignment~\cite{sergeyuk2024reassessing,ouedraogo2025human}. In this study, we adopt Scalabrino's model as a stable and reproducible large-scale baseline, and complement it with structural metrics and a 
human evaluation rather than relying on LLMs both for refactoring and for judging readability.

Understandability is commonly approximated via structural metrics such as Cyclomatic Complexity~\cite{mccabe1976complexity}. However, generic cognitive complexity measures were designed for production code and overlook test-specific constructs~\cite{campbell2018cognitive,ouedraogo2025rethinking}. 
Structured testing patterns such as \emph{Arrange--Act--Assert} (AAA) and its Behavior-Driven Development (BDD) variant \emph{Given--When--Then} (GWT) address this gap: AAA organizes each test into setup, execution, and verification phases, while GWT expresses the same structure in natural-language terms that are accessible to both technical and non-technical stakeholders. Both patterns help developers quickly locate setup, action, and expected outcome within a test, reducing cognitive load when reading test suites. To better reflect the cognitive structure of tests, we combine Cyclomatic Complexity~\cite{mccabe1976complexity} with CCTR, a test-aware cognitive complexity metric tailored to unit tests~\cite{ouedraogo2025rethinking}, which accounts for assertion density, annotations, and test structuring patterns.

\subsection{Test Smells}
\label{subsec:test-smells}

Test smells, introduced by Van Deursen et al.~\cite{van2001refactoring}, denote recurring design flaws in test code that hinder readability and maintainability, analogous to code smells in production systems. Common examples include \emph{Assertion Roulette}, \emph{General Fixture}, \emph{Eager Test}, and \emph{Magic Number Test}. Empirical studies show that such smells negatively affect comprehension and maintenance effort, and that high coverage alone does not guarantee test quality~\cite{bavota2012empirical,bavota2015test}.

Automated test generation is particularly prone to smell diffusion. Prior work reports that EvoSuite-generated suites frequently exhibit smells such as Assertion Roulette and Eager Test~\cite{panichella2020revisiting,palomba2016diffusion}, with large-scale studies confirming substantial smell prevalence across Defects4J and SF110~\cite{ouedraogo2024large}. LLM-generated tests exhibit a different but equally non-trivial smell profile, often introducing patterns such as General Fixture, Duplicate Assert, or Lazy Test, with prevalence varying across prompts and context configurations~\cite{ouedraogo2024test}. Overall, both SBST- and LLM-based approaches produce smell-heavy suites, albeit with distinct distributions. This motivates refactoring strategies that target structural clarity and maintainability at the suite level and may indirectly reshape smell profiles as a by-product of improved design.

\subsection{Code Summaries as LLM Context}
\label{subsec:code-summarizing}

Source code summarization aims to produce concise natural-language descriptions of code elements to support comprehension and maintenance~\cite{iyer2016summarizing,ahmed2022few,sun2024source}. While early approaches relied on neural sequence models~\cite{iyer2016summarizing}, recent work leverages LLMs that generate context-aware summaries in zero- or few-shot settings~\cite{chen2021evaluating,ahmed2024automatic}. Summary quality 
is commonly evaluated using lexical and semantic similarity metrics, as well as code-aware measures such as SIDE~\cite{mastropaolo2024evaluating}.

In this study, we use SIDE as a code-aware alignment signal to select summaries that better reflect the functionality of the corresponding classes. These summaries serve as compact, semantics-aware context in 
our \emph{summary-based} configuration: instead of providing the full class under test, we supply a condensed behavioral description. This design mitigates known limitations of long-context prompting---such as 
the ``lost-in-the-middle'' effect and position-dependent performance degradation~\cite{liu2024lost,hsieh2024found}---while preserving essential semantics. By grounding refactoring in high-quality summaries rather than raw full-class code, we aim to improve 
robustness and focus the LLM on the most relevant behavioral aspects.

\section{Experimental Setup}
\label{sec:setup}

\subsection{Approach Overview}
\label{subsec:approach}

\begin{figure*}[]
    \centering
    \includegraphics[width=\textwidth]{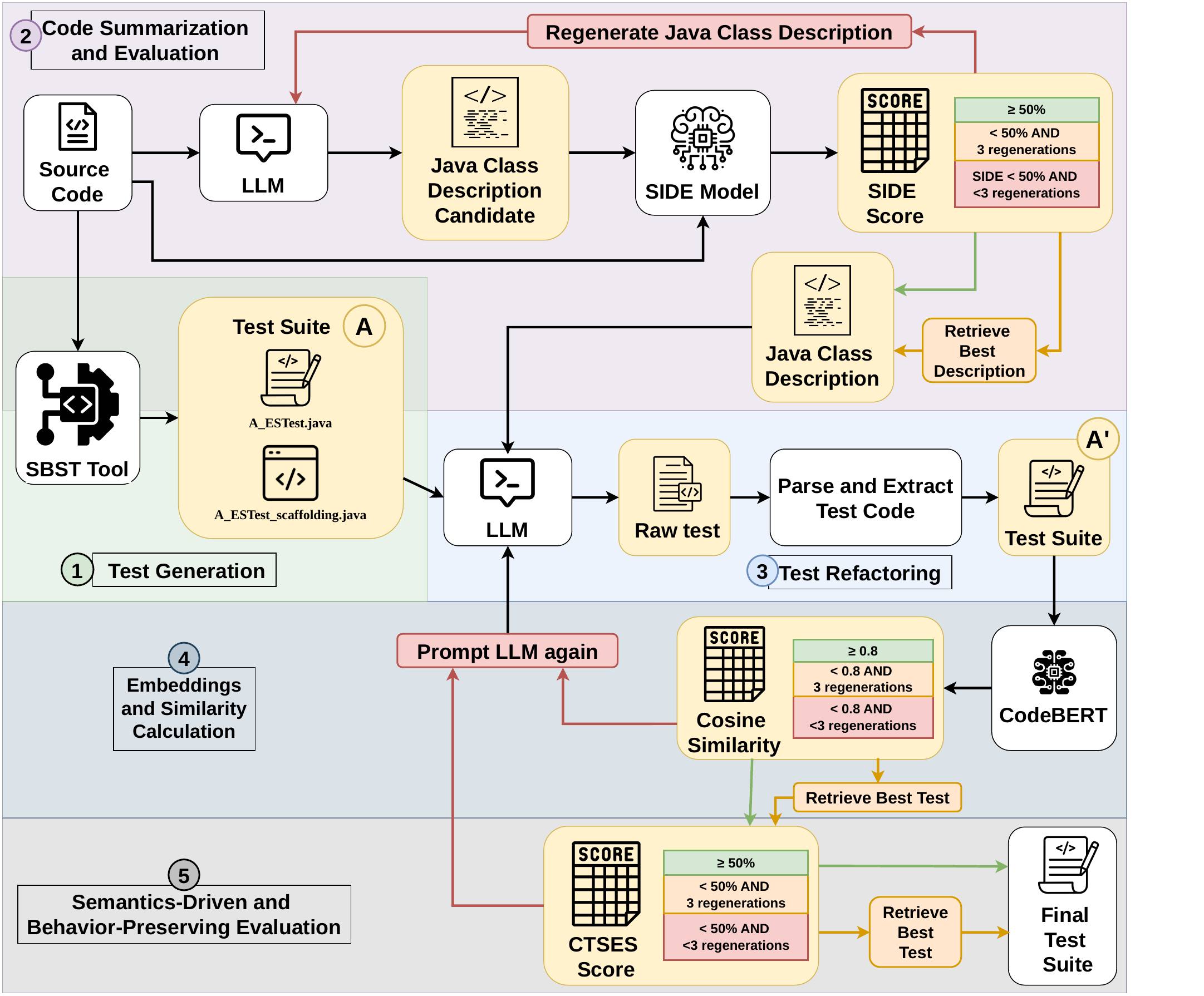}
    \caption{Overview of the general workflow of TestHumanizer.}
    \label{fig:approach-overview}
\end{figure*}

TestHumanizer is designed as a post-processing layer on top of SBST tools such as EvoSuite, incorporating validation gates at each step to ensure that refactored suites remain syntactically correct and semantically faithful. Given an automatically generated test suite for a target class $A$, it produces a refactored suite $A'$ that aims to be more readable and maintainable while preserving behavior 
(Figure~\ref{fig:approach-overview}).
The EvoSuite-generated suite (\texttt{A\_ESTest.java}) and its companion scaffolding file (\texttt{A\_ESTest\_scaffolding.java}), which contains EvoSuite-specific JUnit runners and configuration 
annotations, serve as input to TestHumanizer. The pipeline then proceeds as follows.

\ding{182}~\textbf{Parse and Extract.} TestHumanizer parses \texttt{A\_ESTest.java} and extracts two components: the \emph{immutable static elements} (package and import declarations, annotations, class declaration) that must remain unchanged throughout refactoring, and the \emph{refactorable test body} containing the individual test methods.
\ding{183}~\textbf{Code Summarization} (summary-based configuration only). The class $A$ is summarized by the same LLM used for refactoring. We compute SIDE~\cite{mastropaolo2024evaluating} between the candidate summary and the class source code as a code-aware alignment signal; if SIDE$<50$, we regenerate up to three times and 
retain the highest-scoring candidate, preferring one with SIDE$\ge50$ when available. This yields a compact and semantically aligned description of $A$.
\ding{184}~\textbf{Test Refactoring.} We build a refactoring prompt that includes the immutable static elements and the original test body, and supply one of three context granularities: the tests alone (\emph{tests-only}), the tests with the full class source code (\emph{code-centric}), or the tests with the class summary (\emph{summary-based}). The LLM is instructed to refactor the entire suite for readability, maintainability, and modularity---including \emph{Given--When--Then} structuring, descriptive renaming, and 
grouping of related logic---while strictly preserving the static elements and observable test behavior.
\ding{185}~\textbf{Embedding-Based Similarity Check.} To discourage drastic semantic drift, we compute cosine similarity between embeddings of $A$ and $A'$ using CodeBERT. If similarity$<0.8$, we retry up to three times and retain the candidate with the highest similarity, even if it remains below the threshold.
\ding{186}~\textbf{Composite Semantic Screening.} We compute CTSES~\cite{ouedraogo2025beyond} as a composite semantic screening signal integrating CodeBLEU, METEOR, and ROUGE-L. If CTSES$<50$, 
we similarly retry up to three times and retain the best-scoring candidate.

Overall, TestHumanizer couples SBST robustness with LLM-driven suite-level refactoring. The SIDE-guided summarization, embedding-based similarity check, and CTSES screening together form a multi-layered validation pipeline that balances developer-oriented improvements with behavioral faithfulness.

\subsection{Research Questions}
\label{subsec:research-questions}

\vspace{0.2cm}
\noindent\textbf{RQ1 (Feasibility and robustness): How feasible and robust is TestHumanizer in producing syntactically correct and compilable refactored unit test suites?}
This research question assesses the viability of TestHumanizer as an LLM-based refactoring layer on top of EvoSuite. We distinguish \ding{182}~\emph{feasibility}, by measuring the proportion of EvoSuite test suites that can be successfully refactored without syntax or compilation errors across Defects4J and SF110, and 
\ding{183}~\emph{robustness}, by analyzing how compilability varies across LLMs and context configurations. The goal is to determine whether TestHumanizer can reliably produce behavior-preserving 
refactorings at scale, rather than fragile or ad hoc transformations.

\vspace{0.2cm}
\noindent\textbf{RQ2 (Readability and understandability): To what extent does TestHumanizer improve the readability and understandability of EvoSuite-generated test suites?}
This research question examines the developer-perceived quality of refactored tests along two dimensions. We quantify \ding{182}~\emph{readability} using the machine-learned readability model of Scalabrino et al.~\cite{scalabrino2018comprehensive}, previously validated as a strong proxy for developer-perceived readability, and \ding{183}~\emph{understandability} using two complementary structural indicators: Cyclomatic Complexity to capture control-flow complexity, and CCTR, a test-aware cognitive complexity metric specifically designed for unit tests~\cite{ouedraogo2025rethinking}. By comparing EvoSuite tests with 
their TestHumanizer counterparts, we investigate whether LLM-based refactoring yields test suites that are easier to read and interpret.

\vspace{0.2cm}
\noindent\textbf{RQ3 (Structural and semantic similarity): To what extent do TestHumanizer refactorings remain structurally and semantically close to the original EvoSuite tests?}
Since TestHumanizer is intended to improve readability without altering test intent, this research question examines how closely refactored tests align with their originals at the lexical, structural, and semantic levels. We combine \ding{182}~\emph{lexical and structural similarity} metrics (CodeBLEU, METEOR, and ROUGE-L) with \ding{183}~\emph{embedding-based semantic similarity} using CodeBERT, GraphCodeBERT, and OpenAI's \texttt{text-embedding-3-small}, and \ding{184}~the composite CTSES score~\cite{ouedraogo2025beyond}, which integrates CodeBLEU, METEOR, and ROUGE-L into a human-aligned composite signal specifically designed to evaluate LLM-based test refactorings. Together, these metrics allow us to assess whether structural improvements come at the cost of semantic drift. Note that behavioral preservation in terms of code coverage is examined separately in RQ4, which provides a dynamic, execution-based complement to the static similarity analysis conducted here.

\vspace{0.2cm}
\noindent\textbf{RQ4 (Code coverage impact): How does TestHumanizer impact the structural coverage of the original EvoSuite-generated tests?}
While refactoring should ideally be behavior-preserving, it may inadvertently affect the exercised paths or assertions. This research question compares structural coverage (line, instruction, and method coverage) achieved by EvoSuite suites before and after refactoring with TestHumanizer, across all context configurations. We examine whether LLM-based refactoring maintains, improves, or degrades coverage, and how coverage changes relate to the readability and understandability improvements observed in RQ2.

\vspace{0.2cm}
\noindent\textbf{RQ5 (Test smells): Does LLM-based refactoring reduce test smells in EvoSuite-generated test suites?}
Building on prior evidence that EvoSuite tests often exhibit smells such as Assertion Roulette, General Fixture, and Redundant Assertions, this research question investigates whether TestHumanizer effectively mitigates these issues. We measure \ding{182}~\emph{overall smell prevalence} and \ding{183}~\emph{per-smell trends} (e.g., reduction, preservation, or introduction of specific smell types) before and after refactoring, using a state-of-the-art smell detector. The results indicate whether TestHumanizer not only restructures tests syntactically but also improves their design quality.

\vspace{0.2cm}
\noindent\textbf{RQ6 (Human perception and practical usability): How do developers perceive TestHumanizer-refactored tests compared to original EvoSuite tests in terms of readability and willingness to adopt them in practice?} This research question is addressed through a developer study where participants compare EvoSuite and TestHumanizer versions of the same test suites, rating \ding{182}~\emph{perceived readability and clarity of intent} and \ding{183}~\emph{willingness to adopt the tests in real projects}. These ratings complement RQ1--RQ5 by grounding our findings in human perception and practical usability.

\subsection{Large Language Models and Prompt design}
\label{subsec:llm-prompts}

\paragraph{Models and decoding settings.}

We use \texttt{gpt-4o} as our primary LLM across the full evaluation pipeline (class summarization, suite refactoring, and direct test generation baselines). For RQ1 (feasibility and compilability), we additionally include \texttt{mistral-large-2407} to contrast provider behavior under the same prompting and pipeline settings. We set temperature to 0.1 and keep the provider-default \texttt{top\_p} to favor low-variance outputs while allowing limited lexical variation (e.g., naming and phrasing). This controlled setup isolates the effect of our pipeline and prompt design from decoding-specific variability.

\paragraph{Chain-of-Thought prompting.}

Our prompts follow a \emph{Chain-of-Thought} (CoT) structure that explicitly decomposes each task into intermediate steps (e.g., identify intent, analyze dependencies, refactor, and verify behavior) rather than requesting a direct output. Prior work shows that such stepwise prompting can improve multi-step reasoning and transformations~\cite{wei2022chain,wang2022self}, and it has been adopted in LLM-based test generation/refactoring~\cite{gay2023improving,deljouyi2024leveraging,gao2024context} as well as code summarization~\cite{sun2024source}. For scalability, we use single-sample CoT (no self-consistency), retaining the step-by-step structure to steer the model toward structured, behavior-preserving edits.

\paragraph{Class summarization prompt.}
For the summary-based configuration of TestHumanizer, we generate a natural-language summary for each class under test. The summarization prompt is inspired by Sun et al.~\cite{sun2024source} and instructs the model to progressively analyze the class (overall purpose, key methods and fields, relationships between methods, and section-wise descriptions). We report the full template below; placeholders such as \texttt{\{class\_code\}} are instantiated at runtime.

{\scriptsize
\begin{tcolorbox}[colback=gray!5, colframe=gray!40, title=\textbf{Prompt: Class summarization for TestHumanizer},left=2pt, right=2pt, top=2pt, bottom=2pt, boxrule=0.4pt, breakable]
You are an expert software engineer with advanced knowledge of Java programming. Your task is to summarize the following Java class, focusing on its functionality, the relationships between methods, and the roles of key variables. Approach the task in a structured way, addressing each element step-by-step.

1. \textbf{Overall Purpose}: Briefly explain the primary function of the class, including its use case within a larger project if identifiable.

2. \textbf{Key Elements Identification}: Identify and describe the purpose of:
   - Main methods: Explain the role of each significant method, focusing on its purpose, input parameters, and return type.
   - Important variables: Highlight any essential fields, their roles, and how they interact within the methods.

3. \textbf{Method Relationships}: For methods that are closely related (e.g., helper methods, overloaded methods), group and describe their relationship to provide context on how they collectively support the class’s functionality.

4. \textbf{Summarize by Sections}:
   - For each logical section of the class, generate a concise, human-readable summary in 2--3 sentences, ensuring clarity on how to use the class effectively.
   - If the class contains distinct groups of related methods, summarize them together to clarify their combined role in achieving a particular function.

Each step is designed to build toward a coherent and context-rich explanation of the class’s functionality.

Here is the class to summarize:
\{class\_code\}

Generate the summary in 2--3 sentences for each method or logical section of the class. If the class contains multiple related methods, group them in the summary to explain their relationship. Please ensure the complete summary is placed between triple backticks ``` ``` for easy extraction.
\end{tcolorbox}
}

These summaries serve as compact, semantics-aware context for the \emph{summary-based} refactoring scenario (Section~\ref{subsec:code-summarizing}), mitigating token-limit and ``lost-in-the-middle'' issues compared to prompting on the full class code.

\paragraph{Refactoring prompts for TestHumanizer.}

TestHumanizer uses three closely related CoT refactoring prompts that differ only in the provided context: \emph{tests-only} (EvoSuite suite only), \emph{code-centric} (suite + full class source code), and \emph{summary-based} (suite + class summary). All variants share the same core design: a Java-testing expert persona, explicit constraints to preserve EvoSuite scaffolding and observable behavior, and objectives to refactor the \emph{entire} suite for readability, maintainability, and modularity, including \emph{Given--When--Then} comments to make setup, action, and expected outcomes explicit. We report the full \emph{summary-based} prompt below; the other variants follow the same template and are available in our replication package.

{\scriptsize
\begin{tcolorbox}[colback=gray!5, colframe=gray!40, title=\textbf{Prompt: Summary-based refactoring with TestHumanizer},left=2pt, right=2pt, top=2pt, bottom=2pt, boxrule=0.4pt, breakable]
You are an expert software engineer with advanced knowledge of Java testing and refactoring. The following test suite was generated by EvoSuite and must be refactored for readability, maintainability, and modularity without altering functionality. For each class, you must refactor the entire test suite (all the test methods) and provide the complete refactored suite at the end. The class summary is provided to support contextual refactoring. Leverage this summary to refine the test suite effectively.

\textbf{Class Summary}: \{source\_code\_summary\}

Follow the steps below:

\textbf{Constraints}:

- Do Not Alter: Retain EvoSuite-specific elements (package/import statements, annotations, and class declaration): \{static\_part\}

- Preserve Functionality: Ensure no changes to the test behavior.

- Add Given-When-Then Comments: Use structured comments to clarify each test’s context and purpose.

\textbf{Steps}:

1. \textbf{Understand Test Intent}: Briefly describe the test class’s purpose, focal functions, and relevant context.

2. \textbf{Analyze Dependencies}: Identify dependencies and group related methods to assess opportunities for reducing complexity.

3. \textbf{Refine Test Methods}:
   - Rename Methods and Variables: Use descriptive names for methods and variables to reflect their role clearly.
   - Add Given-When-Then Comments: Structure comments as follows:
     - \textbf{Given}: Describe setup/preconditions.
     - \textbf{When}: Specify the action or method being tested.
     - \textbf{Then}: Summarize expected results.

4. \textbf{Verify and Review}: Ensure refactoring aligns with EvoSuite constraints and retains functional integrity.

\textbf{Test Suite to Refactor}:
\{test\_code\}

Generate the refactored test code within triple backticks ```  for easy extraction.
\end{tcolorbox}
}

The corresponding \emph{tests-only} and \emph{code-centric} prompts differ only in the context block that precedes the steps (\emph{no additional context} vs.\ \emph{full class source code}). Their full templates are included in our replication package.

\paragraph{Direct test generation prompts.}
To establish a direct LLM generation baseline for comparison with TestHumanizer refactorings, we design CoT prompts that treat the model as a professional Java tester, following the prompt design validated in our prior large-scale study on LLM-based test generation~\cite{ouedraogo2024large}. The prompt instructs the model to (i) extract and list public methods, (ii) generate basic JUnit~4 tests for each method, (iii) identify edge cases and exception scenarios from the class source code, and (iv) merge all test cases into a single \texttt{\{class\_name\}Test.java} file, returned between triple backticks. This baseline reflects a realistic one-shot LLM generation scenario and allows us to isolate the added value of the TestHumanizer refactoring pipeline over standalone LLM generation. The full prompt is available in the replication package alongside example instances.

\paragraph{On model and prompting choices.}
We rely on \texttt{gpt-4o} and \texttt{mistral-large-2407}, two general-purpose commercial LLMs, with a Chain-of-Thought prompting strategy rather than native multi-step reasoning (``thinking'') modes or agentic orchestration. This design choice reflects both the state of LLM tooling at the time of our large-scale experimental campaign (31{,}500 refactorings across two datasets and three context configurations) and the computational budget required to run such a campaign reliably and reproducibly. It is consistent with recent 
empirical studies of LLM-based refactoring, which similarly evaluate established general-purpose models under standard prompting~\cite{zhang2026empirical}.
\subsection{The SBST tool}
\label{subsec:sbst-evosuite}

For the SBST baseline, we use EvoSuite~\cite{fraser2011evosuite}, a widely adopted and empirically validated tool for automated unit test generation in Java. EvoSuite is consistently among the strongest baselines in SBST evaluations and competitions, and it is commonly used as a reference point in large-scale studies and recent SBST--LLM pipelines when benchmarking feasibility and structural coverage~\cite{jahangirova2023sbft,shamshiri2015automatically,pan2025aster,lemieux2023codamosa}. These results motivate our choice of EvoSuite as the baseline generator for TestHumanizer.
We run EvoSuite with its standard configuration, including DynaMOSA~\cite{panichella2017automated}, and allocate three minutes per class. Following prior work~\cite{fraser2015does,arcuri2013parameter,shamshiri2015automatically,tang2024chatgpt}, we execute 15 independent runs per class to mitigate Genetic Algorithm randomness; each run produces one JUnit suite, so all reported counts and rates are computed at the suite level for consistency with our LLM-based baselines (one suite per model--prompt instance).
To characterize raw SBST output and avoid confounding effects, we disable EvoSuite’s optional post-generation optimizations (e.g., minimization and clean-up). This ensures that TestHumanizer operates on unrefined suites and that comparisons reflect the underlying generation strategies rather than tool-specific post-processing.

\subsection{Datasets}
\label{subsec:datasets}

We evaluate TestHumanizer on two widely used Java benchmarks: Defects4J~\cite{just2014defects4j}, which contains real faults from open-source projects and is a standard baseline for automated testing research, and SF110~\cite{fraser2014large}, a curated collection of 110 Java projects widely used to study SBST at scale. Both datasets are commonly adopted in prior work on SBST and LLM-based testing~\cite{fraser2015does,shamshiri2015automatically,panichella2017automated,ouedraogo2024large}.
We analyze a subset of production classes selected to span diverse sizes and complexities while remaining within practical LLM context limits: 147 classes from 15 Defects4J projects and 203 classes from 69 SF110 projects. Table~\ref{tab:dataset} summarizes token and LOC statistics and the average number of methods per class, capturing the variability in class granularity that TestHumanizer must handle.

\begin{table*}[ht]
\vspace{0.5em}
\caption{Overview of the sampled classes from Defects4J and SF110.}
\label{tab:dataset}
\centering
\scalebox{0.6}
{
\begin{tabular}{ccccccccccc}
\toprule
\textbf{Dataset} & \textbf{\#Projects} & \textbf{\#Classes} & \textbf{\begin{tabular}[c]{@{}c@{}}Max\\ Tokens\end{tabular}} & \textbf{\begin{tabular}[c]{@{}c@{}}Min\\ Tokens\end{tabular}} & \textbf{\begin{tabular}[c]{@{}c@{}}Average\\ Tokens\end{tabular}} & \textbf{\begin{tabular}[c]{@{}c@{}}Max\\ Loc\end{tabular}} & \textbf{\begin{tabular}[c]{@{}c@{}}Min\\ Loc\end{tabular}} & \textbf{\begin{tabular}[c]{@{}c@{}}Average\\ Loc\end{tabular}} & \textbf{\begin{tabular}[c]{@{}c@{}}Total\\ Locs\end{tabular}} & \textbf{\begin{tabular}[c]{@{}c@{}}Average Methods\\ Per Class\end{tabular}} \\
\midrule
Defects4J & 15 & 147 & 27896 & 156 & 1958.19 & 945 & 4 & 169.40 & 373530 & 17.38 \\
SF110 & 69 & 203 & 27968 & 151 & 3462.57 & 2879 & 4 & 294.14 & 895664 & 27.82 \\
\bottomrule
\end{tabular}
}
\vspace{0.5em}
\end{table*}

\subsection{Implementation and Configuration}
\label{subsec:implementation_and_Configuration}

All experiments are executed on a dedicated server (AMD EPYC 7552, 48 cores, 640\,GB RAM) to ensure that observed timeouts and failures reflect tool or API behavior rather than resource constraints.
EvoSuite is configured as described in Section~\ref{subsec:sbst-evosuite}. For each selected class (Section~\ref{subsec:dataset}), we run EvoSuite 15 times with DynaMOSA and a three-minute budget per run, producing 15 independent suites per class.
LLM-based components are implemented as Python clients calling \texttt{gpt-4o} with a fixed decoding setup (temperature = 0.1, provider-default \texttt{top\_p}; Section~\ref{subsec:llms-prompt}). We generate direct LLM-based suites from the class source code (15 runs per class) and refactor every EvoSuite suite under the three TestHumanizer context configurations (\emph{tests-only}, \emph{code-centric}, \emph{summary-based}), yielding 15$\times$3 = 45 refactorings per class and 31{,}500 refactored suites overall across 350 classes. For RQ1 only, we additionally run the same feasibility/compilability pipeline with \texttt{mistral-large-2407} to contrast provider behavior under identical prompting and configuration.

\subsection{Metrics and Evaluation}
\label{sub:metrics}

To assess the impact of TestHumanizer on structural quality, semantic preservation, behavioral correctness, and perceived readability, we employ a comprehensive multi-dimensional evaluation protocol combining static metrics, semantic similarity, dynamic coverage analysis, smell detection, and human assessment.

\subsubsection{Cyclomatic and Cognitive Complexity}

Cyclomatic Complexity (CC)~\cite{mccabe1976complexity} estimates the number of linearly independent execution paths in a program and is defined as:
\begin{equation}
CC = E - N + 2P
\end{equation}
where $E$ is the number of edges in the control-flow graph, $N$ the number of nodes, and $P$ the number of connected components.
In our experiments, we compute cyclomatic complexity using PMD’s\footnote{\url{https://pmd.github.io/}} implementation (aligned with the McCabe definition), as commonly used in static-analysis toolchains.

\subsubsection{CCTR: Test-Aware Cognitive Complexity}

To better capture test-specific structural complexity, we use CCTR~\cite{ouedraogo2025rethinking}, a cognitive complexity metric tailored for unit tests. CCTR is defined as:
\begin{equation}
    \text{CCTR} = \alpha N + \beta A + \gamma M + \delta T
\end{equation}
where $N$ denotes control-flow nesting complexity (as in Sonar Cognitive Complexity\footnote{\url{https://www.sonarsource.com/resources/cognitive-complexity/}}), $A$ counts assertions and \texttt{fail()} statements, $M$ captures mocking-related constructs (e.g., \texttt{mock()}, \texttt{verify()}, \texttt{when()}), and $T$ represents annotation-based signaling (e.g., \texttt{@Test}, \texttt{@BeforeEach}, \texttt{@Parameterized}\allowbreak\texttt{Test}). Following the original formulation~\cite{ouedraogo2025rethinking}, we adopt uniform weights ($\alpha=\beta=\gamma=\delta=1$). Overall, CCTR complements traditional control-flow metrics by incorporating test-specific signals that affect comprehension effort.

\subsubsection{CTSES: Composite Refactoring Similarity}

To quantify semantic and structural preservation between an EvoSuite test suite and its refactored counterpart, we use CTSES~\cite{ouedraogo2025beyond}, a composite similarity score defined as:
\begin{equation}
\text{CTSES} = \alpha \cdot \text{CodeBLEU} + \beta \cdot \text{METEOR} + \gamma \cdot \text{ROUGE-L}, 
\qquad \alpha + \beta + \gamma = 1.
\end{equation}
In our experiments, we report three variants: the uniform baseline $\text{CTSES}_{\text{AVG}}$ with $(\alpha,\beta,\gamma)=(\tfrac{1}{3},\tfrac{1}{3},\tfrac{1}{3})$, as well as the two profiles proposed in~\cite{ouedraogo2025beyond}, namely $\text{CTSES}\_1$ (semantic-prioritized, $(0.5,0.3,0.2)$) and $\text{CTSES}\_2$ (readability-aware, $(0.4,0.3,0.3)$). CodeBLEU captures syntactic and data-flow similarity, METEOR provides synonym-aware lexical matching, and ROUGE-L measures structural alignment via longest common subsequence.

\subsubsection{Embedding-Based Cosine Similarity}

To further assess semantic equivalence, we compute cosine similarity between each original test suite ($A_i$) and its refactored counterpart ($B_i$) using embeddings from three models: CodeBERT, GraphCodeBERT, and OpenAI \texttt{text-embedding-3-small}.
Given embedding vectors $A_i$ and $B_i$, cosine similarity is defined as:

\begin{equation}
\text{Cosine Similarity} = \frac{\sum_{i=1}^{n} A_i B_i}{\sqrt{\sum_{i=1}^{n} A_i^2} \times \sqrt{\sum_{i=1}^{n} B_i^2}}
\end{equation}

High cosine values indicate strong semantic alignment while being robust to renaming and formatting changes.

\subsubsection{SIDE for Class Summarization Filtering}

For the \emph{summary-based} configuration, we use SIDE (Summary alIgnment to coDe sEmantics)~\cite{mastropaolo2024evaluating} to assess whether a generated class-level summary is \emph{suitable} for the code it describes, without requiring any reference description. SIDE is a contrastive-learning metric that returns an alignment score in $[-1,1]$, where higher values indicate stronger semantic agreement between the summary and the code. For each class, we generate up to three candidate summaries and compute SIDE between each candidate and the class source code. We normalize SIDE to a percentage scale and accept a summary once it reaches \textbf{50\%}; if none does, we retain the highest-scoring candidate. This best-effort selection reduces noisy context while ensuring that the summary-based pipeline always has a usable summary for downstream refactoring.

\section{Experimental Results}
\label{sec:results}

\subsection{[RQ1]:  Feasibility and compilability of TestHumanizer refactorings.}

\noindent\textbf{Experiment Design:} 

For each class sampled from Defects4J and SF110 (Section~\ref{subsec:datasets}), EvoSuite generates 15 test suites (Section~\ref{subsec:sbst-evosuite}). Each suite is refactored with TestHumanizer under three context configurations (\emph{tests-only}, \emph{code-centric}, \emph{summary-based}) using both \texttt{gpt-4o} 
and \texttt{mistral-large-2407}, yielding $15 \times 3 \times 2$ refactored suites per class. In parallel, we generate 15 direct LLM-based test suites per class and per model as LLM-only baselines (Section~\ref{subsec:llm-prompts}). For the \emph{summary-based} configuration, each class is summarized by the same LLM used for refactoring; up to three candidate summaries are generated and the one 
with the highest SIDE score is retained (Section~\ref{subsec:approach}). All LLM-produced suites are first 
parsed with Tree-sitter\footnote{\url{https://github.com/tree-sitter/tree-sitter} [Accessed: June 2026]} to ensure syntactic validity, then compiled within the corresponding project build environment. The unit of 
analysis is a single test suite file. For each dataset, model, and configuration, we report: (i) the \emph{feasibility rate} (percentage of prompts yielding syntactically valid suites), (ii) the \emph{compilation rate} (percentage of syntactically valid suites that compile successfully), and (iii) the average number of test methods per suite as a structural proxy.

\noindent\textbf{Experiment Results:}

Table~\ref{tab:synthesis-feasibility} summarizes feasibility and compilation rates across paradigms, models, and datasets (detailed statistics in Tables~\ref{tab:test-generated} and~\ref{tab:refactoring-stats}).

\begin{table*}[t]
\caption{Generation and refactoring success rates and suite complexity.}
\label{tab:synthesis-feasibility}
\centering
\scalebox{0.60}{
\begin{tabular}{ccccccc}
\toprule
\textbf{Dataset} &
\textbf{Paradigm} &
\textbf{Approach} &
\textbf{Model} &
\textbf{\begin{tabular}[c]{@{}c@{}}Feasibility\\ Rate (\%)\end{tabular}} &
\textbf{\begin{tabular}[c]{@{}c@{}}Compilation\\ Rate (\%)\end{tabular}} &
\textbf{\begin{tabular}[c]{@{}c@{}}Test Methods\\ Average\end{tabular}} \\
\midrule

\multirow{9}{*}{Defects4J}
& \multirow{3}{*}{Generation}
& \multirow{2}{*}{Raw generation}
& gpt-4o             & 99.77  & 68.48  & 8.58 \\
& 
& 
& mistral-large-2407 & 95.24  & 51.50  & 8.06 \\
&
& \multicolumn{1}{c}{EvoSuite}
& EvoSuite           & 100.00 & 100.00 & 17.38 \\
\cmidrule(lr){2-7}
& \multirow{6}{*}{Refactoring}
& \multirow{2}{*}{Tests-only}
& gpt-4o             & 85.09  & 90.66  & 14.37 \\
&
&
& mistral-large-2407 & 85.46  & 90.64  & 15.50 \\
&
& \multirow{2}{*}{Code-centric}
& gpt-4o             & 84.49  & 89.91  & 14.29 \\
&
&
& mistral-large-2407 & 83.93  & 88.52  & 15.65 \\
&
& \multirow{2}{*}{Summary-based}
& gpt-4o             & \textbf{95.00} & \textbf{97.24} & 14.16 \\
&
&
& mistral-large-2407 & 92.94  & 92.92  & 15.54 \\

\midrule

\multirow{9}{*}{SF110}
& \multirow{3}{*}{Generation}
& \multirow{2}{*}{Raw generation}
& gpt-4o             & 98.52  & 78.00  & 8.07 \\
&
&
& mistral-large-2407 & 97.87  & 77.83  & 5.52 \\
&
& \multicolumn{1}{c}{EvoSuite}
& EvoSuite           & 100.00 & 100.00 & 27.82 \\
\cmidrule(lr){2-7}
& \multirow{6}{*}{Refactoring}
& \multirow{2}{*}{Tests-only}
& gpt-4o             & 77.09  & 92.41  & 18.63 \\
&
&
& mistral-large-2407 & 76.58  & 91.95  & 18.85 \\
&
& \multirow{2}{*}{Code-centric}
& gpt-4o             & 75.58  & 90.10  & 18.32 \\
&
&
& mistral-large-2407 & 74.87  & 89.02  & 18.60 \\
&
& \multirow{2}{*}{Summary-based}
& gpt-4o             & \textbf{95.82} & \textbf{98.69} & 18.59 \\
&
&
& mistral-large-2407 & 93.16  & 96.99  & 18.77 \\

\bottomrule
\end{tabular}
}
\end{table*}

\smallskip
As expected, EvoSuite achieves 100\% feasibility and compilation on both datasets, confirming its robustness as an SBST baseline, though at the cost of comparatively large suites (17.38 test methods per class on Defects4J, 27.82 on SF110). Direct LLM-based generation, by contrast, exhibits very high syntactic feasibility but substantially lower compilability. On Defects4J, \texttt{gpt-4o} and \texttt{mistral-large-2407} produce syntactically valid tests in 99.77\% and 95.24\% of cases, yet only 68.48\% and 51.50\% compile successfully. On SF110, feasibility remains high (98.52--97.87\%) while compilation drops to 78.00\% and 77.83\%, respectively. These results confirm that one-shot LLM generation alone does not scale reliably, motivating TestHumanizer's refactoring-based approach as a more robust alternative.

TestHumanizer substantially improves compilation stability over direct LLM generation across all configurations. In the \emph{tests-only} and \emph{code-centric} settings, feasibility ranges between 75--85\% with compilation rates between 88--92\%. The \emph{summary-based} configuration consistently achieves the best results: with \texttt{gpt-4o}, feasibility reaches 95.00\% and compilation 97.24\% on Defects4J, and 95.82\% and 98.69\% on SF110. \texttt{mistral-large-2407} follows the same trend, exceeding 
92--93\% feasibility and 96--97\% compilation in the summary-based setting. Importantly, refactored suites remain smaller than EvoSuite originals (typically 14--19 methods versus 17--28), showing that structural simplification does not come at the expense of compilability.
A practical threat to feasibility in the \emph{code-centric} configuration with \texttt{gpt-4o} is repeated API-level input rejection (\texttt{Error code: 400}, ``issue with repetitive patterns 
in your prompt'', \texttt{invalid\_prompt}).\footnote{\url{https://community.openai.com/t/error-code-400-for-repetitive-prompt-patterns/627157} [Accessed: June 2026]}\footnote{\url{https://community.openai.com/t/repetitive-prompt-error-400/716240} [Accessed: June 2026]} Such rejections arise when long prompts contain strongly templated or repetitive content. Despite preprocessing, full-class prompts (class source + structured refactoring instructions) remain the longest and most repetitive inputs, making them most susceptible to this failure mode. This phenomenon was not observed with \texttt{mistral-large-2407}, suggesting provider-specific validation heuristics. Overall, the \emph{summary-based} configuration is the most robust: compressing class context into a compact semantic representation reduces API-level rejections and achieves the highest compilation stability across datasets and models.

\vspace{0.5em}
\noindent\colorbox{gray!20}{{\parbox{0.98\linewidth}{
\textbf{Finding 1:} TestHumanizer successfully refactors compilable SBST suites in the vast majority of cases, achieving compilation rates of 88--98\% across configurations and datasets---substantially higher than direct LLM generation (51--78\%). The \emph{summary-based} configuration is consistently the most feasible and robust across both LLMs and datasets. In the \emph{code-centric} setting, \texttt{gpt-4o} is additionally affected by API-level prompt rejections (HTTP~400) triggered by repetitive prompt patterns, a failure mode not observed with \texttt{mistral-large-2407}.
}}}
\vspace{0.5em}

\begin{figure*}[ht]
    \centering
    \includegraphics[width=\textwidth,scale=3]{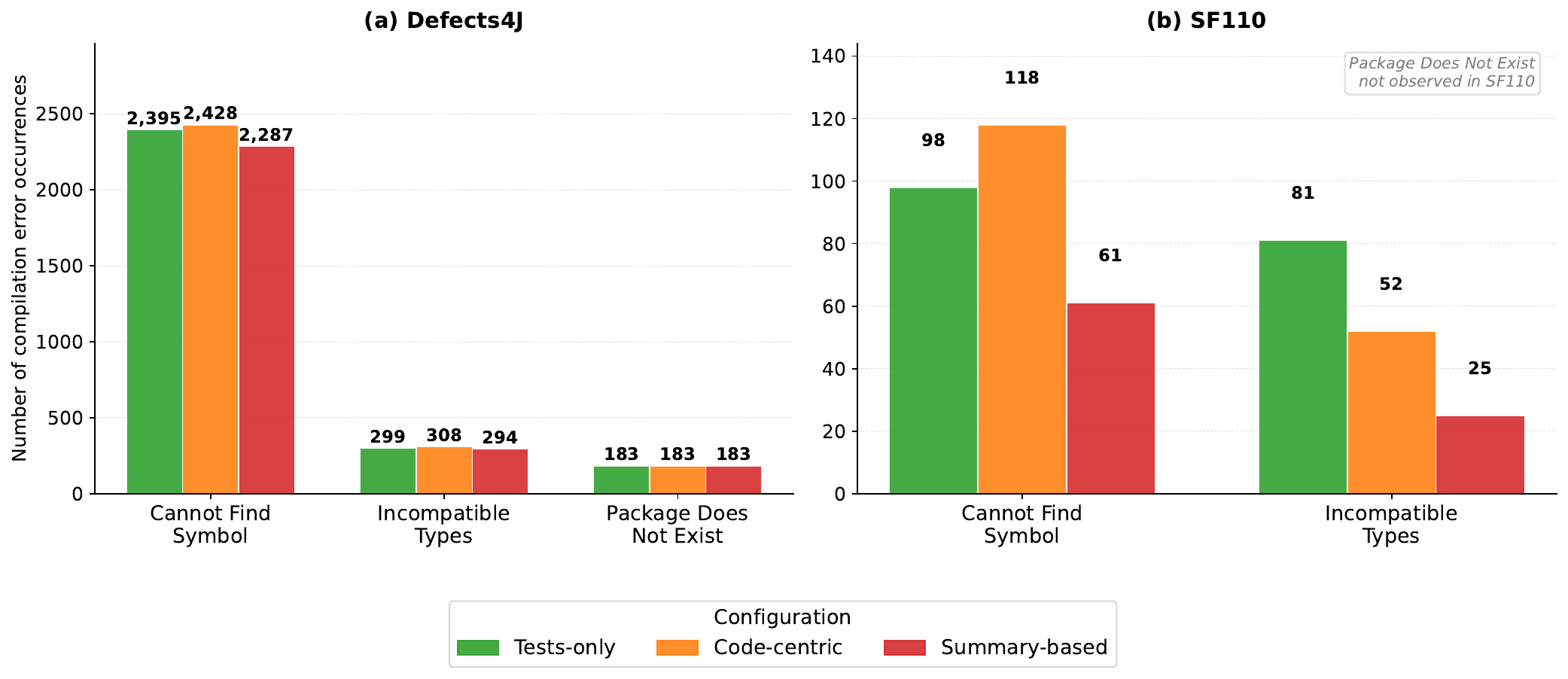}
    \caption{Compilation error breakdown across context configurations 
    (\texttt{gpt-4o}).}
    \label{fig:compilation-errors}
\end{figure*}

To characterize compilation failures, we analyzed their error categories across all non-compiling refactorings produced by \texttt{gpt-4o}, focusing on three interpretable error types that directly reflect LLM hallucination patterns: \emph{Cannot Find Symbol}, \emph{Incompatible Types}, and \emph{Package Does Not Exist}. Figure~\ref{fig:compilation-errors} reports their occurrence counts across the three context configurations.

On Defects4J, \emph{Cannot Find Symbol} is the dominant error category across all three configurations (2,287--2,428 occurrences), followed by \emph{Incompatible Types} (294--308) and \emph{Package 
Does Not Exist} (183, stable across configurations). Crucially, the \emph{summary-based} configuration consistently yields the lowest counts across all three categories, with \emph{Cannot Find Symbol} dropping from 2,428 (code-centric) to 2,287, and \emph{Incompatible Types} from 308 to 294. On SF110, the reduction is more pronounced: \emph{Cannot Find Symbol} falls from 118 (code-centric) to 61 in the 
summary-based setting, and \emph{Incompatible Types} drops from 81 (tests-only) to 25---a threefold decrease. \emph{Package Does Not Exist} was not observed in SF110.
These patterns are consistent with identifier-level LLM hallucinations: \emph{Cannot Find Symbol} arises when the LLM alters class or constructor identifiers and method calls that are present in the original suite; \emph{Incompatible Types} reflects hallucinated type signatures or return types; and \emph{Package Does Not Exist} indicates invented or incorrectly modified import statements. Importantly, providing the full class source in the \emph{code-centric} configuration does not prevent these failures---some errors involve identifiers explicitly present in the input, suggesting over-aggressive transformation rather than missing context. The consistently lower error counts of the \emph{summary-based} configuration indicate that a compact, abstract class representation implicitly constrains the LLM's transformation space, reducing opportunities for identifier-level drift.

\vspace{0.5em}
\noindent\colorbox{gray!20}{{\parbox{0.98\linewidth}{
\textbf{Finding 2:} Compilation failures are predominantly caused by identifier-level LLM hallucinations. \emph{Cannot Find Symbol} is the dominant error category on Defects4J (2,287--2,428 occurrences across configurations), while on SF110 both \emph{Cannot Find Symbol} and \emph{Incompatible Types} drop substantially in the summary-based setting (from 118 to 61 and from 81 to 25, respectively). Providing the full class context (\emph{code-centric}) does not reduce these failures; the compact \emph{summary-based} context consistently yields the fewest hallucination-induced errors across both datasets.
}}}
\vspace{0.5em}

Beyond compilability, we analyze the quality of the class summaries used in the \emph{summary-based} configuration through SIDE, a code-aware summary--code alignment metric~\cite{mastropaolo2024evaluating}. As shown in Table~\ref{tab:summary_side_stats}, \texttt{gpt-4o} achieves average SIDE scores of 0.56 on Defects4J and 0.61 on SF110, whereas \texttt{mistral-large-2407} reaches lower averages of 0.28 and 0.26, respectively. In our pipeline, each class is summarized up to three times and we retain the highest-scoring candidate, making SIDE a \emph{relative selection signal} rather than an absolute quality threshold.
SIDE was originally trained on short, Javadoc-style method summaries; our setting involves multi-sentence class-level summaries, an out-of-distribution scenario in which we observed occasional instance-level inconsistencies (e.g., similar summaries receiving divergent scores). We therefore interpret SIDE in aggregate as a lightweight alignment filter rather than a ground-truth oracle. This cautious use is consistent with recent findings by Sun et al.~\cite{sun2024source}, who report weak correlations between 
traditional lexical metrics and human judgments for LLM-generated summaries. The high feasibility and compilation rates of the summary-based configuration confirm that SIDE-guided selection provides sufficiently coherent context for reliable refactoring.

\begin{table}[t]
\caption{Summary length statistics and SIDE score distribution across 
datasets and models.}
\label{tab:summary_side_stats}
\centering
\scalebox{0.60}{
\begin{tabular}{l l r r r r r}
\hline
\textbf{Dataset} & \textbf{Model} & \textbf{Token Min} & 
\textbf{Token Max} & \textbf{SIDE Min} & \textbf{SIDE Max} & 
\textbf{SIDE Avg} \\
\hline
\multirow{2}{*}{Defects4J} 
 & gpt-4o             & 285 & 1262 & -0.504 & 0.998 & 0.563 \\
 & mistral-large-2407 & 134 & 1412 & -0.736 & 0.999 & 0.282 \\
\hline
\multirow{2}{*}{SF110} 
 & gpt-4o             & 321 & 1208 & -0.625 & 0.990 & 0.605 \\
 & mistral-large-2407 & 187 & 1964 & -0.799 & 0.983 & 0.261 \\
\hline
\end{tabular}
}
\end{table}

\vspace{0.5em}
\noindent\colorbox{gray!20}{{\parbox{0.98\linewidth}{
\textbf{Finding 3:} \texttt{gpt-4o} produces class summaries more semantically aligned with the source code than \texttt{mistral-large-2407} (SIDE avg: 0.56--0.61 vs.\ 0.26--0.28). Used as a relative selection filter rather than an absolute threshold, SIDE-guided summarization provides sufficiently coherent context for 
the summary-based configuration to remain the most stable and compilable variant of TestHumanizer across both datasets.
}}}

\vspace{0.5em}
Our RQ1 findings are consistent with prior evidence that LLM-based post-processing of generated tests can improve their quality, yet still produces a non-negligible fraction of non-compiling outputs requiring explicit safeguards. Biagiola et al.~\cite{biagiola2025improving} show that even conservative transformations focused on naming can occasionally break compilation, motivating feasibility filters in refactoring pipelines. Deljouyi et al.~\cite{deljouyi2024leveraging} report similar issues in an SBST+LLM setting and rely on systematic compilation verification with repair/rollback. Our study extends these observations by showing that broader, suite-level refactoring starting from compilable EvoSuite suites can achieve high compilation rates (88--98\%)---well above direct LLM generation (51--78\%) and approaching EvoSuite's compilability as an upper-bound reference---while also exposing practical feasibility threats specific to long, code-centric inputs (API-level prompt rejections and hallucination-induced edits). These threats further motivate compact context designs such as our summary-based configuration.

\highlight{
\textbf{Summary of RQ1:}
EvoSuite achieves 100\% feasibility and compilation on both datasets but produces comparatively large suites. Direct one-shot LLM generation remains brittle, with compilation rates as low as 51--78\%. TestHumanizer, operating as a refactoring layer over compilable EvoSuite suites, achieves substantially higher compilation stability: refactored suites compile in 88--98\% of cases across configurations and datasets, with the \emph{summary-based} variant consistently the most stable. Two main threats to feasibility are identified: (i)~API-level prompt rejections (HTTP~400) affecting long \emph{code-centric} prompts with \texttt{gpt-4o} (not observed with \texttt{mistral-large-2407}), and (ii)~hallucination-induced edits in non-compiling refactorings (e.g., unintended changes to referenced identifiers or imports), which are less frequent in the compact \emph{summary-based} setting. SIDE-guided summarization yields higher average summary--code alignment for \texttt{gpt-4o} than \texttt{mistral-large-2407}; used as a relative selection filter, it contributes to the robustness of the summary-based configuration in practice.
}

\subsection{[RQ2]: Impact of TestHumanizer on readability and understandability.}

\noindent\textbf{Experiment Design:} 

For each EvoSuite suite and its corresponding TestHumanizer-refactored variants, we assess two dimensions of test quality as defined in RQ2 (Readability and Understandability).
%
We use the machine-learned readability model of Scalabrino et al.~\cite{scalabrino2018comprehensive} as a proxy for developer-perceived readability~\cite{sergeyuk2024reassessing,ouedraogo2025human}. This model combines structural and lexical features and has been validated as a strong predictor of human readability judgments.
%
We use two complementary structural indicators as proxies for cognitive effort: (i)~Cyclomatic Complexity (CC)~\cite{mccabe1976complexity}, which captures control-flow complexity, and (ii)~CCTR, a test-aware cognitive complexity metric specifically designed for unit 
tests~\cite{ouedraogo2025rethinking}, accounting for assertion density, annotations, and test structuring patterns. Lower values on both indicators reflect suites that place less cognitive demand on 
developers.
We compute all three metrics on: (a)~original EvoSuite suites, (b)~TestHumanizer-refactored suites for each context configuration (\emph{tests-only}, \emph{code-centric}, \emph{summary-based}) using 
\texttt{gpt-4o}, and (c)~direct \texttt{gpt-4o}-generated test suites. The unit of analysis is a single test suite file. 

\noindent\textbf{Experiment Results:}

\begin{figure*}[ht]
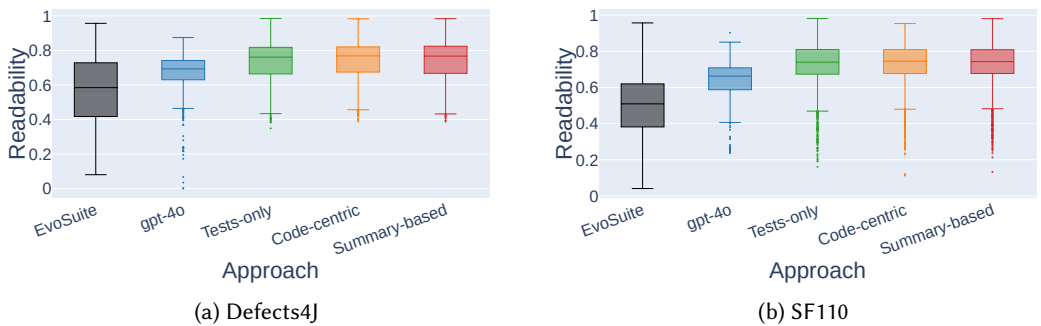

\centering

\begin{subfigure}{0.48\textwidth}
  \centering
  \includegraphics[width=\textwidth]{Images/Readability_defects4j.pdf}
  \caption{(a) Defects4J}
  \label{readability-Defects4J}
\end{subfigure}
\hfill
\begin{subfigure}{0.48\textwidth}
  \centering
  \includegraphics[width=\textwidth]{Images/Readability_sf110.pdf}
  \caption{(b) SF110}
  \label{readability-SF110}
\end{subfigure}

\caption{Readability with Scalabrino.}
\label{fig:readability}
\end{figure*}

\begin{figure*}[ht]
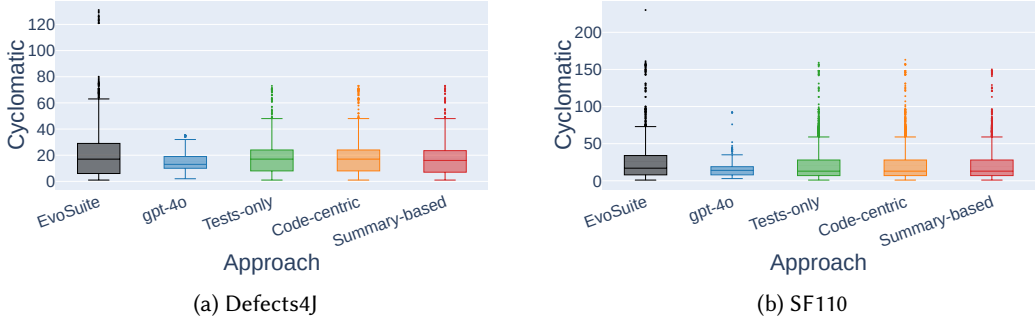

\centering

\begin{subfigure}{0.48\textwidth}
  \centering
  \includegraphics[width=\textwidth]{Images/Cyclomatic_defects4j.pdf}
  \caption{(a) Defects4J}
  \label{cyclomatic-Defects4J}
\end{subfigure}
\hfill
\begin{subfigure}{0.48\textwidth}
  \centering
  \includegraphics[width=\textwidth]{Images/Cyclomatic_sf110.pdf}
  \caption{(b) SF110}
  \label{cyclomatic-SF110}
\end{subfigure}

\caption{Cyclomatic Complexity.}
\label{fig:cyclomatic}
\end{figure*}

\begin{figure*}[ht]
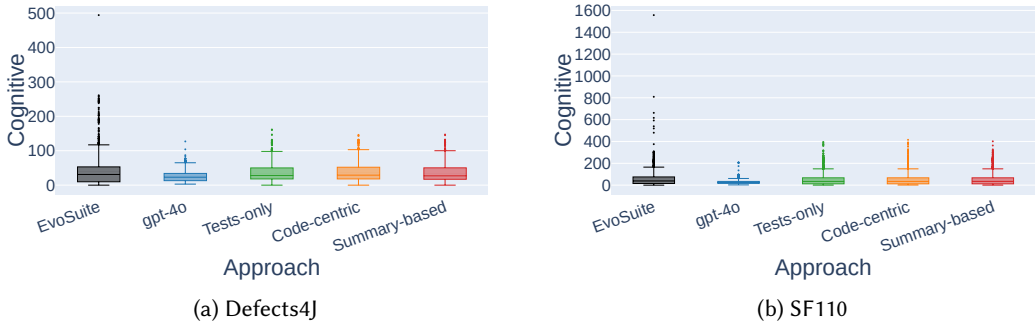

\centering

\begin{subfigure}{0.48\textwidth}
  \centering
  \includegraphics[width=\textwidth]{Images/Cognitive_defects4j.pdf}
  \caption{(a) Defects4J}
  \label{cctr-Defects4J}
\end{subfigure}
\hfill
\begin{subfigure}{0.48\textwidth}
  \centering
  \includegraphics[width=\textwidth]{Images/Cognitive_sf110.pdf}
  \caption{(b) SF110}
  \label{cctr-SF110}
\end{subfigure}

\caption{Test-Aware Cognitive Complexity.}
\label{fig:cctr}
\end{figure*}

Figures~\ref{fig:readability},~\ref{fig:cyclomatic}, and~\ref{fig:cctr} show the distributions of readability, Cyclomatic Complexity (CC), and CCTR across paradigms on Defects4J and SF110.

All three TestHumanizer configurations with \texttt{gpt-4o} outperform both EvoSuite and direct LLM-generated tests on predicted readability. On Defects4J, the mean Scalabrino score increases from 0.56 (EvoSuite) to 0.67 (direct \texttt{gpt-4o}), and further to 0.74--0.75 for TestHumanizer, with the \emph{summary-based} variant achieving the highest mean (0.75). A consistent pattern holds on SF110: EvoSuite averages 0.51, direct \texttt{gpt-4o} reaches 0.64, and all TestHumanizer configurations exceed 0.73, again with \emph{summary-based} slightly ahead. 

EvoSuite suites exhibit the highest CC and CCTR on both datasets (19.8 CC and 39.5 CCTR on Defects4J; 25.6 CC and 58.6 CCTR on SF110). Direct \texttt{gpt-4o} tests are structurally much simpler 
(CC~$\approx$~15--16, CCTR~$\approx$~26--28), reflecting shorter, less branch-heavy suites that sacrifice structural richness for simplicity. TestHumanizer refactorings strike an intermediate balance: they consistently reduce both CC and CCTR relative to EvoSuite (e.g., \emph{summary-based} CC~$\approx$~17.1 vs.\ 19.8 and CCTR~$\approx$~34.2 vs.\ 39.5 on Defects4J; CC~$\approx$~19.8 vs.\ 25.6 and CCTR~$\approx$~46.7 vs.\ 58.6 on SF110), while remaining more structurally rich than direct LLM outputs. Among the three configurations, \emph{summary-based} yields the lowest complexity on both datasets.

Taken together, these results show that TestHumanizer improves predicted readability while reducing the cognitive load of refactored suites, without collapsing into the overly minimal structure that characterizes direct LLM generation. The \emph{summary-based} configuration consistently achieves the best trade-off across both dimensions.

\vspace{0.5em}
\noindent\colorbox{gray!20}{{\parbox{0.98\linewidth}{
\textbf{Finding 4:} TestHumanizer systematically improves predicted readability (Scalabrino scores: 0.74--0.75 vs.\ 0.56 for EvoSuite on Defects4J) and reduces cognitive load (CC and CCTR) relative to EvoSuite across both datasets, while preserving more structural richness than direct LLM generation. The \emph{summary-based} configuration of \texttt{gpt-4o} achieves the best overall balance between readability gains and understandability improvements.
}}}

\vspace{0.5em}
These results complement recent work by Biagiola et al.~\cite{biagiola2025improving}, who target the readability of automatically generated tests but restrict LLM transformations to identifiers and test names. Their study shows that naming-only transformations can reach readability levels comparable to developer written tests. Our findings indicate that allowing the LLM to refactor entire test structures---bodies, control flow, and assertions---yields suites that are substantially more readable and cognitively lighter than EvoSuite originals, while retaining high compilability (RQ1) and strong semantic similarity (RQ3).

\highlight{
\textbf{Summary of RQ2:} TestHumanizer consistently improves predicted readability compared to EvoSuite across both datasets, with all three configurations increasing Scalabrino scores substantially and the 
\emph{summary-based} variant of \texttt{gpt-4o} achieving the highest mean on both Defects4J and SF110. 
On understandability, EvoSuite suites exhibit the highest CC and CCTR, while direct LLM-generated tests are the simplest but often overly minimal. TestHumanizer refactorings strike an intermediate balance: they significantly reduce control-flow and cognitive complexity relative to EvoSuite, yet remain structurally richer than direct LLM outputs. The \emph{summary-based} configuration provides the best trade-off between readability gains and understandability improvements across both datasets.
}

\subsection{[RQ3]: Structural and Semantic Similarity of TestHumanizer Refactorings}

\noindent\textbf{Experiment Design:}

To assess how closely TestHumanizer refactorings align with their original EvoSuite suites at the lexical, structural, and semantic levels, we compare each compilable refactoring with its EvoSuite counterpart using a set of complementary similarity metrics. Following RQ1, we focus on the \texttt{gpt-4o} configuration as the most stable and compilable setting. Note that behavioral preservation in terms of code execution is examined separately in RQ4 through dynamic coverage analysis; the present RQ focuses on static similarity signals.
For each EvoSuite--TestHumanizer pair across the three context configurations (\emph{tests-only}, \emph{code-centric}, \emph{summary-based}), we compute three families of similarity 
signals. First, \emph{lexical and structural similarity} via CodeBLEU~\cite{ren2020codebleu}, METEOR~\cite{banerjee2005meteor}, and ROUGE-L~\cite{lin2004rouge,lin2004automatic}, which capture token-level overlap, synonym-aware matching, and longest common subsequence alignment, respectively. Second, \emph{embedding-based semantic similarity} as cosine similarity using 
CodeBERT~\cite{feng2020codebert}, GraphCodeBERT~\cite{guo2020graphcodebert}, and OpenAI's 
\texttt{text-embedding-3-small}\footnote{\url{https://openai.com/index/new-embedding-models-and-api-updates/} [Accessed: June 2026]}, which are robust to surface-level renaming 
and reformatting. Third, \emph{CTSES}~\cite{ouedraogo2025beyond}, a composite human-aligned similarity metric that integrates CodeBLEU, METEOR, and ROUGE-L into a single score specifically designed to evaluate LLM-based test refactorings; both CTSES variants 
(semantic-prioritized CTSES\_1 and readability-aware CTSES\_2) are reported.
The unit of analysis is a single suite pair, restricted to successfully compiled outputs (RQ1). For each dataset and configuration, we report distributional statistics (min, quartiles, mean, max) and the proportion of suites above or below key thresholds 
(e.g., 0.5). These similarity results are interpreted jointly with the readability and understandability findings from RQ2, to examine whether structural improvements come at the cost of semantic drift or whether refactoring remains closely aligned with the original suite.

\noindent\textbf{Experiment Results:}

Table~\ref{tab:rq3-sim-stats-per-model-approach} reports cosine similarities between original EvoSuite suites and their TestHumanizer refactorings for \texttt{gpt-4o} across CodeBERT, GraphCodeBERT, and OpenAI embeddings. Across both datasets and all configurations, 
embedding similarities are near saturation. With CodeBERT, mean similarities range from 0.9948 to 0.9962, with third quartiles above 0.9978 and minima above 0.93--0.95. GraphCodeBERT shows a comparable pattern (means 0.9857--0.9892, Q3 above 0.9922). OpenAI's \texttt{text-embedding-3-small} yields slightly lower means ($\approx$0.925--0.928) and a small tail of lower-similarity cases (minima $\approx$0.62--0.79), yet the overall distribution remains strongly concentrated in the high-similarity region. Differences between configurations are negligible: on Defects4J the \emph{summary-based} variant attains the highest average similarity, while on SF110 \emph{code-centric} is marginally ahead, but these gaps are minor relative to the overall saturation effect. Taken together, embedding-based similarity confirms that TestHumanizer's structural and readability improvements (RQ2) do not induce substantial semantic shifts, and that code-aware embeddings alone are not discriminative enough to separate the three refactoring configurations.

\begin{table*}[ht]
\vspace{0.5em}
\caption{Cosine similarity statistics per embedding model and approach.}
\label{tab:rq3-sim-stats-per-model-approach}
\centering
\scalebox{0.60}{
\begin{tabular}{lclrrrrr}
\toprule
\textbf{Dataset} & \textbf{Approach} & \textbf{Embedding} &
\textbf{Min} & \textbf{Q1} & \textbf{Mean} & \textbf{Q3} & \textbf{Max} \\
\midrule
\multirow{9}{*}{Defects4J}
& \multirow{3}{*}{Tests-only} & CodeBERT       & 0.9496 & 0.9940 & 0.9948 & 0.9978 &  0.9999 \\
&                            & GraphCodeBERT  & 0.8909 & 0.9819 & 0.9857 & 0.9922 &  0.9999 \\
&                            & OpenAI         & 0.6191 & 0.9110 & 0.9277 & 0.9471 & 1.0000 \\
\cmidrule(lr){2-8}
& \multirow{3}{*}{Code-centric} & CodeBERT      & 0.9377 & 0.9955 & 0.9953 & 0.9982 & 0.9999 \\
&                               & GraphCodeBERT & 0.8893 & 0.9851 & 0.9871 & 0.9936 & 0.9999 \\
&                               & OpenAI        & 0.6340 & 0.9085 & 0.9251 & 0.9459 & 1.0000 \\
\cmidrule(lr){2-8}
& \multirow{3}{*}{Summary-based} & CodeBERT      & 0.9307 & 0.9977 & 0.9962 & 0.9992 & 1.0000 \\
&                                & GraphCodeBERT & 0.8903 & 0.9939 & 0.9920 & 0.9976 & 1.0000 \\
&                                & OpenAI        & 0.6250 & 0.9083 & 0.9249 & 0.9455 & 1.0000 \\
\midrule
\multirow{9}{*}{SF110}
& \multirow{3}{*}{Tests-only} & CodeBERT       & 0.9326 & 0.9943 & 0.9952 & 0.9979 & 1.0000 \\
&                            & GraphCodeBERT  & 0.9045 & 0.9853 & 0.9880 & 0.9932 & 1.0000 \\
&                            & OpenAI         & 0.7897 & 0.9123 & 0.9279 & 0.9472 & 0.9970 \\
\cmidrule(lr){2-8}
& \multirow{3}{*}{Code-centric} & CodeBERT      & 0.9457 & 0.9951 & 0.9958 & 0.9982 & 1.0000 \\
&                               & GraphCodeBERT & 0.9003 & 0.9869 & 0.9892 & 0.9944 & 1.0000 \\
&                               & OpenAI        & 0.7445 & 0.9104 & 0.9263 & 0.9463 & 0.9846 \\
\cmidrule(lr){2-8}
& \multirow{3}{*}{Summary-based} & CodeBERT      & 0.9441 & 0.9948 & 0.9957 & 0.9981 & 1.0000 \\
&                                & GraphCodeBERT & 0.8885 & 0.9861 & 0.9888 & 0.9941 & 1.0000 \\
&                                & OpenAI        & 0.7819 & 0.9094 & 0.9258 & 0.9461 & 1.0000 \\
\bottomrule
\end{tabular}
}
\vspace{0.5em}
\end{table*}

\vspace{0.5em}
\noindent\colorbox{gray!20}{{\parbox{0.98\linewidth}{
\textbf{Finding 5:} Across Defects4J and SF110, TestHumanizer refactorings for \texttt{gpt-4o} remain extremely close to the original EvoSuite suites in embedding space (CodeBERT, GraphCodeBERT, OpenAI), with cosine similarity means above 0.92 and often above 0.99. Embedding-based similarity thus indicates strong semantic preservation for all three context configurations, and is not discriminative enough on its own to distinguish between tests-only, code-centric, and summary-based refactorings.
}}}
\vspace{0.5em}

Table~\ref{tab:rq3-metrics-stats} reports lexical and structural similarity. Across both datasets, METEOR and ROUGE-L remain consistently high (means $\approx$0.68--0.70; $Q1 > 0.59$), indicating substantial overlap in structure and wording. CodeBLEU is 
more conservative (means $\approx$0.55--0.58), flagging 18\% of refactorings below 0.5 on Defects4J and up to 37\% on SF110. The composite CTSES metric~\cite{ouedraogo2025beyond}---which integrates CodeBLEU, METEOR, and ROUGE-L into a score specifically designed for 
evaluating LLM-based test refactorings---provides a more stable assessment. On Defects4J, CTSES\_2 averages $\approx$0.65, with only 4.6--6.3\% of refactorings below 0.5, a threshold defined in CTSES~\cite{ouedraogo2025beyond} to identify refactorings that maintain sufficient structural and lexical alignment with the original. On SF110, CTSES\_2 remains around 0.62--0.63, with 86--89\% of refactorings above this threshold despite the larger and more complex classes. Differences across configurations are marginal, with \emph{summary-based} slightly ahead on SF110. Together with near-saturated embedding similarities, these results indicate that TestHumanizer performs non-trivial structural improvements---renaming, reorganization, and control-flow simplification---while remaining closely aligned with the original suites in the vast majority of cases.

To better characterize the limits of similarity-based evaluation, we manually inspected refactorings that simultaneously received low scores from CodeBLEU, METEOR, and ROUGE-L---cases where all three lexical signals suggest substantial surface-level divergence. This inspection revealed that low lexical similarity does not always indicate poor refactoring quality: in some instances, the LLM performs a structurally divergent yet beneficial restructuring that preserves test intent while departing strongly from the original at the token level. As a concrete example, one refactoring of the \texttt{XPathLexer} test suite in SF110 received CodeBLEU of 0.21, METEOR of 0.08, and ROUGE-L of 0.14---all below 0.22---because the LLM radically reorganized an original suite of over 100 noisy, duplicated methods by introducing named constants, descriptive method names, and structured comments while eliminating redundancy. The resulting suite is substantially more readable, yet its surface form is too distant from the original for lexical metrics to recognize the alignment. This case illustrates a systematic limitation: even CTSES, which mitigates some of CodeBLEU's over-pessimism, can still penalize broad-scope refactorings that are beneficial, motivating 
complementary evaluation dimensions such as naming quality, documentation coverage, and smell reduction.

\begin{table*}[ht]
\caption{Metric statistics per dataset and approach.}
\label{tab:rq3-metrics-stats}
\centering
\scalebox{0.60}{
\begin{tabular}{lllrrrrrcc}
\toprule
\textbf{Dataset} & \textbf{Approach} & \textbf{Metric} & \textbf{Min} & \textbf{Q1} & \textbf{Mean} & \textbf{Q3} & \textbf{Max} & \textbf{<0.5\%} & \textbf{$\ge$0.5\%} \\
\midrule
\multirow{18}{*}{Defects4J} & \multirow{6}{*}{Tests-only} & METEOR & 0.1418 & 0.6192 & 0.6935 & 0.7588 & 1.0000 & 4.57 & 95.43 \\
 &  & ROUGE-L & 0.1291 & 0.6259 & 0.6964 & 0.7428 & 1.0000 & 2.40 & 97.60 \\
 &  & CodeBLEU & 0.1900 & 0.5139 & 0.5808 & 0.6092 & 1.0000 & 17.55 & 82.45 \\
 &  & Average\_score & 0.1827 & 0.5913 & 0.6569 & 0.6990 & 1.0000 & 4.03 & 95.97 \\
 &  & CTSES\_1 & 0.1808 & 0.5715 & 0.6377 & 0.6766 & 1.0000 & 6.07 & 93.93 \\
 &  & CTSES\_2 & 0.1834 & 0.5841 & 0.6493 & 0.6898 & 1.0000 & 4.57 & 95.43 \\
\cmidrule(lr){2-10}
 & \multirow{6}{*}{Code-centric} & METEOR & 0.1457 & 0.6226 & 0.6841 & 0.7542 & 1.0000 & 5.49 & 94.51 \\
 &  & ROUGE-L & 0.1157 & 0.6286 & 0.6910 & 0.7500 & 1.0000 & 4.45 & 95.55 \\
 &  & CodeBLEU & 0.2114 & 0.5188 & 0.5786 & 0.6249 & 1.0000 & 18.12 & 81.88 \\
 &  & Average\_score  & 0.1953 & 0.5955 & 0.6512 & 0.7048 & 1.0000 & 5.56 & 94.44 \\
 &  & CTSES\_1 & 0.1952 & 0.5796 & 0.6327 & 0.6836 & 1.0000 & 8.01 & 91.99 \\
 &  & CTSES\_2 & 0.1969 & 0.5906 & 0.6440 & 0.6959 & 1.0000 & 6.28 & 93.72 \\
\cmidrule(lr){2-10}
 & \multirow{6}{*}{Summary-based} & METEOR & 0.1693 & 0.6237 & 0.6875 & 0.7564 & 1.0000 & 5.39 & 94.61 \\
 &  & ROUGE-L & 0.1212 & 0.6338 & 0.6933 & 0.7518 & 1.0000 & 4.33 & 95.67 \\
 &  & CodeBLEU & 0.2186 & 0.5219 & 0.5846 & 0.6321 & 1.0000 & 18.01 & 81.99 \\
 &  & Average\_score  & 0.2121 & 0.6001 & 0.6551 & 0.7051 & 1.0000 & 5.58 & 94.42 \\
 &  & CTSES\_1 & 0.2098 & 0.5822 & 0.6372 & 0.6865 & 1.0000 & 7.39 & 92.61 \\
 &  & CTSES\_2 & 0.2128 & 0.5940 & 0.6481 & 0.6974 & 1.0000 & 6.07 & 93.93 \\
\midrule
\multirow{18}{*}{SF110} & \multirow{6}{*}{Tests-only} & METEOR & 0.0757 & 0.5994 & 0.6626 & 0.7488 & 0.9761 & 9.74 & 90.26 \\
 &  & ROUGE-L & 0.1056 & 0.6134 & 0.6760 & 0.7426 & 1.0000 & 6.48 & 93.52 \\
 &  & CodeBLEU & 0.1445 & 0.4753 & 0.5527 & 0.6208 & 0.9693 & 35.67 & 64.33 \\
 &  & Average\_score  & 0.1109 & 0.5674 & 0.6305 & 0.6987 & 0.9756 & 10.84 & 89.16 \\
 &  & CTSES\_1 & 0.1182 & 0.5443 & 0.6104 & 0.6784 & 0.9738 & 13.28 & 86.72 \\
 &  & CTSES\_2 & 0.1143 & 0.5588 & 0.6227 & 0.6911 & 0.9750 & 11.47 & 88.53 \\
\cmidrule(lr){2-10}
 & \multirow{6}{*}{Code-centric} & METEOR & 0.0474 & 0.5893 & 0.6551 & 0.7504 & 0.9458 & 12.00 & 88.00 \\
 &  & ROUGE-L & 0.0964 & 0.6111 & 0.6804 & 0.7603 & 0.9904 & 8.00 & 92.00 \\
 &  & CodeBLEU & 0.1380 & 0.4706 & 0.5553 & 0.6380 & 0.9300 & 37.51 & 62.49 \\
 &  & Average\_score  & 0.0979 & 0.5611 & 0.6303 & 0.7102 & 0.9516 & 12.75 & 87.25 \\
 &  & CTSES\_1 & 0.1084 & 0.5376 & 0.6103 & 0.6918 & 0.9430 & 15.85 & 84.15 \\
 &  & CTSES\_2 & 0.1031 & 0.5522 & 0.6228 & 0.7030 & 0.9487 & 13.61 & 86.39 \\
\cmidrule(lr){2-10}
 & \multirow{6}{*}{Summary-based} & METEOR & 0.0584 & 0.5943 & 0.6623 & 0.7529 & 0.9716 & 10.88 & 89.12 \\
 &  & ROUGE-L & 0.1070 & 0.6129 & 0.6838 & 0.7575 & 0.9904 & 7.65 & 92.35 \\
 &  & CodeBLEU & 0.1185 & 0.4781 & 0.5628 & 0.6398 & 0.9626 & 35.00 & 65.00 \\
 &  & Average\_score  & 0.1102 & 0.5668 & 0.6363 & 0.7101 & 0.9749 & 11.61 & 88.39 \\
 &  & CTSES\_1 & 0.1182 & 0.5434 & 0.6169 & 0.6922 & 0.9709 & 14.18 & 85.82 \\
 &  & CTSES\_2 & 0.1145 & 0.5581 & 0.6290 & 0.7032 & 0.9736 & 12.37 & 87.63 \\
\bottomrule
\end{tabular}
}
\vspace{0.5em}
\end{table*}

\vspace{0.5em}
\noindent\colorbox{gray!20}{{\parbox{0.98\linewidth}{
\textbf{Finding 6:} Across Defects4J and SF110, TestHumanizer refactorings maintain high lexical and structural similarity to the original EvoSuite suites. Using CTSES---a composite metric designed for LLM-based test refactoring evaluation---roughly 86--95\% of refactorings remain above the 0.5 alignment threshold despite non-trivial structural changes, whereas CodeBLEU alone would flag substantially more refactorings as problematic. Manual inspection further reveals that cases with simultaneously low lexical scores do not always correspond to poor refactorings: structurally divergent but beneficial restructurings can be penalized by surface-level metrics, reinforcing the need to complement similarity signals with additional quality dimensions such as naming, documentation, and smell reduction.
}}}

\vspace{0.5em}
Our analysis relates to recent work on LLM-assisted improvement of generated tests. Biagiola et al.~\cite{biagiola2025improving} restrict LLMs to renaming identifiers and test names, assessing semantic stability via cosine similarity between improved variants (OpenAI embeddings, average $\approx$0.9). Deljouyi et al.~\cite{deljouyi2024leveraging} use CodeBLEU as a safeguard: refactorings scoring below 0.5 trigger re-prompting, and those repeatedly failing are reverted to the original. In contrast, TestHumanizer applies a broader scope of transformation, allowing the LLM to modify not only identifiers and comments but also test 
bodies, control flow, and assertions, and therefore requires stricter semantic safeguards. We filter first on compilability (RQ1), then rely on a richer set of signals---METEOR, ROUGE-L, CodeBLEU, CTSES, and embedding-based similarity (CodeBERT, GraphCodeBERT, OpenAI)---to identify and discard refactorings with substantial semantic divergence. Despite this broader transformation scope, the vast majority of suites remain above the CTSES alignment threshold 
($\ge$90--95\% on Defects4J, $\ge$85--90\% on SF110; Table~\ref{tab:rq3-metrics-stats}) while achieving near-perfect code-aware cosine similarity (Table~\ref{tab:rq3-sim-stats-per-model-approach}), suggesting that large-scale structural refactoring of SBST tests is 
feasible in practice when multi-metric safeguards are in place.

\highlight{
\textbf{Summary of RQ3:} Embedding-based similarity (CodeBERT, GraphCodeBERT, OpenAI) indicates extremely strong alignment across both datasets, with cosine similarities near saturation and minimal differences between configurations. Lexical and structural metrics confirm this trend: METEOR and ROUGE-L remain consistently high, while CodeBLEU is more conservative. The composite CTSES metric provides a more stable estimate, with 86--95\% of refactorings remaining above the 0.5 alignment threshold. Manual inspection reveals that cases with simultaneously low lexical scores do not always correspond to undesirable changes: broad-scope restructurings that preserve intent but diverge strongly at the token level (e.g., the \texttt{XPathLexer} case) can be penalized by surface-level metrics. Overall, TestHumanizer maintains close alignment with the original suites in the vast majority of cases, while highlighting the limits of purely similarity-based evaluation for broad-scope refactorings.
}

\subsection{[RQ4]: Effect of TestHumanizer on code coverage.}

\noindent\textbf{Experiment Design:} 

For RQ4, we measure the structural coverage achieved by each test suite (original EvoSuite, TestHumanizer refactorings, and direct LLM-generated tests) when executed against the class under test within its project build environment---i.e., the fixed project version in which the class is defined, as provided by Defects4J and SF110. We use JaCoCo\footnote{\url{https://www.jacoco.org/jacoco/} [Accessed: June 2026]} to collect \emph{instruction}, \emph{line}, and \emph{method} coverage. As in previous RQs, we restrict the analysis to suites that compile successfully (RQ1) and focus on \texttt{gpt-4o} as the best-performing LLM identified in RQ1.
For every EvoSuite suite, we measure coverage for (i)~the original EvoSuite suite, (ii)~the three TestHumanizer variants with \texttt{gpt-4o} (\emph{tests-only}, \emph{code-centric}, \emph{summary-based}), and (iii)~the direct \texttt{gpt-4o} test suite generated from the class source code. The unit of analysis is a single test suite file. 

\noindent\textbf{Experiment Results:}       

\begin{figure*}[ht]
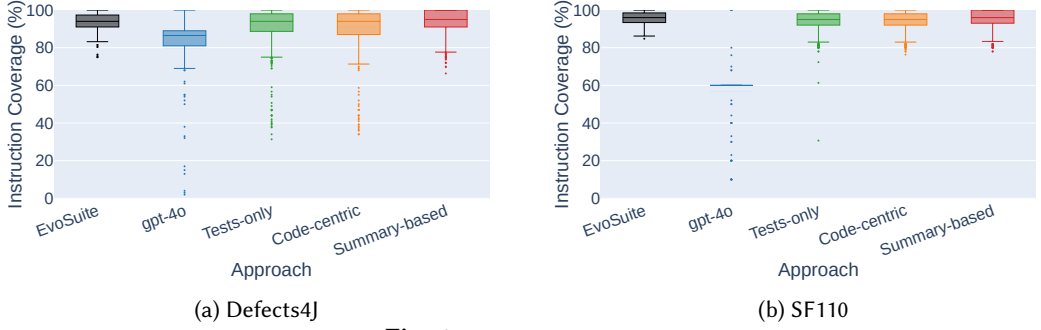

\centering

\begin{subfigure}{0.48\textwidth}
  \centering
  \includegraphics[width=\textwidth]{Images/boxplot_Defects4J_Instruction.pdf}
  \caption{(a) Defects4J}
  \label{coverage-Defects4J}
\end{subfigure}
\hfill
\begin{subfigure}{0.48\textwidth}
  \centering
  \includegraphics[width=\textwidth]{Images/boxplot_SF110_Instruction.pdf}
  \caption{(b) SF110}
  \label{coverage-SF110}
\end{subfigure}

\caption{Instruction coverage.}
\label{fig:instruction-coverage}
\end{figure*}

Figure~\ref{fig:instruction-coverage} shows the distribution of \emph{instruction} coverage. EvoSuite provides a strong SBST baseline, with tightly clustered high coverage on both Defects4J and SF110. Direct \texttt{gpt-4o} suites exhibit substantially lower and more dispersed coverage, with medians around 80\% on Defects4J and close to 60\% on SF110. In contrast, all three TestHumanizer configurations closely track EvoSuite, with the \emph{summary-based} variant showing the strongest overlap and, on Defects4J, slightly 
exceeding EvoSuite's median and upper quartile instruction coverage.

\begin{figure*}[ht]
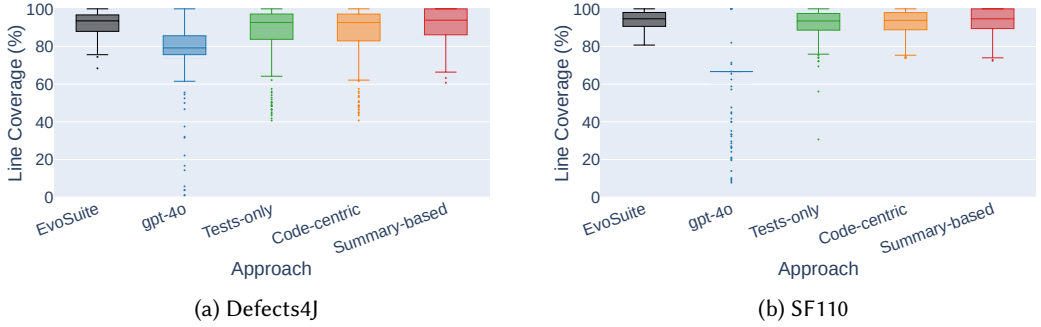

\centering

\begin{subfigure}{0.48\textwidth}
  \centering
  \includegraphics[width=\textwidth]{Images/boxplot_Defects4J_Line.pdf}
  \caption{(a) Defects4J}
  \label{coverage-Defects4J}
\end{subfigure}
\hfill
\begin{subfigure}{0.48\textwidth}
  \centering
  \includegraphics[width=\textwidth]{Images/boxplot_SF110_Line.pdf}
  \caption{(b) SF110}
  \label{coverage-SF110}
\end{subfigure}

\caption{Line coverage.}
\label{fig:line-coverage}
\end{figure*}

A similar trend emerges for \emph{line} coverage (Figure~\ref{fig:line-coverage}). The \emph{tests-only} and \emph{code-centric} variants incur only modest reductions relative 
to EvoSuite (typically a few percentage points), while the \emph{summary-based} configuration almost entirely recovers EvoSuite's line coverage, with medians above 90\% on Defects4J and 
above 93\% on SF110. Coverage regressions below this range are rare in the summary-based setting, in sharp contrast to the wide dispersion observed for direct LLM generation.

\begin{figure*}[ht]
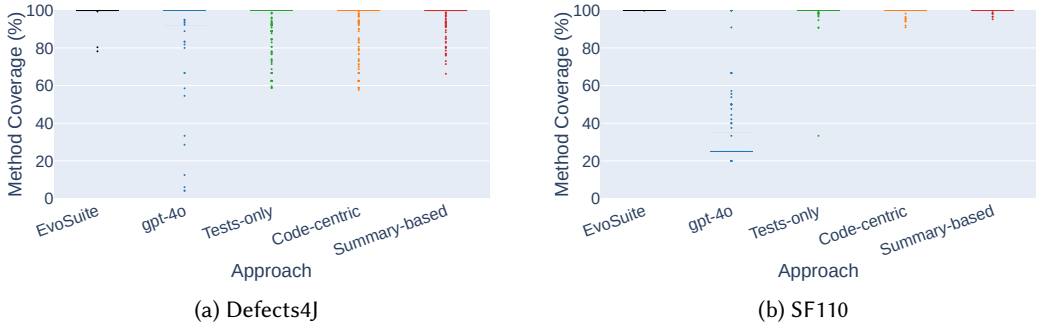

\centering

\begin{subfigure}{0.48\textwidth}
  \centering
  \includegraphics[width=\textwidth]{Images/boxplot_Defects4J_Method.pdf}
  \caption{(a) Defects4J}
  \label{coverage-Defects4J}
\end{subfigure}
\hfill
\begin{subfigure}{0.48\textwidth}
  \centering
  \includegraphics[width=\textwidth]{Images/boxplot_SF110_Method.pdf}
  \caption{(b) SF110}
  \label{coverage-SF110}
\end{subfigure}

\caption{Method coverage.}
\label{fig:method-coverage}
\end{figure*}

Figure~\ref{fig:method-coverage} reports \emph{method} coverage. EvoSuite achieves near-perfect coverage on both datasets (medians at 100\%). Direct \texttt{gpt-4o} tests cover far fewer methods, particularly on SF110 where the distribution centers well below 50\%. TestHumanizer almost entirely preserves EvoSuite's method coverage: all three configurations maintain medians at 100\% on both datasets, with mean method coverage above 99.4\% on Defects4J and 99.9\% on SF110 for the \emph{summary-based} variant (detailed statistics in 
Table~\ref{tab:rq4-coverage-stats}).

\vspace{0.5em}
\noindent\colorbox{gray!20}{{\parbox{0.98\linewidth}{
\textbf{Finding 7:} TestHumanizer preserves EvoSuite's high structural coverage across all three coverage types, with deviations typically within 1--2 percentage points and near-100\% method coverage. Direct \texttt{gpt-4o} generation loses 15--40 coverage points on average depending on the metric and dataset. The \emph{summary-based} configuration delivers the strongest coverage preservation, complementing the readability and understandability improvements observed in RQ2.
}}}

\vspace{0.5em}
Our coverage results also position TestHumanizer relative to recent LLM-based enhancements of generated tests. Biagiola et al.~\cite{biagiola2025improving} apply a strict coverage-preservation regime: transformations are limited to renaming, and any JaCoCo coverage change leads to rejection of the transformed suite. UTGen~\cite{deljouyi2024leveraging} allows richer changes (improved comments and bodies) and monitors coverage and mutation score, but uses CodeBLEU as the main safeguard without enforcing exact coverage equality. In contrast, TestHumanizer applies broader-scope refactorings---restructuring test bodies, control flow, and comments---yet the \emph{summary-based} configuration maintains instruction, line, and method coverage essentially at EvoSuite levels on both datasets. This suggests that broad-scope structural refactoring of SBST tests can preserve high coverage at least as reliably as more conservative LLM transformations, provided that strong semantic and compilability 
safeguards (RQ1--RQ3) are in place.

\highlight{
\textbf{Summary of RQ4:} EvoSuite provides a strong coverage baseline with consistently high 
instruction, line, and method coverage across both datasets. Direct \texttt{gpt-4o} generation exhibits substantially lower and more dispersed coverage, often losing 15--40 percentage points depending on the metric and dataset. TestHumanizer refactorings are largely coverage-neutral: instruction and line coverage closely track EvoSuite's distributions across all three configurations, with only marginal deviations, and method coverage remains near-perfect in the vast majority of cases. The \emph{summary-based} configuration shows the strongest alignment with EvoSuite, occasionally matching or slightly exceeding its instruction and line coverage. Overall, these results indicate that LLM-based refactoring can improve readability and understandability (RQ2) while preserving the strong coverage achieved by SBST---an advantage that one-shot LLM generation, which sacrifices substantial execution reach, does not offer. Note that LLM-based generation approaches specifically designed to maximize coverage~\cite{lemieux2023codamosa,pan2025aster} operate under a different paradigm and are not compared here; our claim applies specifically to the one-shot generation setting evaluated 
in this study.
}

\subsection{[RQ5]: Impact of TestHumanizer on test smells.}

\noindent\textbf{Experiment Design:} 

To quantify how TestHumanizer affects the smell profile of EvoSuite-generated tests, we run TsDetect~\cite{peruma2020tsdetect} on all original EvoSuite suites and their \texttt{gpt-4o} 
TestHumanizer refactorings, and include direct \texttt{gpt-4o} generation as an additional baseline. Following recent large-scale analyses~\cite{panichella2020revisiting,ouedraogo2024large,ouedraogo2024test}, we focus on smells that are particularly prevalent or characteristic in SBST- and LLM-based tests: \emph{Assertion Roulette} (AR), \emph{General Fixture} (GF), \emph{Eager Test} (EaT), \emph{Lazy Test} (LT), \emph{Duplicate Assert} (DA), \emph{Magic Number} (MG), and \emph{Dependent Test} (DpT), together with the remaining TsDetect categories.
TsDetect reports smells at the test-method level. We aggregate results by dataset (Defects4J, SF110) and by approach (EvoSuite, direct \texttt{gpt-4o}, and TestHumanizer under \emph{tests-only}, \emph{code-centric}, and \emph{summary-based}), and compute for each 
(dataset, approach, smell) tuple the percentage of tests affected. We then compare EvoSuite vs.\ TestHumanizer to assess how refactoring alters the SBST smell distribution, and TestHumanizer vs.\ direct LLM generation to determine whether refactoring existing SBST suites 
yields a more favorable smell profile than generating tests from scratch. Since our prompts do not include explicit smell-removal instructions, any observed smell reductions are interpreted as emergent effects of the structural and readability improvements targeted by TestHumanizer.

\noindent\textbf{Experiment Results:} 
Table~\ref{tab:rq5-test-smells} summarizes how the smell profile changes when moving from original EvoSuite suites to TestHumanizer refactorings and to direct \texttt{gpt-4o}-generated tests.

\begin{table*}[ht]
\caption{Percentage of tests affected by each test smell (TsDetect).}
\label{tab:rq5-test-smells}
\centering
\scalebox{0.57}{
\begin{tabular}{llcccccccccccccccccc}
\toprule
\textbf{Dataset} & \textbf{Approach} 
& \textbf{AR} & \textbf{CLT} & \textbf{CI} & \textbf{DfT}
& \textbf{ECT} & \textbf{GF} & \textbf{MG} & \textbf{PS}
& \textbf{SE} & \textbf{VT}
& \textbf{EaT} & \textbf{LT} & \textbf{DA} & \textbf{UT}
& \textbf{IT} & \textbf{RO} & \textbf{MT} & \textbf{DpT} \\
\midrule

\multirow{5}{*}{Defects4J}
& Test-only     
& 100.00 & 5.87 & 1.10 & \textbf{0.00} & 5.46 
& 95.43 & \textbf{0.00} & 4.27 & 0.69 & 17.96 
& \textbf{0.00} & 63.39 & 89.53 & 0.53 & 5.46 
& 0.09 & 3.58 & 100.00 \\

& Code-centric  
& 100.00 & 10.66 & 0.89 & \textbf{0.00} & 2.41 
& 96.11 & \textbf{0.00} & 4.19 & 0.32 & 17.23 
& \textbf{0.00} & 63.83 & 89.88 & 0.54 & 2.41 
& 0.03 & 3.55 & 100.00 \\

& Summary-based 
& \textbf{99.85} & \textbf{5.02} & 0.58 & \textbf{0.00} & \textbf{2.15} 
& \textbf{92.81} & \textbf{0.00} & 4.11 & \textbf{0.16} & \textbf{16.80} 
& \textbf{0.00} & \textbf{62.76} & \textbf{88.38} & \textbf{0.39} & \textbf{2.18} 
& \textbf{0.00} & \textbf{3.47} & 100.00 \\

& gpt-4o        
& 100.00 & 53.76 & 4.12 & 0.15 & 16.77 
& 28.91 & 3.96 & \textbf{1.93} & 0.86 & 17.58 
& 0.91 & 48.58 & 60.47 & 5.95 & 21.39 
& \textbf{0.00} & 2.29 & \textbf{99.80} \\

& EvoSuite      
& 100.00 & 36.97 & \textbf{0.00} & \textbf{0.00} & 3.94 
& 96.06 & \textbf{0.00} & 4.67 & \textbf{0.00} & 19.70 
& \textbf{0.00} & 55.41 & 80.72 & \textbf{0.11} & 3.94 
& \textbf{0.00} & 4.15 & 100.00 \\

\midrule

\multirow{5}{*}{SF110}
& Test-only     
& 100.00 & 8.11 & 1.50 & \textbf{0.00} & 0.11 
& 99.85 & \textbf{0.00} & 1.93 & 0.11 & 11.18 
& \textbf{0.00} & 29.37 & 44.04 & 4.88 & 0.11 
& 0.03 & 1.93 & 100.00 \\

& Code-centric  
& 100.00 & 11.20 & 1.38 & \textbf{0.00} & 0.04 
& 99.72 & \textbf{0.00} & 1.78 & 0.17 & 11.19 
& \textbf{0.00} & 28.73 & 43.38 & 4.52 & 0.04 
& \textbf{0.00} & 1.78 & 100.00 \\

& Summary-based 
& \textbf{99.82} & \textbf{4.37} & 1.18 & \textbf{0.00} & \textbf{0.00} 
& \textbf{99.13} & \textbf{0.00} & \textbf{1.73} & \textbf{0.11} & \textbf{11.12} 
& \textbf{0.00} & \textbf{28.69} & \textbf{43.22} & \textbf{4.27} & \textbf{0.00} 
& \textbf{0.00} & \textbf{1.73} & 100.00 \\

& gpt-4o        
& 100.00 & 41.05 & 1.51 & 0.09 & 24.49 
& 14.87 & 5.34 & 2.58 & 1.96 & 13.62 
& 0.27 & 52.00 & 46.75 & 5.16 & 32.15 
& \textbf{0.00} & 3.21 & \textbf{99.73} \\

& EvoSuite      
& 100.00 & 48.31 & \textbf{0.00} & \textbf{0.00} & 0.32 
& 99.59 & 0.05 & 2.38 & \textbf{0.00} & 9.52 
& \textbf{0.00} & 63.92 & 84.48 & 4.21 & 0.37 
& \textbf{0.00} & 2.38 & 100.00 \\

\bottomrule
\end{tabular}
}
\vspace{0.5em}
\end{table*}

On both datasets, the most pervasive EvoSuite smells remain largely unchanged after refactoring. \emph{Assertion Roulette} (AR) and \emph{Magic Number Test} (MT) still affect essentially all tests ($\approx$100\%) in every configuration, and \emph{General Fixture} 
(GF) stays very high: around 96\% for EvoSuite and 93--96\% for the three TestHumanizer variants on Defects4J, and $\approx$99\% on SF110. This is expected, as our prompts do not explicitly target assertion redesign, fixture restructuring, or test independence.

TestHumanizer does, however, substantially reduce several structural smells characteristic of SBST output. On Defects4J, \emph{Conditional Logic Test} (CLT) drops from 36.97\% in EvoSuite to 5.87--10.66\% in the \emph{tests-only} and \emph{code-centric} variants, and further to 5.02\% in the \emph{summary-based} configuration. By contrast, direct \texttt{gpt-4o} generation actually exhibits a higher CLT prevalence (53.76\%), showing that TestHumanizer 
simplifies control flow without introducing the heavy branching that often characterizes tests written from scratch by the LLM. A consistent pattern holds on SF110, where CLT falls from 48.31\% (EvoSuite) to 4.37\% (summary-based) versus 41.05\% for direct generation.

For smells related to test focus and duplication, the effect is dataset-dependent. On Defects4J, \emph{Lazy Test} (LT) and \emph{Duplicate Assert} (DA) are slightly more frequent after refactoring than in EvoSuite (LT: 55.41\% $\to$ 62.76\%; DA: 80.72\% $\to$ 88.38\% in the summary-based variant), while direct \texttt{gpt-4o} tests exhibit lower DA (60.47\%) and LT (48.58\%). On SF110, the picture reverses: TestHumanizer clearly mitigates both 
smells, with LT dropping from 63.92\% to $\approx$29\% across all configurations (vs.\ 52.00\% for direct generation) and DA from 84.48\% to $\approx$43--44\% (vs.\ 46.75\%). The 
\emph{summary-based} configuration consistently performs best on SF110, suggesting that richer contextual guidance helps the LLM reorganize long, repetitive suites into more focused tests.

TestHumanizer also avoids LLM-specific smell patterns introduced by direct generation. \emph{Exception Catching Test} (ECT) and \emph{Magic Number} (MG) are noticeably more frequent in direct \texttt{gpt-4o}-generated suites (Defects4J: ECT 16.77\%, MG 3.96\%) than in any TestHumanizer variant (ECT $\approx$2--5\%, MG = 0\%), while TestHumanizer still delivers the readability and understandability improvements observed in RQ2. Overall, TestHumanizer shifts EvoSuite suites toward a smell profile that is less branch-heavy and, on larger systems (SF110), substantially less lazy and less assertion-duplicated, while preserving the structural richness and coverage of SBST.

\vspace{0.5em}
\noindent\colorbox{gray!20}{{\parbox{0.98\linewidth}{
\textbf{Finding 8:} Since our prompts do not include explicit smell-removal instructions, any smell reductions observed in TestHumanizer refactorings emerge as side effects of structural and readability improvements. The dominant EvoSuite smells---Assertion Roulette, General Fixture, and Dependent Test---remain pervasive across all configurations, as they require deep fixture redesign beyond the scope of our prompts. However, TestHumanizer substantially reduces structural smells: Conditional Logic Test drops from 36.97\% to 5.02\% on Defects4J and from 48.31\% to 4.37\% on SF110 in the \emph{summary-based} setting, and Lazy Test and Duplicate Assert are markedly reduced on SF110. Compared to direct LLM generation, TestHumanizer avoids LLM-specific smell patterns such as excessive exception handling and magic numbers.
}}}

\vspace{0.5em}
Our smell results can be contextualized against prior analyses of LLM-generated tests. Siddiq et al.~\cite{siddiq2024using} show using TsDetect on Defects4J that LLM-generated tests exhibit substantial levels of classic smells such as \emph{Assertion Roulette} and \emph{Magic Number Test}, often comparable to or higher than automatically generated and developer-written tests. Ouédraogo et al.~\cite{ouedraogo2024test} extend this picture to Defects4J and SF110, reporting that EvoSuite suites are dominated by \emph{Assertion Roulette} and \emph{Eager Test}, while LLM-generated tests introduce relatively more \emph{General Fixture}, \emph{Duplicate Assert}, \emph{Lazy Test}, and \emph{Empty Test}. In our study, direct \texttt{gpt-4o}-generated suites show a similar smell-heavy profile, with near-universal \emph{Magic Number Test}, pervasive \emph{Assertion Roulette}, and non-trivial 
levels of \emph{Lazy Test} and \emph{Duplicate Assert}. TestHumanizer, operating as a refactoring layer on top of EvoSuite, largely preserves the underlying SBST smell profile but avoids the additional smell shifts observed in fully LLM-generated suites, and improves specific dimensions such as \emph{Conditional Logic Test} and, on SF110, \emph{Lazy Test}. These results suggest that LLM-based structural refactoring can improve test readability (RQ2) 
without amplifying smell diffusion beyond what is already present in EvoSuite, and without introducing the LLM-specific smell patterns characteristic of direct generation.

\highlight{
\textbf{Summary of RQ5:} Since our prompts do not explicitly target smell removal, all 
observed smell reductions emerge as side effects of structural and readability improvements. The dominant EvoSuite smells---Assertion Roulette and General Fixture---remain pervasive across all configurations, as they require deep fixture and assertion redesign beyond the scope of the refactoring prompts. TestHumanizer does, however, substantially reduce structural smells: Conditional Logic Test is dramatically reduced in all configurations relative to EvoSuite (from 36.97\% to 5.02\% on Defects4J and from 48.31\% to 4.37\% on SF110 in the \emph{summary-based} setting), while direct LLM generation actually worsens this smell. On larger systems (SF110), TestHumanizer also markedly reduces Lazy Test and Duplicate Assert, suggesting improved test focus and reduced redundancy. Compared to direct LLM generation, TestHumanizer avoids LLM-specific smell patterns such as excessive exception handling and magic numbers, while preserving the structural richness and coverage of SBST. The 
\emph{summary-based} configuration offers the most consistent improvements across both datasets.
}

\subsection{[RQ6]: Developer perception and practical usability of TestHumanizer.}

\noindent\textbf{Experiment Design:} 

To complement the metric-based analyses in RQ1--RQ5, we conduct a developer study comparing EvoSuite suites with their \emph{summary-based} \texttt{gpt-4o} TestHumanizer refactorings, selected as the most robust configuration across all prior RQs. We sample 30 classes (15 Defects4J, 15 SF110), stratified by class size into three LOC bins---small ($<$80 LOC), medium (80--179 LOC), and large ($\geq$180 LOC)---under the constraint that both the original and refactored suites compile successfully. The final evaluation set consists of 15 Defects4J classes (230 EvoSuite test methods) and 15 SF110 classes (214 EvoSuite test methods), for a total of 30 classes and 444 test methods (Table~\ref{tab:rq6-sample}). The complete list 
of selected (dataset, project, class, LOC) tuples is provided in Table~\ref{app:rq6-sample-details}.
For each class, one EvoSuite suite and its matched TestHumanizer refactoring are lightly cleaned of boilerplate while preserving all \texttt{@Test} methods. Four evaluators assess the suites: two senior software developers with industrial experience and two graduate students with advanced testing knowledge. Tasks are split into two balanced sets (15 classes each), with every class evaluated independently by one senior and one graduate evaluator (within-group pairing). For each class, evaluators receive a short natural-language description of the class under test and two anonymized suite versions presented in randomized order. They are asked to rate two dimensions on a 1--5 Likert scale: (i)~\emph{perceived readability and clarity of intent}---how easily they can understand what each test is verifying; and (ii)~\emph{willingness to adopt}---whether they would include the suite in a real project. Evaluators may also provide free-text comments to explain their ratings.
We analyze paired ratings using Wilcoxon signed-rank tests with effect sizes ($r$). Inter-annotator agreement is assessed using Krippendorff's $\alpha$ (ordinal) and quadratic weighted Cohen's $\kappa$ for each evaluator group, and reported overall across all 
four raters. Free-text comments are thematically analyzed to contextualize quantitative findings.

\noindent\textbf{Experiment Results:}                

In the following, we refer to the summary-based \texttt{gpt-4o} refactoring simply as \emph{TestHumanizer} for brevity.

\begin{figure}
  \centering
  \includegraphics[width=0.6\textwidth]{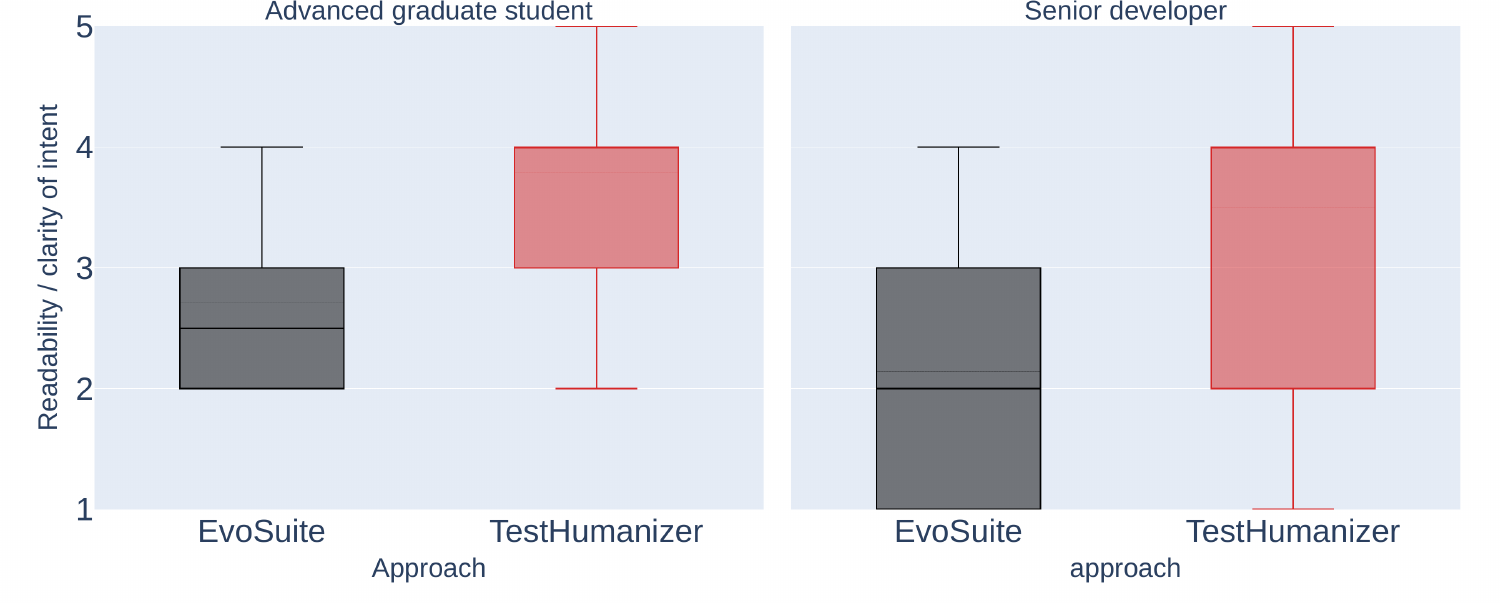} 
   \caption{Distributions of readability scores from Human evaluation. }
   \label{human_eval:readability}
   \vspace{0.5em}
\end{figure}

\begin{table}[t]
\caption{Overview of the human-evaluation sample by dataset and class size.}
\label{tab:rq6-sample}
\centering
\scalebox{0.6}{
\begin{tabular}{llrr}
\toprule
\textbf{Dataset} & \textbf{Size bin} & \textbf{\#Classes} & \textbf{\#Test methods} \\
\midrule
\multirow{4}{*}{Defects4J}
  & Small        & 5  & 30  \\
  & Medium       & 5  & 65  \\
  & Large        & 5  & 135 \\
  & \textbf{Total} & \textbf{15} & \textbf{230} \\
\midrule
\multirow{4}{*}{SF110}
  & Small        & 5  & 29  \\
  & Medium       & 5  & 90  \\
  & Large        & 5  & 95  \\
  & \textbf{Total} & \textbf{15} & \textbf{214} \\
\midrule
\multicolumn{2}{l}{\textbf{Overall total}} & \textbf{30} & \textbf{444} \\
\bottomrule
\end{tabular}
}
\vspace{0.5em}
\end{table}
%
Figure~\ref{human_eval:readability} and Table~\ref{tab:rq6-human-stats} show a consistent shift in perceived readability from EvoSuite to TestHumanizer for both evaluator profiles. Graduate evaluators rate EvoSuite at a mean of 2.71 (median 2.5, IQR [2,3]) and TestHumanizer at 3.79 (median 4.0, IQR [3,4]). Senior developers are more critical of EvoSuite (mean 2.14, 
median 2) but similarly rate TestHumanizer higher (mean 3.50, median 4). Overall, EvoSuite suites are judged low to neutral, whereas TestHumanizer consistently shifts ratings toward neutral to high. These gains are statistically significant: graduate evaluators 
$W=30$, $p=0.0005$, $r=0.71$; senior developers $W=63$, $p=0.0022$, $r=0.59$ (Table~\ref{tab:rq6_wilcoxon}).

\begin{figure}[t]
  \centering
  \includegraphics[width=0.6\textwidth]{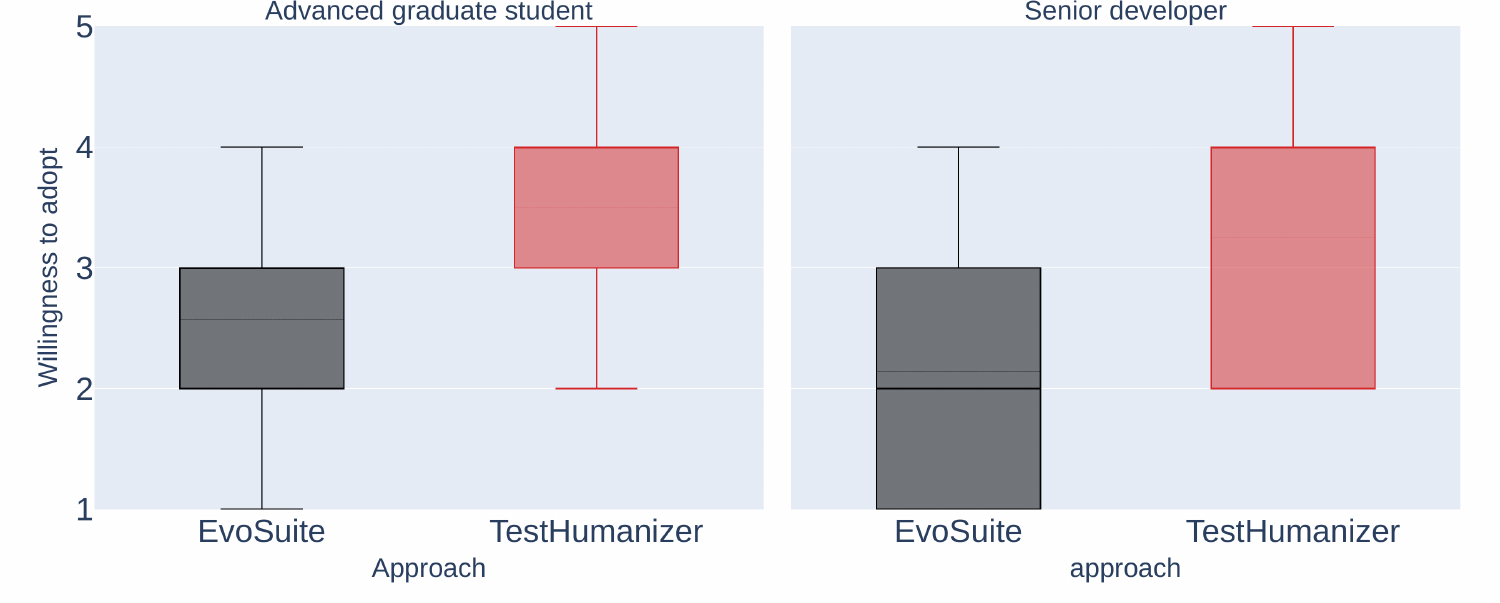}
  \caption{Distributions of adoption scores from the human 
  evaluation. Same layout as Figure~\ref{human_eval:readability}.}
  \label{human_eval:adoption}
\end{figure}

A consistent pattern holds for adoption (Figure~\ref{human_eval:adoption}). Graduate evaluators report adoption scores rising from mean 2.57 (median 3.0, IQR [2,3]) to 3.50 (median 4.0, IQR [3,4]). Senior developers rate EvoSuite lower (mean 2.14, median 2) but report substantially higher adoption for TestHumanizer (mean 3.25, median 4.0). EvoSuite suites cluster around low to neutral adoption, whereas TestHumanizer shifts ratings toward neutral to high without lowering medians or quartiles. These gains are significant: graduate evaluators $W=36$, $p=0.0008$, $r=0.68$; senior developers $W=81$, $p=0.0049$, $r=0.53$.

\begin{table*}[t]
\caption{Wilcoxon signed-rank test comparing \textit{TestHumanizer} (summary-based refactoring) and \textit{EvoSuite} human ratings.}

\label{tab:rq6_wilcoxon}
\centering
\scalebox{0.6}{
\begin{tabular}{llrrrrrr}
\toprule
\textbf{Metric} &
\textbf{Evaluator} &
\textbf{\begin{tabular}[c]{@{}r@{}}Median\\ EvoSuite\end{tabular}} &
\textbf{\begin{tabular}[c]{@{}r@{}}Median\\ TestHumanizer\end{tabular}} &
\textbf{\begin{tabular}[c]{@{}r@{}}Mean diff\\ (S--E)\end{tabular}} &
\textbf{$W$} &
\textbf{\begin{tabular}[c]{@{}r@{}}$p$ (two-\\ sided)\end{tabular}} &
\textbf{\begin{tabular}[c]{@{}r@{}}Effect\\ $r$\end{tabular}} \\
\midrule

Readability &
\multirow{2}{*}{Advanced graduate student} &
2.5 & 4.0 & 1.07 & 30 & 0.0005 & 0.71 \\
Adopt       & 
& 3.0 & 4.0 & 0.93 & 36 & 0.0008 & 0.68 \\

\midrule

Readability &
\multirow{2}{*}{Senior developer} &
2.0 & 4.0 & 1.36 & 63 & 0.0022 & 0.59 \\
Adopt       &
& 2.0 & 4.0 & 1.11 & 81 & 0.0049 & 0.53 \\

\bottomrule
\end{tabular}
}
\vspace{0.5em}
\end{table*}
%

Inter-annotator agreement is substantial overall ($\alpha = 0.71$--$0.85$; Table~\ref{tab:rq6_iaa_compact}), with particularly high agreement for TestHumanizer readability ($\alpha = 0.85$). One group exhibited low agreement on EvoSuite readability ($\alpha = -0.15$), which is consistent with prior evidence that EvoSuite-style tests elicit divergent judgments due to their mechanical structure~\cite{panichella2020revisiting,almasi2017industrial}; the same group reached near-perfect agreement on the corresponding TestHumanizer condition ($\alpha = 0.94$), suggesting that refactored suites yield more stable and consensual judgments.

\begin{table}[t]
\caption{Inter-annotator agreement (Krippendorff’s $\alpha$, ordinal) for human evaluation.}
\label{tab:rq6_iaa_compact}
\centering
\scalebox{0.6}{
\begin{tabular}{llcc}
\toprule
\textbf{Scope} & \textbf{Approach} & \textbf{Readability} & \textbf{Adoption} \\
\midrule
Group A & EvoSuite      & 0.82  & 0.79 \\
        & TestHumanizer & 0.72  & 0.74 \\
\midrule
Group B & EvoSuite      & -0.15 & 0.56 \\
        & TestHumanizer & 0.94  & 0.94 \\
\midrule
Overall & EvoSuite      & 0.71  & 0.80 \\
        & TestHumanizer & 0.85  & 0.82 \\
\bottomrule
\end{tabular}
}
\vspace{0.5em}
\end{table}

\vspace{0.5em}
\noindent\colorbox{gray!20}{{\parbox{0.98\linewidth}{
\textbf{Finding 9:} Across both datasets and evaluator profiles, TestHumanizer is consistently preferred over EvoSuite on both dimensions: perceived readability improves from low/neutral to neutral/high (graduate: $r=0.71$; senior: $r=0.59$), and willingness to adopt follows the same trend (graduate: $r=0.68$; senior: $r=0.53$), with all gains significant at $p<0.01$ (Wilcoxon). Inter-annotator agreement is substantial overall ($\alpha = 
0.71$--$0.85$), and particularly high for TestHumanizer readability ($\alpha = 0.85$), confirming that refactored suites elicit more consensual judgments than raw EvoSuite outputs.
}}}

\vspace{0.5em}
Free-text comments reinforce the quantitative findings. EvoSuite suites are consistently described as mechanical and auto-generated, with evaluators citing generic test names, arbitrary-looking inputs, and heavy exception scaffolding that obscures behavioral 
intent---perceptions well-documented in prior work~\cite{panichella2020revisiting,almasi2017industrial,daka2015modeling}. By contrast, evaluators praise TestHumanizer suites for their explicit scenario structure, descriptive naming, and reduced boilerplate. Several comments specifically highlight the Given--When--Then organization as making tests easier to read, review, and maintain. Senior evaluators in particular emphasize improved reviewability and reduced effort to understand test coverage. A recurring caveat, shared across both evaluator groups, is that improved readability does not substitute for independent coverage and correctness validation: most evaluators state they would still cross-check assertions and edge cases before fully relying on the refactored suite in production. This indicates that TestHumanizer is perceived as a strong aid for comprehension and review, rather than a replacement for rigorous test validation.

\vspace{0.5em}
\noindent\colorbox{gray!20}{{\parbox{0.98\linewidth}{
\textbf{Finding 10:} Qualitative feedback confirms that EvoSuite suites are perceived as mechanical and auto-generated---consistent with prior evidence~\cite{panichella2020revisiting,almasi2017industrial}---whereas TestHumanizer suites are praised for clearer scenario structure, descriptive naming, and Given--When--Then organization. Evaluators across both profiles nonetheless stress that readability gains do not replace independent validation of coverage and correctness, positioning TestHumanizer as a comprehension and review aid rather than a full substitute for rigorous testing.
}}}

\vspace{0.5em}
Our RQ6 results complement prior human-centered evaluations of LLM-enhanced tests. Deljouyi et al.~\cite{deljouyi2024leveraging} assess understandability improvements through a user study linking test quality to downstream bug-fixing performance. Biagiola et al.~\cite{biagiola2025improving} investigate developer perception under conservative, naming-oriented transformations. In contrast, our study isolates the usability impact of suite-level refactoring on paired Likert judgments of readability and adoption (30 classes, 
444 test methods), providing direct evidence that LLM-based post-processing substantially improves developer acceptance of SBST outputs, while independent behavioral validation remains necessary.

\highlight{
\textbf{Summary of RQ6:} In a stratified developer study on 30 classes (444 test methods), 
TestHumanizer significantly improves perceived readability and willingness to adopt compared to EvoSuite for both graduate and senior evaluators (Wilcoxon, $p<0.01$, medium-to-large effects: $r=0.53$--$0.71$). Inter-annotator agreement is substantial overall ($\alpha=0.71$--$0.85$), and particularly high for TestHumanizer readability ($\alpha=0.85$). Qualitative feedback aligns with these results: EvoSuite suites are consistently described as mechanical 
and auto-generated, while TestHumanizer is praised for clearer scenario structure, descriptive naming, and Given--When--Then organization. Evaluators stress that readability gains do not replace independent checks of coverage and correctness, positioning TestHumanizer as a comprehension aid rather than a full validation substitute.
}

\section{Discussion}
\label{sec:discussion}

\subsection{Why Hallucinations Persist and Implications for Test Refactoring Pipelines}
\label{subsec:hallucination-structural}

RQ1 shows that hallucination-induced compilation failures are not isolated occurrences but structural risks inherent to LLM-based refactoring. Even under explicit semantics-preservation constraints and full-class context, some refactorings introduce unintended edits 
to identifiers, method calls, or imports that break compilation. Recent theory frames hallucination as a systemic consequence of current training and evaluation regimes rather than an accidental defect~\cite{kalai2025why}: when abstention is not rewarded, models 
are incentivized to produce plausible-looking transformations under uncertainty, and refactoring errors reflect overconfident optimization rather than prompt misconfiguration.
Crucially, providing more context does not eliminate this risk. The \emph{code-centric} configuration---which supplies the full class source---did not prevent semantic-breaking edits and was additionally exposed to API-level prompt rejections, whereas the more abstract 
\emph{summary-based} configuration achieved higher compilation stability (RQ1). Robustness thus depends less on the volume of context provided than on how that context constrains the LLM's 
transformation space. A compact, abstract representation appears to implicitly bound the set of identifiers and structures the model can modify, reducing opportunities for hallucination-induced drift.
Consistent with observations in agentic repair workflows~\cite{melo2025agentic}, hallucination must be treated as a pipeline-level phenomenon rather than a prompt-engineering problem. 
LLM-based refactoring therefore requires mandatory compilation and validation gates at each transformation step---not as optional safeguards, but as structural requirements of any reliable refactoring pipeline~\cite{kalai2025why,melo2025agentic}.

Our findings indicate that robustness in LLM-based test refactoring is primarily a matter of workflow design rather than prompt engineering. Given the structural hallucination risks identified in RQ1, refactoring should be treated as a staged process that interleaves transformation, validation, and repair---rather than a single-shot LLM call followed by optional checking. TestHumanizer currently follows this single-pass, prompt-driven design: rather than dynamically planning and iterating over refactoring decisions, it relies on a fixed pipeline with post-hoc validation gates (Section~\ref{subsec:approach}). Recent agentic frameworks for 
general-purpose code refactoring, such as RefAgent~\cite{oueslati2025refagent}, suggest that dynamic planner--execution--testing loops can further improve reliability and adaptability over static, single-pass prompting, motivating the extensions discussed below.

\noindent\emph{Agentic pipelines} operationalize this principle through generate--detect--repair loops. Multi-agent systems have been shown to improve smell detection and refactoring reliability, yet 
they do not fully eliminate erroneous edits~\cite{melo2025agentic}. In our setting, similar orchestration could iteratively localize and repair non-compiling transformations, with compilation enforced as a hard acceptance gate at each iteration. The SIDE-based summarization loop and embedding-based similarity checks in TestHumanizer already embody a lightweight version of this principle; extending them with targeted repair steps, in the spirit of RefAgent's compilation-testing feedback loop~\cite{oueslati2025refagent}, would further reduce the fraction of non-compiling outputs observed in RQ1.

\noindent\emph{Retrieval-augmented refactoring} offers complementary grounding. Injecting project-specific resources---such as API usage patterns, existing test suites, or class-level documentation---constrains the transformation space and reduces unsupported identifier edits, directly addressing the dominant 
\emph{Cannot Find Symbol} failures observed in RQ1. Retrieval-enhanced prompting has already been shown to improve coverage in test generation~\cite{shin2024retrieval}; analogous grounding for refactoring could similarly stabilize compilability and, as a side effect, mitigate structural smells (RQ5).

Together, these directions suggest moving from prompt-centric refactoring toward guarded, retrieval-grounded, and potentially agentic architectures that integrate staged decomposition, external evidence, and mandatory verification gates. Such designs would strengthen the reliability gains of TestHumanizer (RQ1) while further consolidating its readability and structural improvements (RQ2, RQ5)---providing a principled path toward making LLM-based test refactoring robust enough for continuous integration pipelines.

\subsection{Rethinking Evaluation Beyond Similarity in LLM-Based Test Refactoring}
\label{subsec:beyond-similarity}

RQ3 shows that similarity-centric evaluation is insufficient for LLM-based refactoring. Embedding-based similarities saturate across configurations and are too weakly discriminative to distinguish meaningfully between \emph{tests-only}, \emph{code-centric}, and 
\emph{summary-based} refactorings. Token-level metrics such as CodeBLEU can penalize structurally divergent yet beneficial restructurings, echoing broader evidence that overlap metrics may mis-rank stylistically or structurally different LLM 
outputs~\cite{sun2024source}. CTSES mitigates some of this over-pessimism as a composite, human-aligned signal, but our manual inspection shows it can still mis-rank broad-scope refactorings that are genuinely beneficial.
These limitations call for \emph{multi-signal evaluation} that goes beyond surface overlap. Alongside similarity metrics, evaluation frameworks for LLM-based test refactoring should incorporate human-oriented indicators that directly capture the goals of refactoring: intent clarity, naming and documentation quality, scenario structure (e.g., Given--When--Then adherence), and smell reduction. Structural feasibility signals---compilation rates and 
code coverage---should serve as hard gates rather than optional complements, since a refactoring that improves readability at the cost of compilation or coverage provides no practical value. Developing \emph{refactoring-aware metrics} that jointly capture 
these dimensions remains an open challenge, and one our study highlights as critical for progress in this area.
A further complication is that readability itself is perception-driven. Developers reach only moderate agreement even under controlled evaluation settings~\cite{sergeyuk2024assessing}, 
and LLM-based readability evaluators, while increasingly human-aligned, can exhibit systematic biases and stability trade-offs~\cite{ouedraogo2025human}. This means that no single 
automated metric---whether similarity-based or LLM-judged---can fully substitute for human assessment. Overall, evaluation should treat similarity as a necessary but insufficient signal, and systematically connect automated metrics to developer preference through human studies, as demonstrated in RQ6.

\subsection{Implications for Researchers} 
\label{subsec:implications-researchers}

Our results suggest several research directions to make SBST outputs practically usable through LLM-based refactoring.
\textbf{LLMs should act as post-processing layers rather than SBST replacements.} Direct one-shot generation remains brittle and coverage-poor at scale, whereas refactoring compilable SBST suites yields near-baseline compilability while largely preserving 
coverage (RQ1, RQ4). Future work should explore how this refactoring-layer paradigm generalizes to other SBST tools and testing frameworks beyond EvoSuite.
\textbf{Context granularity is a first-class design choice.} Full-code prompting does not prevent hallucination-induced failures and can introduce practical feasibility issues such as API-level rejections, while compact summary-based context is consistently the most stable configuration (RQ1). Context compression and semantic abstraction should be treated as 
first-order robustness levers in LLM-based refactoring pipelines, not as implementation details.
\textbf{Compilation and dynamic checks are mandatory safeguards.} Even with explicit semantics-preservation instructions, some refactorings introduce hallucinated identifier edits that break compilation (RQ1). Compile-time gates and, when needed, repair or rollback loops are not optional additions but structural requirements of any reliable refactoring pipeline.
\textbf{Similarity alone is insufficient for evaluation.} Embedding similarities saturate and lexical overlap can misclassify structurally divergent yet beneficial restructurings (RQ3). Composite similarity metrics such as CTSES should be complemented with structural indicators---readability, understandability, and smell profiles (RQ2, RQ5)---and selective manual validation of low-similarity cases. Developing refactoring-aware evaluation 
frameworks that jointly capture these dimensions is an open and pressing research challenge.
\textbf{Refactoring objectives should be made explicit.} Readability and understandability improve consistently as primary objectives, while smell-profile shifts emerge as beneficial side effects (RQ2, RQ5). This suggests that objective-driven refactoring 
prompts---explicitly targeting smell reduction or modularity alongside readability---may yield further quality gains, and that systematic trade-off analysis between readability, structural 
stability, and behavioral preservation deserves dedicated study.
\textbf{Developer preference should inform evaluation design.} Human ratings confirm that readability gains translate into higher willingness to adopt, while still requiring independent validation of behavioral adequacy (RQ6). Future evaluation frameworks should go beyond automated metrics and connect refactoring quality signals to developer-centered outcomes, including maintenance effort, bug-finding effectiveness, and long-term suite 
evolution.

\subsection{Implications for Developers} 
\label{subsec:implication-developers}

Our findings provide actionable guidance for practitioners considering LLM-based refactoring of automatically generated tests.
\textbf{Use LLMs to improve maintainability---not to replace systematic test generation.} Direct LLM-generated tests are more readable but exhibit substantially lower and less stable 
coverage (RQ4). Refactoring SBST suites with TestHumanizer preserves coverage while improving clarity and structure (RQ2, RQ4). In practice, LLMs are best deployed as a refinement layer on top of existing SBST pipelines, not as a replacement for the coverage guarantees that SBST tools provide.
\textbf{Prefer compact contextual guidance over full-code prompting.} Providing the entire class under test does not guarantee higher reliability and may introduce instability---including API-level prompt rejections and hallucination-induced identifier edits (RQ1). Summary-based context offers a more robust and scalable alternative while maintaining strong compilation and coverage guarantees. For practitioners, this means investing in lightweight 
class summarization as a pre-processing step rather than passing raw source code directly to the LLM.
\textbf{Keep compilation and coverage checks in the loop.} Although most refactorings preserve test behavior, a non-negligible fraction introduce hallucinated changes that break identifier 
references or alter API usage (RQ1). Automated compile-time verification and coverage regression checks should be integrated into CI pipelines whenever LLM-based refactoring is used, and refactorings that fail these checks should be reverted rather than manually corrected.
\textbf{Expect structural improvements---but not automatic smell elimination.} TestHumanizer substantially reduces control-flow-heavy patterns such as Conditional Logic Test, and on 
larger systems (SF110) also mitigates Lazy Test and Duplicate Assert (RQ5). However, dominant smells such as Assertion Roulette and General Fixture are not automatically resolved, as they require deeper fixture and assertion redesign beyond the scope of 
readability-oriented refactoring prompts. Developers should treat LLM-based refactoring as structural enhancement, not as a substitute for deliberate test redesign.
\noindent\textbf{Readability gains translate into higher developer acceptance---but independent validation remains necessary.} Human evaluators consistently prefer TestHumanizer-refactored suites in terms of clarity and willingness to adopt (RQ6). Nevertheless, improved readability does not substitute for independent validation of behavioral correctness and coverage adequacy. LLM-refactored tests should be treated as improved starting points for review, not 
as fully validated outputs ready for production use.

\subsection{Threats to Validity}
\label{subsec:threats}

\noindent\textbf{External Validity.} Our study focuses on Java unit tests generated by EvoSuite over 
Defects4J and SF110, refactored via two general-purpose LLMs (\texttt{gpt-4o} and \texttt{mistral-large-2407}). Results may not directly generalize to other languages (e.g., Python), testing 
frameworks, SBST tools, or domains with different API characteristics. We mitigate this threat by using two widely adopted benchmarks covering diverse projects and by analyzing two distinct model families, but further replication on additional ecosystems and generators remains necessary. For RQ6, the developer study covers 30 
classes (444 test methods) rated by four evaluators; while stratified across datasets and LOC bins, and strengthened through paired ratings, Wilcoxon testing, and inter-annotator agreement analysis, it 
remains a bounded sample that may not capture all industrial test styles or evaluation contexts. This scale is nonetheless consistent with prior human-centered evaluations of LLM-based test improvement~\cite{deljouyi2024leveraging,biagiola2025improving}, which similarly rely on small expert panels rather than large-scale crowdsourced studies.

\noindent\textbf{Internal Validity.} LLM outputs are non-deterministic and can vary with decoding 
parameters, system prompts, or model updates. We mitigate this by fixing prompting templates and applying the same pipeline systematically across all configurations, though some variance is unavoidable. Feasibility in the \emph{code-centric} setting for \texttt{gpt-4o} is additionally affected by API-level prompt 
rejections (HTTP~400) triggered by repetitive prompt patterns, which may depend on provider-side input validation policies and can influence comparative feasibility across models and configurations. A non-negligible fraction of non-compiling refactorings is caused by hallucination-induced edits that break referenced program entities (see Figure~\ref{fig:compilation-errors}); we rely on compilation 
filtering and manual inspection to characterize these cases, but cannot fully eliminate them without automated repair or rollback mechanisms. Our results are also reported over compilable suites for metrics requiring execution (e.g., coverage), which may slightly bias comparisons toward more robust configurations; however, this reflects a realistic deployment constraint in which only compilable suites are actionable.

A further internal validity concern is potential data leakage from Defects4J, which prior work has shown can exhibit memorization signals in some LLMs~\cite{ramos2025large}. This risk is most pronounced in smaller, older models trained on limited budgets (e.g., \texttt{codegen-multi}); the same study reports that larger, more recently trained models exhibit substantially weaker memorization signals~\cite{ramos2025large}, suggesting that \texttt{gpt-4o} and \texttt{mistral-large-2407} are less exposed to this threat. 
Critically, TestHumanizer's core task is suite-level \emph{refactoring} of EvoSuite-generated tests, not generation of bug-fixing patches from Defects4J bug descriptions---the task most directly implicated in prior leakage studies---further reducing the likelihood that memorization explains our compilability and coverage 
findings. We complement Defects4J with SF110, a considerably larger and more heterogeneous benchmark unlikely to share this risk profile, to mitigate this threat through dataset diversity.

\noindent\textbf{Construct Validity.} Readability and maintainability are multi-faceted constructs that 
resist single-metric characterization. We mitigate metric bias by combining complementary indicators: a machine-learned readability proxy~\cite{scalabrino2018comprehensive}, structural and cognitive complexity measures (Cyclomatic Complexity and CCTR), composite similarity metrics (CTSES and embedding-based cosine similarity), structural coverage, and test smell prevalence. Nevertheless, automated metrics can mischaracterize broad-scope yet beneficial refactorings, as observed in our manual inspection of low-similarity cases (RQ3). For RQ6, we use two Likert-scale questions (readability/clarity of intent and willingness to adopt) supplemented by free-text feedback; while this captures perceived usability across 
two evaluator profiles, it does not constitute a full longitudinal industrial evaluation. In particular, willingness to adopt as a proxy for practical usefulness should be interpreted cautiously. Future 
studies with longer-term maintenance tasks, defect-detection outcomes, and a larger and more diverse pool of evaluators would provide stronger ecological validity for the human-centered findings reported in RQ6.

A final construct validity consideration concerns our choice of model generation and prompting strategy. We rely on \texttt{gpt-4o} and \texttt{mistral-large-2407} with Chain-of-Thought prompting, rather than native multi-step reasoning (``thinking'') modes or agentic orchestration (Section~\ref{subsec:approach}). Newer models and reasoning-oriented prompting strategies may alter the feasibility, hallucination, and readability trade-offs we report. However, our central finding---that context granularity and 
validation gates, rather than raw model capability, govern refactoring robustness---is a property of the pipeline design rather than of any specific model generation, and is consistent with recent 
empirical studies of LLM-based refactoring conducted under comparable prompting regimes~\cite{zhang2026empirical}. Validating this claim across reasoning-mode and agentic LLMs remains important future work.

\section{Related Work}
\label{sec:relatedwork}

Our work sits at the intersection of three active research threads: LLM-based improvement of automatically generated tests, smell-driven refactoring of unit tests, and analysis-guided LLM test generation. We position TestHumanizer relative to each.

\subsection{Improving Automatically Generated Tests with LLMs}
\label{subsec:rw-llm-generated}

Recent work has explored LLMs as post-processing layers to improve the readability and maintainability of automatically generated tests. Biagiola et al.~\cite{biagiola2025improving} refactor SBST-generated suites through constrained transformations that primarily target 
identifier renaming, and emphasize compilation checking as a key feasibility guard. Deljouyi et al.~\cite{deljouyi2024leveraging} integrate LLM-based rewriting within an SBST+LLM workflow, 
incorporating compilation verification and fallback mechanisms to handle non-compiling outputs and using CodeBLEU as a semantic safeguard. Both approaches apply relatively narrow transformations to individual tests or identifiers.
In contrast, TestHumanizer operates at the level of entire test suites and applies a broader scope of structural transformation---including Given--When--Then documentation, grouping of related logic, and control-flow simplification. We additionally study three context 
granularities and quantify trade-offs across feasibility, semantic similarity, structural coverage, smell profiles, and developer preference, providing a more comprehensive view of the readability--robustness trade-off in LLM-based test refactoring.

\subsection{LLM-Driven Refactoring of Human-Written Tests via Smell-Specific Guidance}
\label{subsec:rw-llm-smell-driven}

Beyond automatically generated tests, Gao et al.~\cite{gao2025automated} propose UTRefactor, an LLM-based framework that targets smell removal in human-written unit tests using explicit smell knowledge encoded in a DSL, together with checkpointed prompting and validation. Their study reports substantial smell reduction on real-world Java projects, while also 
noting that compilation and execution failures can still occur when refactoring rules conflict with project-specific assertion APIs.
UTRefactor and TestHumanizer share the goal of improving test quality through LLM-based refactoring, but differ in scope and target. UTRefactor applies rule-guided, smell-specific 
transformations to developer-written tests, whereas TestHumanizer takes a holistic approach to ``humanizing'' SBST-generated suites---improving structure, naming, and documentation---and evaluates outcomes through a multi-dimensional protocol that includes compilation rates, coverage preservation, similarity metrics, and human judgments. Moreover, any smell reductions in TestHumanizer emerge as side effects of readability-oriented refactoring rather 
than explicit optimization targets.

\subsection{LLM-Based Test Generation with Analysis-Guided Pipelines}
\label{subsec:llm-based-analysis-guided}

A parallel line of work investigates LLMs as \emph{test generators} augmented with program analysis and verification loops to obtain executable, high-coverage suites. Pan et al.~\cite{pan2025aster} propose ASTER, which combines lightweight static analysis to extract 
test-relevant context---including mocking opportunities---with post-generation sanitization, compile/execute checks, iterative repair, and coverage-guided augmentation. ASTER yields tests that professional developers perceive as more natural while maintaining 
coverage competitive with EvoSuite and CodaMOSA.
This generation-centric paradigm is fundamentally different from our refactoring-based approach. Where ASTER and similar pipelines start from scratch and use analysis to guide LLM generation, 
TestHumanizer starts from compilable, high-coverage EvoSuite suites and uses LLMs as a controlled structural improvement layer. This distinction is consequential: by building on SBST outputs, TestHumanizer inherits their coverage guarantees and avoids the 
compilability fragility of one-shot LLM generation, while still achieving the readability and understandability improvements that generation-centric approaches aim for. Our systematic evaluation of context granularity, semantic preservation, and developer preference 
provides complementary evidence on how LLMs can be most effectively 
integrated into test engineering workflows.

\section{Conclusion and Future Work}
\label{sec:conclusion}

Automatically generated tests provide strong behavioral reach but often lack the readability and maintainability that make them practical for developers. Pure LLM-based generation, while producing more natural tests, remains brittle and coverage-poor at scale. Our 
results show that these paradigms are complementary: when used as a controlled refactoring layer over compilable SBST suites, LLMs can substantially improve the developer-perceived quality of generated tests without sacrificing structural robustness or coverage.
Across 350 classes from Defects4J and SF110, TestHumanizer preserves near-EvoSuite compilation rates (88--98\%) and structural coverage while improving predicted readability, reducing control-flow and test-aware cognitive complexity, and substantially mitigating structural smells---most notably, Conditional Logic Test drops from 36.97\% to 5.02\% on Defects4J and from 48.31\% to 4.37\% on SF110 in the summary-based configuration. These quality gains translate into significantly higher developer preference and willingness to adopt, as confirmed by our developer study on 30 classes (444 test methods, Wilcoxon $p<0.01$, medium-to-large effects). At the same time, hallucination-induced compilation failures, context sensitivity, and the limitations of similarity-based evaluation reveal that LLM-based refactoring must be treated as a guarded transformation process, supported by mandatory compilation gates, coverage checks, 
and multi-dimensional quality assessment.
Beyond TestHumanizer, our findings advocate for hybrid SBST+LLM architectures in which SBST ensures behavioral guarantees and LLMs enhance structure and interpretability under explicit validation constraints. The summary-based configuration consistently emerges as the most robust design choice, demonstrating that compact semantic abstractions can constrain LLM transformations more effectively than full source code---a principle with broader implications for context-aware LLM engineering.
Future work should pursue three directions. First, \emph{agentic and retrieval-augmented refactoring pipelines} with generate--detect--repair loops and project-aware grounding would address the hallucination-induced failures identified in RQ1 through principled 
repair rather than simple rejection. Second, \emph{refactoring-aware evaluation frameworks} that move beyond similarity-centric metrics toward developer-aligned signals---capturing intent clarity, naming quality, scenario structure, and smell reduction---would provide more faithful assessments of broad-scope refactorings that current metrics penalize unfairly. Third, \emph{objective-driven refactoring strategies} that explicitly model trade-offs between readability, smell elimination, coverage, and behavioral stability would enable 
more targeted and controllable improvements than the emergent effects 
observed in this study.
Overall, LLMs are most effective not as standalone test generators, but as carefully constrained refinement layers over robust automated testing infrastructures---a paradigm that preserves the coverage guarantees of SBST while closing the usability gap that has long 
limited the practical adoption of automatically generated tests.

\section*{Acknowledgements}{
This research was funded in whole, or in part, by the Luxembourg National Research Fund (FNR), grant reference AFR PhD bilateral, project reference 17185670. This work was also supported by the European Research Council (ERC) under the European Union’s Horizon 2020 research and innovation program (grant agreement No. 949014) and the Fundamental Research Funds for the Central Universities(AE89991/478). For the purpose of open access, and in fulfilment of the obligations arising from the grant agreement, the author has applied a Creative Commons Attribution 4.0 International (CC BY 4.0) license to any Author Accepted Manuscript version arising from this submission.
}


\bibliographystyle{ACM-Reference-Format}
\bibliography{references}


\begin{thebibliography}{72}


\ifx \showCODEN    \undefined \def \showCODEN     #1{\unskip}     \fi
\ifx \showDOI      \undefined \def \showDOI       #1{#1}\fi
\ifx \showISBNx    \undefined \def \showISBNx     #1{\unskip}     \fi
\ifx \showISBNxiii \undefined \def \showISBNxiii  #1{\unskip}     \fi
\ifx \showISSN     \undefined \def \showISSN      #1{\unskip}     \fi
\ifx \showLCCN     \undefined \def \showLCCN      #1{\unskip}     \fi
\ifx \shownote     \undefined \def \shownote      #1{#1}          \fi
\ifx \showarticletitle \undefined \def \showarticletitle #1{#1}   \fi
\ifx \showURL      \undefined \def \showURL       {\relax}        \fi
\providecommand\bibfield[2]{#2}
\providecommand\bibinfo[2]{#2}
\providecommand\natexlab[1]{#1}
\providecommand\showeprint[2][]{arXiv:#2}

\bibitem[Ahmed and Devanbu(2022)]%
        {ahmed2022few}
\bibfield{author}{\bibinfo{person}{Toufique Ahmed} {and} \bibinfo{person}{Premkumar Devanbu}.} \bibinfo{year}{2022}\natexlab{}.
\newblock \showarticletitle{Few-shot training LLMs for project-specific code-summarization}. In \bibinfo{booktitle}{\emph{Proceedings of the 37th IEEE/ACM International Conference on Automated Software Engineering}}. \bibinfo{pages}{1--5}.
\newblock


\bibitem[Ahmed et~al\mbox{.}(2024)]%
        {ahmed2024automatic}
\bibfield{author}{\bibinfo{person}{Toufique Ahmed}, \bibinfo{person}{Kunal~Suresh Pai}, \bibinfo{person}{Premkumar Devanbu}, {and} \bibinfo{person}{Earl~T Barr}.} \bibinfo{year}{2024}\natexlab{}.
\newblock \showarticletitle{Automatic semantic augmentation of language model prompts (for code summarization). In 2024 IEEE/ACM 46th International Conference on Software Engineering (ICSE)}.
\newblock \bibinfo{journal}{\emph{IEEE Computer Society}} (\bibinfo{year}{2024}), \bibinfo{pages}{1004--1004}.
\newblock


\bibitem[Almasi et~al\mbox{.}(2017)]%
        {almasi2017industrial}
\bibfield{author}{\bibinfo{person}{M~Moein Almasi}, \bibinfo{person}{Hadi Hemmati}, \bibinfo{person}{Gordon Fraser}, \bibinfo{person}{Andrea Arcuri}, {and} \bibinfo{person}{Janis Benefelds}.} \bibinfo{year}{2017}\natexlab{}.
\newblock \showarticletitle{An industrial evaluation of unit test generation: Finding real faults in a financial application}. In \bibinfo{booktitle}{\emph{2017 IEEE/ACM 39th International Conference on Software Engineering: Software Engineering in Practice Track (ICSE-SEIP)}}. IEEE, \bibinfo{pages}{263--272}.
\newblock


\bibitem[Arcuri and Fraser(2013)]%
        {arcuri2013parameter}
\bibfield{author}{\bibinfo{person}{Andrea Arcuri} {and} \bibinfo{person}{Gordon Fraser}.} \bibinfo{year}{2013}\natexlab{}.
\newblock \showarticletitle{Parameter tuning or default values? An empirical investigation in search-based software engineering}.
\newblock \bibinfo{journal}{\emph{Empirical Software Engineering}}  \bibinfo{volume}{18} (\bibinfo{year}{2013}), \bibinfo{pages}{594--623}.
\newblock


\bibitem[Bacchelli et~al\mbox{.}(2008)]%
        {bacchelli2008effectiveness}
\bibfield{author}{\bibinfo{person}{Alberto Bacchelli}, \bibinfo{person}{Paolo Ciancarini}, {and} \bibinfo{person}{Davide Rossi}.} \bibinfo{year}{2008}\natexlab{}.
\newblock \showarticletitle{On the effectiveness of manual and automatic unit test generation}. In \bibinfo{booktitle}{\emph{2008 The Third International Conference on Software Engineering Advances}}. IEEE, \bibinfo{pages}{252--257}.
\newblock


\bibitem[Banerjee and Lavie(2005)]%
        {banerjee2005meteor}
\bibfield{author}{\bibinfo{person}{Satanjeev Banerjee} {and} \bibinfo{person}{Alon Lavie}.} \bibinfo{year}{2005}\natexlab{}.
\newblock \showarticletitle{METEOR: An automatic metric for MT evaluation with improved correlation with human judgments}. In \bibinfo{booktitle}{\emph{Proceedings of the acl workshop on intrinsic and extrinsic evaluation measures for machine translation and/or summarization}}. \bibinfo{pages}{65--72}.
\newblock


\bibitem[Bavota et~al\mbox{.}(2012)]%
        {bavota2012empirical}
\bibfield{author}{\bibinfo{person}{Gabriele Bavota}, \bibinfo{person}{Abdallah Qusef}, \bibinfo{person}{Rocco Oliveto}, \bibinfo{person}{Andrea De~Lucia}, {and} \bibinfo{person}{David Binkley}.} \bibinfo{year}{2012}\natexlab{}.
\newblock \showarticletitle{An empirical analysis of the distribution of unit test smells and their impact on software maintenance}. In \bibinfo{booktitle}{\emph{2012 28th IEEE international conference on software maintenance (ICSM)}}. IEEE, \bibinfo{pages}{56--65}.
\newblock


\bibitem[Bavota et~al\mbox{.}(2015)]%
        {bavota2015test}
\bibfield{author}{\bibinfo{person}{Gabriele Bavota}, \bibinfo{person}{Abdallah Qusef}, \bibinfo{person}{Rocco Oliveto}, \bibinfo{person}{Andrea De~Lucia}, {and} \bibinfo{person}{Dave Binkley}.} \bibinfo{year}{2015}\natexlab{}.
\newblock \showarticletitle{Are test smells really harmful? an empirical study}.
\newblock \bibinfo{journal}{\emph{Empirical Software Engineering}}  \bibinfo{volume}{20} (\bibinfo{year}{2015}), \bibinfo{pages}{1052--1094}.
\newblock


\bibitem[Beck(2000)]%
        {beck2000extreme}
\bibfield{author}{\bibinfo{person}{Kent Beck}.} \bibinfo{year}{2000}\natexlab{}.
\newblock \bibinfo{booktitle}{\emph{Extreme programming explained: embrace change}}.
\newblock \bibinfo{publisher}{addison-wesley professional}.
\newblock


\bibitem[Bhatia et~al\mbox{.}(2024)]%
        {bhatia2024unit}
\bibfield{author}{\bibinfo{person}{Shreya Bhatia}, \bibinfo{person}{Tarushi Gandhi}, \bibinfo{person}{Dhruv Kumar}, {and} \bibinfo{person}{Pankaj Jalote}.} \bibinfo{year}{2024}\natexlab{}.
\newblock \showarticletitle{Unit test generation using generative ai: A comparative performance analysis of autogeneration tools}. In \bibinfo{booktitle}{\emph{Proceedings of the 1st International Workshop on Large Language Models for Code}}. \bibinfo{pages}{54--61}.
\newblock


\bibitem[Biagiola et~al\mbox{.}(2025)]%
        {biagiola2025improving}
\bibfield{author}{\bibinfo{person}{Matteo Biagiola}, \bibinfo{person}{Gianluca Ghislotti}, {and} \bibinfo{person}{Paolo Tonella}.} \bibinfo{year}{2025}\natexlab{}.
\newblock \showarticletitle{Improving the readability of automatically generated tests using large language models}. In \bibinfo{booktitle}{\emph{2025 IEEE Conference on Software Testing, Verification and Validation (ICST)}}. IEEE, \bibinfo{pages}{162--173}.
\newblock


\bibitem[Buse and Weimer(2009)]%
        {buse2009learning}
\bibfield{author}{\bibinfo{person}{Raymond~PL Buse} {and} \bibinfo{person}{Westley~R Weimer}.} \bibinfo{year}{2009}\natexlab{}.
\newblock \showarticletitle{Learning a metric for code readability}.
\newblock \bibinfo{journal}{\emph{IEEE Transactions on software engineering}} \bibinfo{volume}{36}, \bibinfo{number}{4} (\bibinfo{year}{2009}), \bibinfo{pages}{546--558}.
\newblock


\bibitem[Campbell(2018)]%
        {campbell2018cognitive}
\bibfield{author}{\bibinfo{person}{G~Ann Campbell}.} \bibinfo{year}{2018}\natexlab{}.
\newblock \showarticletitle{Cognitive complexity: An overview and evaluation}. In \bibinfo{booktitle}{\emph{Proceedings of the 2018 international conference on technical debt}}. \bibinfo{pages}{57--58}.
\newblock


\bibitem[Chen et~al\mbox{.}(2021)]%
        {chen2021evaluating}
\bibfield{author}{\bibinfo{person}{Mark Chen}, \bibinfo{person}{Jerry Tworek}, \bibinfo{person}{Heewoo Jun}, \bibinfo{person}{Qiming Yuan}, \bibinfo{person}{Henrique Ponde de~Oliveira Pinto}, \bibinfo{person}{Jared Kaplan}, \bibinfo{person}{Harri Edwards}, \bibinfo{person}{Yuri Burda}, \bibinfo{person}{Nicholas Joseph}, \bibinfo{person}{Greg Brockman}, {et~al\mbox{.}}} \bibinfo{year}{2021}\natexlab{}.
\newblock \showarticletitle{Evaluating large language models trained on code}.
\newblock \bibinfo{journal}{\emph{arXiv preprint arXiv:2107.03374}} (\bibinfo{year}{2021}).
\newblock


\bibitem[Daka et~al\mbox{.}(2015)]%
        {daka2015modeling}
\bibfield{author}{\bibinfo{person}{Ermira Daka}, \bibinfo{person}{Jos{\'e} Campos}, \bibinfo{person}{Gordon Fraser}, \bibinfo{person}{Jonathan Dorn}, {and} \bibinfo{person}{Westley Weimer}.} \bibinfo{year}{2015}\natexlab{}.
\newblock \showarticletitle{Modeling readability to improve unit tests}. In \bibinfo{booktitle}{\emph{Proceedings of the 2015 10th Joint Meeting on Foundations of Software Engineering}}. \bibinfo{pages}{107--118}.
\newblock


\bibitem[Deljouyi et~al\mbox{.}(2024)]%
        {deljouyi2024leveraging}
\bibfield{author}{\bibinfo{person}{Amirhossein Deljouyi}, \bibinfo{person}{Roham Koohestani}, \bibinfo{person}{Maliheh Izadi}, {and} \bibinfo{person}{Andy Zaidman}.} \bibinfo{year}{2024}\natexlab{}.
\newblock \showarticletitle{Leveraging Large Language Models for Enhancing the Understandability of Generated Unit Tests}.
\newblock \bibinfo{journal}{\emph{arXiv preprint arXiv:2408.11710}} (\bibinfo{year}{2024}).
\newblock


\bibitem[Feng et~al\mbox{.}(2020)]%
        {feng2020codebert}
\bibfield{author}{\bibinfo{person}{Zhangyin Feng}, \bibinfo{person}{Daya Guo}, \bibinfo{person}{Duyu Tang}, \bibinfo{person}{Nan Duan}, \bibinfo{person}{Xiaocheng Feng}, \bibinfo{person}{Ming Gong}, \bibinfo{person}{Linjun Shou}, \bibinfo{person}{Bing Qin}, \bibinfo{person}{Ting Liu}, \bibinfo{person}{Daxin Jiang}, {et~al\mbox{.}}} \bibinfo{year}{2020}\natexlab{}.
\newblock \showarticletitle{Codebert: A pre-trained model for programming and natural languages}.
\newblock \bibinfo{journal}{\emph{arXiv preprint arXiv:2002.08155}} (\bibinfo{year}{2020}).
\newblock


\bibitem[Fraser and Arcuri(2011)]%
        {fraser2011evosuite}
\bibfield{author}{\bibinfo{person}{Gordon Fraser} {and} \bibinfo{person}{Andrea Arcuri}.} \bibinfo{year}{2011}\natexlab{}.
\newblock \showarticletitle{Evosuite: automatic test suite generation for object-oriented software}. In \bibinfo{booktitle}{\emph{Proceedings of the 19th ACM SIGSOFT symposium and the 13th European conference on Foundations of software engineering}}. \bibinfo{pages}{416--419}.
\newblock


\bibitem[Fraser and Arcuri(2014)]%
        {fraser2014large}
\bibfield{author}{\bibinfo{person}{Gordon Fraser} {and} \bibinfo{person}{Andrea Arcuri}.} \bibinfo{year}{2014}\natexlab{}.
\newblock \showarticletitle{A large-scale evaluation of automated unit test generation using evosuite}.
\newblock \bibinfo{journal}{\emph{ACM Transactions on Software Engineering and Methodology (TOSEM)}} \bibinfo{volume}{24}, \bibinfo{number}{2} (\bibinfo{year}{2014}), \bibinfo{pages}{1--42}.
\newblock


\bibitem[Fraser et~al\mbox{.}(2015)]%
        {fraser2015does}
\bibfield{author}{\bibinfo{person}{Gordon Fraser}, \bibinfo{person}{Matt Staats}, \bibinfo{person}{Phil McMinn}, \bibinfo{person}{Andrea Arcuri}, {and} \bibinfo{person}{Frank Padberg}.} \bibinfo{year}{2015}\natexlab{}.
\newblock \showarticletitle{Does automated unit test generation really help software testers? a controlled empirical study}.
\newblock \bibinfo{journal}{\emph{ACM Transactions on Software Engineering and Methodology (TOSEM)}} \bibinfo{volume}{24}, \bibinfo{number}{4} (\bibinfo{year}{2015}), \bibinfo{pages}{1--49}.
\newblock


\bibitem[Gao et~al\mbox{.}(2024)]%
        {gao2024context}
\bibfield{author}{\bibinfo{person}{Yi Gao}, \bibinfo{person}{Xing Hu}, \bibinfo{person}{Xiaohu Yang}, {and} \bibinfo{person}{Xin Xia}.} \bibinfo{year}{2024}\natexlab{}.
\newblock \showarticletitle{Context-Enhanced LLM-Based Framework for Automatic Test Refactoring}.
\newblock \bibinfo{journal}{\emph{arXiv preprint arXiv:2409.16739}} (\bibinfo{year}{2024}).
\newblock


\bibitem[Gao et~al\mbox{.}(2025)]%
        {gao2025automated}
\bibfield{author}{\bibinfo{person}{Yi Gao}, \bibinfo{person}{Xing Hu}, \bibinfo{person}{Xiaohu Yang}, {and} \bibinfo{person}{Xin Xia}.} \bibinfo{year}{2025}\natexlab{}.
\newblock \showarticletitle{Automated Unit Test Refactoring}.
\newblock \bibinfo{journal}{\emph{Proceedings of the ACM on Software Engineering}} \bibinfo{volume}{2}, \bibinfo{number}{FSE} (\bibinfo{year}{2025}), \bibinfo{pages}{713--733}.
\newblock


\bibitem[Gay(2023)]%
        {gay2023improving}
\bibfield{author}{\bibinfo{person}{Gregory Gay}.} \bibinfo{year}{2023}\natexlab{}.
\newblock \showarticletitle{Improving the Readability of Generated Tests Using GPT-4 and ChatGPT Code Interpreter}. In \bibinfo{booktitle}{\emph{International Symposium on Search Based Software Engineering}}. Springer, \bibinfo{pages}{140--146}.
\newblock


\bibitem[Grano et~al\mbox{.}(2018)]%
        {grano2018empirical}
\bibfield{author}{\bibinfo{person}{Giovanni Grano}, \bibinfo{person}{Simone Scalabrino}, \bibinfo{person}{Harald~C Gall}, {and} \bibinfo{person}{Rocco Oliveto}.} \bibinfo{year}{2018}\natexlab{}.
\newblock \showarticletitle{An empirical investigation on the readability of manual and generated test cases}. In \bibinfo{booktitle}{\emph{Proceedings of the 26th Conference on Program Comprehension}}. \bibinfo{pages}{348--351}.
\newblock


\bibitem[Guo et~al\mbox{.}(2020)]%
        {guo2020graphcodebert}
\bibfield{author}{\bibinfo{person}{Daya Guo}, \bibinfo{person}{Shuo Ren}, \bibinfo{person}{Shuai Lu}, \bibinfo{person}{Zhangyin Feng}, \bibinfo{person}{Duyu Tang}, \bibinfo{person}{Shujie Liu}, \bibinfo{person}{Long Zhou}, \bibinfo{person}{Nan Duan}, \bibinfo{person}{Alexey Svyatkovskiy}, \bibinfo{person}{Shengyu Fu}, {et~al\mbox{.}}} \bibinfo{year}{2020}\natexlab{}.
\newblock \showarticletitle{Graphcodebert: Pre-training code representations with data flow}.
\newblock \bibinfo{journal}{\emph{arXiv preprint arXiv:2009.08366}} (\bibinfo{year}{2020}).
\newblock


\bibitem[Hsieh et~al\mbox{.}(2024)]%
        {hsieh2024found}
\bibfield{author}{\bibinfo{person}{Cheng-Yu Hsieh}, \bibinfo{person}{Yung-Sung Chuang}, \bibinfo{person}{Chun-Liang Li}, \bibinfo{person}{Zifeng Wang}, \bibinfo{person}{Long Le}, \bibinfo{person}{Abhishek Kumar}, \bibinfo{person}{James Glass}, \bibinfo{person}{Alexander Ratner}, \bibinfo{person}{Chen-Yu Lee}, \bibinfo{person}{Ranjay Krishna}, {et~al\mbox{.}}} \bibinfo{year}{2024}\natexlab{}.
\newblock \showarticletitle{Found in the middle: Calibrating positional attention bias improves long context utilization}. In \bibinfo{booktitle}{\emph{Findings of the Association for Computational Linguistics: ACL 2024}}. \bibinfo{pages}{14982--14995}.
\newblock


\bibitem[Iyer et~al\mbox{.}(2016)]%
        {iyer2016summarizing}
\bibfield{author}{\bibinfo{person}{Srinivasan Iyer}, \bibinfo{person}{Ioannis Konstas}, \bibinfo{person}{Alvin Cheung}, {and} \bibinfo{person}{Luke Zettlemoyer}.} \bibinfo{year}{2016}\natexlab{}.
\newblock \showarticletitle{Summarizing source code using a neural attention model}. In \bibinfo{booktitle}{\emph{54th Annual Meeting of the Association for Computational Linguistics 2016}}. Association for Computational Linguistics, \bibinfo{pages}{2073--2083}.
\newblock


\bibitem[Jahangirova and Terragni(2023)]%
        {jahangirova2023sbft}
\bibfield{author}{\bibinfo{person}{Gunel Jahangirova} {and} \bibinfo{person}{Valerio Terragni}.} \bibinfo{year}{2023}\natexlab{}.
\newblock \showarticletitle{SBFT tool competition 2023-Java test case generation track}. In \bibinfo{booktitle}{\emph{2023 IEEE/ACM International Workshop on Search-Based and Fuzz Testing (SBFT)}}. IEEE, \bibinfo{pages}{61--64}.
\newblock


\bibitem[Just et~al\mbox{.}(2014)]%
        {just2014defects4j}
\bibfield{author}{\bibinfo{person}{Ren{\'e} Just}, \bibinfo{person}{Darioush Jalali}, {and} \bibinfo{person}{Michael~D Ernst}.} \bibinfo{year}{2014}\natexlab{}.
\newblock \showarticletitle{Defects4J: A database of existing faults to enable controlled testing studies for Java programs}. In \bibinfo{booktitle}{\emph{Proceedings of the 2014 international symposium on software testing and analysis}}. \bibinfo{pages}{437--440}.
\newblock


\bibitem[Kalai et~al\mbox{.}(2025)]%
        {kalai2025why}
\bibfield{author}{\bibinfo{person}{Adam Kalai}, \bibinfo{person}{Ofir Nachum}, \bibinfo{person}{Santosh Vempala}, {and} \bibinfo{person}{Edwin Zhang}.} \bibinfo{year}{2025}\natexlab{}.
\newblock \bibinfo{title}{Why Language Models Hallucinate}.
\newblock
\newblock
\urldef\tempurl%
\url{https://doi.org/10.48550/arXiv.2509.04664}
\showDOI{\tempurl}


\bibitem[Lemieux et~al\mbox{.}(2023)]%
        {lemieux2023codamosa}
\bibfield{author}{\bibinfo{person}{Caroline Lemieux}, \bibinfo{person}{Jeevana~Priya Inala}, \bibinfo{person}{Shuvendu~K Lahiri}, {and} \bibinfo{person}{Siddhartha Sen}.} \bibinfo{year}{2023}\natexlab{}.
\newblock \showarticletitle{Codamosa: Escaping coverage plateaus in test generation with pre-trained large language models}. In \bibinfo{booktitle}{\emph{2023 IEEE/ACM 45th International Conference on Software Engineering (ICSE)}}. IEEE, \bibinfo{pages}{919--931}.
\newblock


\bibitem[Lin(2004)]%
        {lin2004rouge}
\bibfield{author}{\bibinfo{person}{Chin-Yew Lin}.} \bibinfo{year}{2004}\natexlab{}.
\newblock \showarticletitle{Rouge: A package for automatic evaluation of summaries}. In \bibinfo{booktitle}{\emph{Text summarization branches out}}. \bibinfo{pages}{74--81}.
\newblock


\bibitem[Lin and Och(2004)]%
        {lin2004automatic}
\bibfield{author}{\bibinfo{person}{Chin-Yew Lin} {and} \bibinfo{person}{Franz~Josef Och}.} \bibinfo{year}{2004}\natexlab{}.
\newblock \showarticletitle{Automatic evaluation of machine translation quality using longest common subsequence and skip-bigram statistics}. In \bibinfo{booktitle}{\emph{Proceedings of the 42nd annual meeting of the association for computational linguistics (ACL-04)}}. \bibinfo{pages}{605--612}.
\newblock


\bibitem[Liu et~al\mbox{.}(2024)]%
        {liu2024lost}
\bibfield{author}{\bibinfo{person}{Nelson~F Liu}, \bibinfo{person}{Kevin Lin}, \bibinfo{person}{John Hewitt}, \bibinfo{person}{Ashwin Paranjape}, \bibinfo{person}{Michele Bevilacqua}, \bibinfo{person}{Fabio Petroni}, {and} \bibinfo{person}{Percy Liang}.} \bibinfo{year}{2024}\natexlab{}.
\newblock \showarticletitle{Lost in the middle: How language models use long contexts}.
\newblock \bibinfo{journal}{\emph{Transactions of the association for computational linguistics}}  \bibinfo{volume}{12} (\bibinfo{year}{2024}), \bibinfo{pages}{157--173}.
\newblock


\bibitem[Lukasczyk and Fraser(2022)]%
        {lukasczyk2022pynguin}
\bibfield{author}{\bibinfo{person}{Stephan Lukasczyk} {and} \bibinfo{person}{Gordon Fraser}.} \bibinfo{year}{2022}\natexlab{}.
\newblock \showarticletitle{Pynguin: Automated unit test generation for python}. In \bibinfo{booktitle}{\emph{Proceedings of the ACM/IEEE 44th International Conference on Software Engineering: Companion Proceedings}}. \bibinfo{pages}{168--172}.
\newblock


\bibitem[Mastropaolo et~al\mbox{.}(2024)]%
        {mastropaolo2024evaluating}
\bibfield{author}{\bibinfo{person}{Antonio Mastropaolo}, \bibinfo{person}{Matteo Ciniselli}, \bibinfo{person}{Massimiliano Di~Penta}, {and} \bibinfo{person}{Gabriele Bavota}.} \bibinfo{year}{2024}\natexlab{}.
\newblock \showarticletitle{Evaluating Code Summarization Techniques: A New Metric and an Empirical Characterization}. In \bibinfo{booktitle}{\emph{Proceedings of the IEEE/ACM 46th International Conference on Software Engineering}}. \bibinfo{pages}{1--13}.
\newblock


\bibitem[McCabe(1976)]%
        {mccabe1976complexity}
\bibfield{author}{\bibinfo{person}{Thomas~J McCabe}.} \bibinfo{year}{1976}\natexlab{}.
\newblock \showarticletitle{A complexity measure}.
\newblock \bibinfo{journal}{\emph{IEEE Transactions on software Engineering}} \bibinfo{number}{4} (\bibinfo{year}{1976}), \bibinfo{pages}{308--320}.
\newblock


\bibitem[Melo et~al\mbox{.}(2025)]%
        {melo2025agentic}
\bibfield{author}{\bibinfo{person}{Rian Melo}, \bibinfo{person}{Pedro Sim{\~o}es}, \bibinfo{person}{Rohit Gheyi}, \bibinfo{person}{Marcelo d’Amorim}, \bibinfo{person}{M{\'a}rcio Ribeiro}, \bibinfo{person}{Gustavo Soares}, \bibinfo{person}{Eduardo Almeida}, {and} \bibinfo{person}{Elvys Soares}.} \bibinfo{year}{2025}\natexlab{}.
\newblock \showarticletitle{Agentic LMs: Hunting Down Test Smells}.
\newblock \bibinfo{journal}{\emph{IEEE Software}} (\bibinfo{year}{2025}).
\newblock


\bibitem[Oliveira et~al\mbox{.}(2022)]%
        {oliveira2022systematic}
\bibfield{author}{\bibinfo{person}{Delano Oliveira}, \bibinfo{person}{Reyde Bruno}, \bibinfo{person}{Fernanda Madeiral}, \bibinfo{person}{Hidehiko Masuhara}, {and} \bibinfo{person}{Fernando Castor}.} \bibinfo{year}{2022}\natexlab{}.
\newblock \showarticletitle{A systematic literature review on the impact of formatting elements on program understandability}.
\newblock \bibinfo{journal}{\emph{Available at SSRN 4182156}} (\bibinfo{year}{2022}).
\newblock


\bibitem[Ou{\'e}draogo et~al\mbox{.}(2024)]%
        {ouedraogo2024large}
\bibfield{author}{\bibinfo{person}{Wendk{\^u}uni~C Ou{\'e}draogo}, \bibinfo{person}{Kader Kabor{\'e}}, \bibinfo{person}{Haoye Tian}, \bibinfo{person}{Yewei Song}, \bibinfo{person}{Anil Koyuncu}, \bibinfo{person}{Jacques Klein}, \bibinfo{person}{David Lo}, {and} \bibinfo{person}{Tegawend{\'e}~F Bissyand{\'e}}.} \bibinfo{year}{2024}\natexlab{}.
\newblock \showarticletitle{Large-scale, Independent and Comprehensive study of the power of LLMs for test case generation}.
\newblock \bibinfo{journal}{\emph{arXiv preprint arXiv:2407.00225}} (\bibinfo{year}{2024}).
\newblock


\bibitem[Ouedraogo et~al\mbox{.}(2024)]%
        {ouedraogo2024llms}
\bibfield{author}{\bibinfo{person}{Wendkuuni~C Ouedraogo}, \bibinfo{person}{Kader Kabore}, \bibinfo{person}{Haoye Tian}, \bibinfo{person}{Yewei Song}, \bibinfo{person}{Anil Koyuncu}, \bibinfo{person}{Jacques Klein}, \bibinfo{person}{David Lo}, {and} \bibinfo{person}{Tegawende~F Bissyande}.} \bibinfo{year}{2024}\natexlab{}.
\newblock \showarticletitle{Llms and prompting for unit test generation: A large-scale evaluation}. In \bibinfo{booktitle}{\emph{Proceedings of the 39th IEEE/ACM International Conference on Automated Software Engineering}}. \bibinfo{pages}{2464--2465}.
\newblock


\bibitem[Ou{\'e}draogo et~al\mbox{.}(2025a)]%
        {ouedraogo2025human}
\bibfield{author}{\bibinfo{person}{Wendk{\^u}uni~C Ou{\'e}draogo}, \bibinfo{person}{Yinghua Li}, \bibinfo{person}{Xueqi Dang}, \bibinfo{person}{Pawel Borsukiewicz}, \bibinfo{person}{Xin Zhou}, \bibinfo{person}{Anil Koyuncu}, \bibinfo{person}{Jacques Klein}, \bibinfo{person}{David Lo}, {and} \bibinfo{person}{Tegawend{\'e}~F Bissyand{\'e}}.} \bibinfo{year}{2025}\natexlab{a}.
\newblock \showarticletitle{Human-Aligned Code Readability Assessment with Large Language Models}.
\newblock \bibinfo{journal}{\emph{arXiv preprint arXiv:2510.16579}} (\bibinfo{year}{2025}).
\newblock


\bibitem[Ou{\'e}draogo et~al\mbox{.}(2024)]%
        {ouedraogo2024test}
\bibfield{author}{\bibinfo{person}{Wendk{\^u}uni~C Ou{\'e}draogo}, \bibinfo{person}{Yinghua Li}, \bibinfo{person}{Xueqi Dang}, \bibinfo{person}{Xunzhu Tang}, \bibinfo{person}{Anil Koyuncu}, \bibinfo{person}{Jacques Klein}, \bibinfo{person}{David Lo}, {and} \bibinfo{person}{Tegawend{\'e}~F Bissyand{\'e}}.} \bibinfo{year}{2024}\natexlab{}.
\newblock \showarticletitle{Test smells in llm-generated unit tests}.
\newblock \bibinfo{journal}{\emph{arXiv preprint arXiv:2410.10628}} (\bibinfo{year}{2024}).
\newblock


\bibitem[Ou{\'e}draogo et~al\mbox{.}(2025b)]%
        {ouedraogo2025beyond}
\bibfield{author}{\bibinfo{person}{Wendk{\^u}uni~C Ou{\'e}draogo}, \bibinfo{person}{Yinghua Li}, \bibinfo{person}{Xueqi Dang}, \bibinfo{person}{Xin Zhou}, \bibinfo{person}{Anil Koyuncu}, \bibinfo{person}{Jacques Klein}, \bibinfo{person}{David Lo}, {and} \bibinfo{person}{Tegawend{\'e}~F Bissyand{\'e}}.} \bibinfo{year}{2025}\natexlab{b}.
\newblock \showarticletitle{Beyond Surface Similarity: Evaluating LLM-Based Test Refactorings with Structural and Semantic Awareness}.
\newblock \bibinfo{journal}{\emph{arXiv preprint arXiv:2506.06767}} (\bibinfo{year}{2025}).
\newblock


\bibitem[Ou{\'e}draogo et~al\mbox{.}(2025c)]%
        {ouedraogo2025rethinking}
\bibfield{author}{\bibinfo{person}{Wendk{\^u}uni~C Ou{\'e}draogo}, \bibinfo{person}{Yinghua Li}, \bibinfo{person}{Xueqi Dang}, \bibinfo{person}{Xin Zhou}, \bibinfo{person}{Anil Koyuncu}, \bibinfo{person}{Jacques Klein}, \bibinfo{person}{David Lo}, {and} \bibinfo{person}{Tegawend{\'e}~F Bissyand{\'e}}.} \bibinfo{year}{2025}\natexlab{c}.
\newblock \showarticletitle{Rethinking Cognitive Complexity for Unit Tests: Toward a Readability-Aware Metric Grounded in Developer Perception}. In \bibinfo{booktitle}{\emph{2025 IEEE International Conference on Software Maintenance and Evolution (ICSME)}}. IEEE, \bibinfo{pages}{797--802}.
\newblock


\bibitem[Ou{\'e}draogo et~al\mbox{.}(2025d)]%
        {ouedraogo2025enriching}
\bibfield{author}{\bibinfo{person}{Wendk{\^u}uni~C Ou{\'e}draogo}, \bibinfo{person}{Laura Plein}, \bibinfo{person}{Kader Kabore}, \bibinfo{person}{Andrew Habib}, \bibinfo{person}{Jacques Klein}, \bibinfo{person}{David Lo}, {and} \bibinfo{person}{Tegawend{\'e}~F Bissyand{\'e}}.} \bibinfo{year}{2025}\natexlab{d}.
\newblock \showarticletitle{Enriching automatic test case generation by extracting relevant test inputs from bug reports}.
\newblock \bibinfo{journal}{\emph{Empirical Software Engineering}} \bibinfo{volume}{30}, \bibinfo{number}{3} (\bibinfo{year}{2025}), \bibinfo{pages}{85}.
\newblock


\bibitem[Oueslati et~al\mbox{.}(2025)]%
        {oueslati2025refagent}
\bibfield{author}{\bibinfo{person}{Khouloud Oueslati}, \bibinfo{person}{Maxime Lamothe}, {and} \bibinfo{person}{Foutse Khomh}.} \bibinfo{year}{2025}\natexlab{}.
\newblock \showarticletitle{RefAgent: A Multi-agent LLM-based Framework for Automatic Software Refactoring}.
\newblock \bibinfo{journal}{\emph{arXiv preprint arXiv:2511.03153}} (\bibinfo{year}{2025}).
\newblock


\bibitem[Palomba et~al\mbox{.}(2016)]%
        {palomba2016diffusion}
\bibfield{author}{\bibinfo{person}{Fabio Palomba}, \bibinfo{person}{Dario Di~Nucci}, \bibinfo{person}{Annibale Panichella}, \bibinfo{person}{Rocco Oliveto}, {and} \bibinfo{person}{Andrea De~Lucia}.} \bibinfo{year}{2016}\natexlab{}.
\newblock \showarticletitle{On the diffusion of test smells in automatically generated test code: An empirical study}. In \bibinfo{booktitle}{\emph{Proceedings of the 9th international workshop on search-based software testing}}. \bibinfo{pages}{5--14}.
\newblock


\bibitem[Pan et~al\mbox{.}(2025)]%
        {pan2025aster}
\bibfield{author}{\bibinfo{person}{Rangeet Pan}, \bibinfo{person}{Myeongsoo Kim}, \bibinfo{person}{Rahul Krishna}, \bibinfo{person}{Raju Pavuluri}, {and} \bibinfo{person}{Saurabh Sinha}.} \bibinfo{year}{2025}\natexlab{}.
\newblock \showarticletitle{Aster: Natural and multi-language unit test generation with llms}. In \bibinfo{booktitle}{\emph{2025 IEEE/ACM 47th International Conference on Software Engineering: Software Engineering in Practice (ICSE-SEIP)}}. IEEE, \bibinfo{pages}{413--424}.
\newblock


\bibitem[Panichella et~al\mbox{.}(2017)]%
        {panichella2017automated}
\bibfield{author}{\bibinfo{person}{Annibale Panichella}, \bibinfo{person}{Fitsum~Meshesha Kifetew}, {and} \bibinfo{person}{Paolo Tonella}.} \bibinfo{year}{2017}\natexlab{}.
\newblock \showarticletitle{Automated test case generation as a many-objective optimisation problem with dynamic selection of the targets}.
\newblock \bibinfo{journal}{\emph{IEEE Transactions on Software Engineering}} \bibinfo{volume}{44}, \bibinfo{number}{2} (\bibinfo{year}{2017}), \bibinfo{pages}{122--158}.
\newblock


\bibitem[Panichella et~al\mbox{.}(2020)]%
        {panichella2020revisiting}
\bibfield{author}{\bibinfo{person}{Annibale Panichella}, \bibinfo{person}{Sebastiano Panichella}, \bibinfo{person}{Gordon Fraser}, \bibinfo{person}{Anand~Ashok Sawant}, {and} \bibinfo{person}{Vincent~J Hellendoorn}.} \bibinfo{year}{2020}\natexlab{}.
\newblock \showarticletitle{Revisiting test smells in automatically generated tests: limitations, pitfalls, and opportunities}. In \bibinfo{booktitle}{\emph{2020 IEEE international conference on software maintenance and evolution (ICSME)}}. IEEE, \bibinfo{pages}{523--533}.
\newblock


\bibitem[Peruma et~al\mbox{.}(2020)]%
        {peruma2020tsdetect}
\bibfield{author}{\bibinfo{person}{Anthony Peruma}, \bibinfo{person}{Khalid Almalki}, \bibinfo{person}{Christian~D Newman}, \bibinfo{person}{Mohamed~Wiem Mkaouer}, \bibinfo{person}{Ali Ouni}, {and} \bibinfo{person}{Fabio Palomba}.} \bibinfo{year}{2020}\natexlab{}.
\newblock \showarticletitle{Tsdetect: An open source test smells detection tool}. In \bibinfo{booktitle}{\emph{Proceedings of the 28th ACM joint meeting on european software engineering conference and symposium on the foundations of software engineering}}. \bibinfo{pages}{1650--1654}.
\newblock


\bibitem[Ramos et~al\mbox{.}(2025)]%
        {ramos2025large}
\bibfield{author}{\bibinfo{person}{Daniel Ramos}, \bibinfo{person}{Claudia Mamede}, \bibinfo{person}{Kush Jain}, \bibinfo{person}{Paulo Canelas}, \bibinfo{person}{Catarina Gamboa}, {and} \bibinfo{person}{Claire Le~Goues}.} \bibinfo{year}{2025}\natexlab{}.
\newblock \showarticletitle{Are large language models memorizing bug benchmarks?}. In \bibinfo{booktitle}{\emph{2025 IEEE/ACM International Workshop on Large Language Models for Code (LLM4Code)}}. IEEE, \bibinfo{pages}{1--8}.
\newblock


\bibitem[Ren et~al\mbox{.}(2020)]%
        {ren2020codebleu}
\bibfield{author}{\bibinfo{person}{Shuo Ren}, \bibinfo{person}{Daya Guo}, \bibinfo{person}{Shuai Lu}, \bibinfo{person}{Long Zhou}, \bibinfo{person}{Shujie Liu}, \bibinfo{person}{Duyu Tang}, \bibinfo{person}{Neel Sundaresan}, \bibinfo{person}{Ming Zhou}, \bibinfo{person}{Ambrosio Blanco}, {and} \bibinfo{person}{Shuai Ma}.} \bibinfo{year}{2020}\natexlab{}.
\newblock \showarticletitle{Codebleu: a method for automatic evaluation of code synthesis}.
\newblock \bibinfo{journal}{\emph{arXiv preprint arXiv:2009.10297}} (\bibinfo{year}{2020}).
\newblock


\bibitem[Roy et~al\mbox{.}(2020)]%
        {roy2020deeptc}
\bibfield{author}{\bibinfo{person}{Devjeet Roy}, \bibinfo{person}{Ziyi Zhang}, \bibinfo{person}{Maggie Ma}, \bibinfo{person}{Venera Arnaoudova}, \bibinfo{person}{Annibale Panichella}, \bibinfo{person}{Sebastiano Panichella}, \bibinfo{person}{Danielle Gonzalez}, {and} \bibinfo{person}{Mehdi Mirakhorli}.} \bibinfo{year}{2020}\natexlab{}.
\newblock \showarticletitle{DeepTC-Enhancer: Improving the readability of automatically generated tests}. In \bibinfo{booktitle}{\emph{Proceedings of the 35th IEEE/ACM International Conference on Automated Software Engineering}}. \bibinfo{pages}{287--298}.
\newblock


\bibitem[Scalabrino et~al\mbox{.}(2018)]%
        {scalabrino2018comprehensive}
\bibfield{author}{\bibinfo{person}{Simone Scalabrino}, \bibinfo{person}{Mario Linares-V{\'a}squez}, \bibinfo{person}{Rocco Oliveto}, {and} \bibinfo{person}{Denys Poshyvanyk}.} \bibinfo{year}{2018}\natexlab{}.
\newblock \showarticletitle{A comprehensive model for code readability}.
\newblock \bibinfo{journal}{\emph{Journal of Software: Evolution and Process}} \bibinfo{volume}{30}, \bibinfo{number}{6} (\bibinfo{year}{2018}), \bibinfo{pages}{e1958}.
\newblock


\bibitem[Sergeyuk et~al\mbox{.}(2024a)]%
        {sergeyuk2024assessing}
\bibfield{author}{\bibinfo{person}{Agnia Sergeyuk}, \bibinfo{person}{Olga Lvova}, \bibinfo{person}{Sergey Titov}, \bibinfo{person}{Anastasiia Serova}, \bibinfo{person}{Farid Bagirov}, {and} \bibinfo{person}{Timofey Bryksin}.} \bibinfo{year}{2024}\natexlab{a}.
\newblock \showarticletitle{Assessing Consensus of Developers' Views on Code Readability}.
\newblock \bibinfo{journal}{\emph{arXiv preprint arXiv:2407.03790}} (\bibinfo{year}{2024}).
\newblock


\bibitem[Sergeyuk et~al\mbox{.}(2024b)]%
        {sergeyuk2024reassessing}
\bibfield{author}{\bibinfo{person}{Agnia Sergeyuk}, \bibinfo{person}{Olga Lvova}, \bibinfo{person}{Sergey Titov}, \bibinfo{person}{Anastasiia Serova}, \bibinfo{person}{Farid Bagirov}, \bibinfo{person}{Evgeniia Kirillova}, {and} \bibinfo{person}{Timofey Bryksin}.} \bibinfo{year}{2024}\natexlab{b}.
\newblock \showarticletitle{Reassessing Java Code Readability Models with a Human-Centered Approach}. In \bibinfo{booktitle}{\emph{Proceedings of the 32nd IEEE/ACM International Conference on Program Comprehension}}. \bibinfo{pages}{225--235}.
\newblock


\bibitem[Shamshiri et~al\mbox{.}(2015)]%
        {shamshiri2015automatically}
\bibfield{author}{\bibinfo{person}{Sina Shamshiri}, \bibinfo{person}{Ren{\'e} Just}, \bibinfo{person}{Jos{\'e}~Miguel Rojas}, \bibinfo{person}{Gordon Fraser}, \bibinfo{person}{Phil McMinn}, {and} \bibinfo{person}{Andrea Arcuri}.} \bibinfo{year}{2015}\natexlab{}.
\newblock \showarticletitle{Do automatically generated unit tests find real faults? an empirical study of effectiveness and challenges (t)}. In \bibinfo{booktitle}{\emph{2015 30th IEEE/ACM International Conference on Automated Software Engineering (ASE)}}. IEEE, \bibinfo{pages}{201--211}.
\newblock


\bibitem[Shin et~al\mbox{.}(2024)]%
        {shin2024retrieval}
\bibfield{author}{\bibinfo{person}{Jiho Shin}, \bibinfo{person}{Nima~Shiri Harzevili}, \bibinfo{person}{Reem Aleithan}, \bibinfo{person}{Hadi Hemmati}, {and} \bibinfo{person}{Song Wang}.} \bibinfo{year}{2024}\natexlab{}.
\newblock \showarticletitle{Retrieval-augmented test generation: How far are we?}
\newblock \bibinfo{journal}{\emph{arXiv preprint arXiv:2409.12682}} (\bibinfo{year}{2024}).
\newblock


\bibitem[Shore and Warden(2021)]%
        {shore2021art}
\bibfield{author}{\bibinfo{person}{James Shore} {and} \bibinfo{person}{Shane Warden}.} \bibinfo{year}{2021}\natexlab{}.
\newblock \bibinfo{booktitle}{\emph{The art of agile development}}.
\newblock \bibinfo{publisher}{" O'Reilly Media, Inc."}.
\newblock


\bibitem[Siddiq et~al\mbox{.}(2024)]%
        {siddiq2024using}
\bibfield{author}{\bibinfo{person}{Mohammed~Latif Siddiq}, \bibinfo{person}{Joanna~CS Santos}, \bibinfo{person}{Ridwanul~Hasan Tanvir}, \bibinfo{person}{Noshin Ulfat}, \bibinfo{person}{Fahmid Al~Rifat}, {and} \bibinfo{person}{Vin{\'\i}cius~Carvalho Lopes}.} \bibinfo{year}{2024}\natexlab{}.
\newblock \showarticletitle{Using Large Language Models to Generate JUnit Tests: An Empirical Study}.
\newblock  (\bibinfo{year}{2024}).
\newblock


\bibitem[Siddiqui(2021)]%
        {siddiqui2021learning}
\bibfield{author}{\bibinfo{person}{Saleem Siddiqui}.} \bibinfo{year}{2021}\natexlab{}.
\newblock \bibinfo{booktitle}{\emph{Learning Test-Driven Development}}.
\newblock \bibinfo{publisher}{" O'Reilly Media, Inc."}.
\newblock


\bibitem[Sun et~al\mbox{.}(2024)]%
        {sun2024source}
\bibfield{author}{\bibinfo{person}{Weisong Sun}, \bibinfo{person}{Yun Miao}, \bibinfo{person}{Yuekang Li}, \bibinfo{person}{Hongyu Zhang}, \bibinfo{person}{Chunrong Fang}, \bibinfo{person}{Yi Liu}, \bibinfo{person}{Gelei Deng}, \bibinfo{person}{Yang Liu}, {and} \bibinfo{person}{Zhenyu Chen}.} \bibinfo{year}{2024}\natexlab{}.
\newblock \showarticletitle{Source Code Summarization in the Era of Large Language Models}. In \bibinfo{booktitle}{\emph{2025 IEEE/ACM 47th International Conference on Software Engineering (ICSE)}}. IEEE Computer Society, \bibinfo{pages}{419--431}.
\newblock


\bibitem[Tang et~al\mbox{.}(2024)]%
        {tang2024chatgpt}
\bibfield{author}{\bibinfo{person}{Yutian Tang}, \bibinfo{person}{Zhijie Liu}, \bibinfo{person}{Zhichao Zhou}, {and} \bibinfo{person}{Xiapu Luo}.} \bibinfo{year}{2024}\natexlab{}.
\newblock \showarticletitle{Chatgpt vs sbst: A comparative assessment of unit test suite generation}.
\newblock \bibinfo{journal}{\emph{IEEE Transactions on Software Engineering}} (\bibinfo{year}{2024}).
\newblock


\bibitem[Van~Deursen et~al\mbox{.}(2001)]%
        {van2001refactoring}
\bibfield{author}{\bibinfo{person}{Arie Van~Deursen}, \bibinfo{person}{Leon Moonen}, \bibinfo{person}{Alex Van Den~Bergh}, {and} \bibinfo{person}{Gerard Kok}.} \bibinfo{year}{2001}\natexlab{}.
\newblock \showarticletitle{Refactoring test code}. In \bibinfo{booktitle}{\emph{Proceedings of the 2nd international conference on extreme programming and flexible processes in software engineering (XP2001)}}. Citeseer, \bibinfo{pages}{92--95}.
\newblock


\bibitem[Wang et~al\mbox{.}(2024)]%
        {wang2024software}
\bibfield{author}{\bibinfo{person}{Junjie Wang}, \bibinfo{person}{Yuchao Huang}, \bibinfo{person}{Chunyang Chen}, \bibinfo{person}{Zhe Liu}, \bibinfo{person}{Song Wang}, {and} \bibinfo{person}{Qing Wang}.} \bibinfo{year}{2024}\natexlab{}.
\newblock \showarticletitle{Software testing with large language models: Survey, landscape, and vision}.
\newblock \bibinfo{journal}{\emph{IEEE Transactions on Software Engineering}} (\bibinfo{year}{2024}).
\newblock


\bibitem[Wang et~al\mbox{.}(2022)]%
        {wang2022self}
\bibfield{author}{\bibinfo{person}{Xuezhi Wang}, \bibinfo{person}{Jason Wei}, \bibinfo{person}{Dale Schuurmans}, \bibinfo{person}{Quoc Le}, \bibinfo{person}{Ed Chi}, \bibinfo{person}{Sharan Narang}, \bibinfo{person}{Aakanksha Chowdhery}, {and} \bibinfo{person}{Denny Zhou}.} \bibinfo{year}{2022}\natexlab{}.
\newblock \showarticletitle{Self-consistency improves chain of thought reasoning in language models}.
\newblock \bibinfo{journal}{\emph{arXiv preprint arXiv:2203.11171}} (\bibinfo{year}{2022}).
\newblock


\bibitem[Wei et~al\mbox{.}(2022)]%
        {wei2022chain}
\bibfield{author}{\bibinfo{person}{Jason Wei}, \bibinfo{person}{Xuezhi Wang}, \bibinfo{person}{Dale Schuurmans}, \bibinfo{person}{Maarten Bosma}, \bibinfo{person}{Fei Xia}, \bibinfo{person}{Ed Chi}, \bibinfo{person}{Quoc~V Le}, \bibinfo{person}{Denny Zhou}, {et~al\mbox{.}}} \bibinfo{year}{2022}\natexlab{}.
\newblock \showarticletitle{Chain-of-thought prompting elicits reasoning in large language models}.
\newblock \bibinfo{journal}{\emph{Advances in neural information processing systems}}  \bibinfo{volume}{35} (\bibinfo{year}{2022}), \bibinfo{pages}{24824--24837}.
\newblock


\bibitem[Yuan et~al\mbox{.}(2023)]%
        {yuan2023no}
\bibfield{author}{\bibinfo{person}{Zhiqiang Yuan}, \bibinfo{person}{Yiling Lou}, \bibinfo{person}{Mingwei Liu}, \bibinfo{person}{Shiji Ding}, \bibinfo{person}{Kaixin Wang}, \bibinfo{person}{Yixuan Chen}, {and} \bibinfo{person}{Xin Peng}.} \bibinfo{year}{2023}\natexlab{}.
\newblock \showarticletitle{No more manual tests? evaluating and improving chatgpt for unit test generation}.
\newblock \bibinfo{journal}{\emph{arXiv preprint arXiv:2305.04207}} (\bibinfo{year}{2023}).
\newblock


\bibitem[Zhang et~al\mbox{.}(2026)]%
        {zhang2026empirical}
\bibfield{author}{\bibinfo{person}{Yang Zhang}, \bibinfo{person}{Lijie Yuan}, {and} \bibinfo{person}{Chunhao Dong}.} \bibinfo{year}{2026}\natexlab{}.
\newblock \showarticletitle{An empirical study of LLM-based refactoring consistency}.
\newblock \bibinfo{journal}{\emph{Empirical Software Engineering}} \bibinfo{volume}{31}, \bibinfo{number}{6} (\bibinfo{year}{2026}), \bibinfo{pages}{178}.
\newblock


\bibitem[Zhou et~al\mbox{.}(2024)]%
        {zhou2024llm}
\bibfield{author}{\bibinfo{person}{Zhichao Zhou}, \bibinfo{person}{Yutian Tang}, \bibinfo{person}{Yun Lin}, {and} \bibinfo{person}{Jingzhu He}.} \bibinfo{year}{2024}\natexlab{}.
\newblock \showarticletitle{An LLM-based readability measurement for unit tests' context-aware inputs}.
\newblock \bibinfo{journal}{\emph{arXiv preprint arXiv:2407.21369}} (\bibinfo{year}{2024}).
\newblock


\end{thebibliography}

\appendix

\section{Additional Tables}
\label{app:additional-tables}


\begin{table*}[ht]
\caption{Test generation statistics across datasets and tools.}
\label{tab:test-generated}
\centering
\scalebox{0.6}{
\begin{tabular}{ccccccccccccc}
\toprule
\textbf{Dataset} &
\textbf{Model} &
\textbf{\#Projects} &
\textbf{\#Classes} &
\textbf{\begin{tabular}[c]{@{}c@{}}Test Generation\\ Rate (\%)\end{tabular}} &
\textbf{\begin{tabular}[c]{@{}c@{}}Compilability\\ Rate (\%)\end{tabular}} &
\textbf{\begin{tabular}[c]{@{}c@{}}Tokens\\ Max\end{tabular}} &
\textbf{\begin{tabular}[c]{@{}c@{}}Tokens\\ Min\end{tabular}} &
\textbf{\begin{tabular}[c]{@{}c@{}}Tokens\\ Average\end{tabular}} &
\textbf{\begin{tabular}[c]{@{}c@{}}LOC\\ Max\end{tabular}} &
\textbf{\begin{tabular}[c]{@{}c@{}}LOC\\ Min\end{tabular}} &
\textbf{\begin{tabular}[c]{@{}c@{}}LOC\\ Average\end{tabular}} &
\textbf{\begin{tabular}[c]{@{}c@{}}Test Methods\\ Average\end{tabular}} \\
\midrule
\multirow{3}{*}{Defects4J} &
gpt-4o &
\multirow{3}{*}{15} &
\multirow{3}{*}{147} &
99.77 &
68.48 & 1883 & 49 & 517.33 & 301 & 4 & 62.11 & 8.58 \\
&
mistral-large-2407 &
& &
95.24 &
51.50 & 3223 & 48 & 680.45 & 359 & 4 & 71.80 & 8.06 \\
&
EvoSuite &
& &
\textbf{100.00} &
\textbf{100.00} & 27896 & 156 & 1958.19 & 945 & 4 & 169.40 & \textbf{17.38} \\
\midrule
\multirow{3}{*}{SF110} &
gpt-4o &
\multirow{3}{*}{69} &
\multirow{3}{*}{203} &
98.52 &
78.00 & 2211 & 20 & 476.31 & 316 & 4 & 54.11 & 8.07 \\
&
mistral-large-2407 &
& &
97.87 &
77.83 & 2821 & 27 & 461.79 & 319 & 4 & 43.89 & 5.52 \\
&
EvoSuite &
& &
\textbf{100.00} &
\textbf{100.00} & 27968 & 151 & 3462.57 & 2879 & 4 & 294.14 & \textbf{27.82} \\
\bottomrule
\end{tabular}
}
\end{table*}

\begin{table*}[ht]
\caption{Refactoring effectiveness and structural characteristics of refactored test suites.}
\label{tab:refactoring-stats}
\centering
\scalebox{0.6}{
\begin{tabular}{ccccccccccccc}
\toprule
\textbf{Dataset} &
\textbf{Approach} &
\textbf{Model} &
\textbf{\begin{tabular}[c]{@{}c@{}}Refactoring\\ Rate (\%)\end{tabular}} &
\textbf{\begin{tabular}[c]{@{}c@{}}Compilation\\ Rate (\%)\end{tabular}} &
\textbf{\begin{tabular}[c]{@{}c@{}}Tokens\\ Max\end{tabular}} &
\textbf{\begin{tabular}[c]{@{}c@{}}Tokens\\ Min\end{tabular}} &
\textbf{\begin{tabular}[c]{@{}c@{}}Tokens\\ Average\end{tabular}} &
\textbf{\begin{tabular}[c]{@{}c@{}}LOC\\ Max\end{tabular}} &
\textbf{\begin{tabular}[c]{@{}c@{}}LOC\\ Min\end{tabular}} &
\textbf{\begin{tabular}[c]{@{}c@{}}LOC\\ Average\end{tabular}} &
\textbf{\begin{tabular}[c]{@{}c@{}}Test Methods\\ Average\end{tabular}} \\
\midrule
\multirow{6}{*}{Defects4J}
 & \multirow{2}{*}{Tests-only}
 & gpt-4o             & 85.09 & 90.66 & 13057 & 156 & 1950.33 & 1024 & 4 & 183.25 & 14.37 \\
 & 
 & mistral-large-2407 & 85.46 & 90.64 & 12360 & 153 & 2173.74 & 1440 & 4 & 203.67 & 15.50 \\
\cmidrule(lr){2-12}
 & \multirow{2}{*}{Code-centric}
 & gpt-4o             & 84.49 & 89.91 & 12193 & 156 & 1909.53 & 984  & 4 & 182.77 & 14.29 \\
 &
 & mistral-large-2407 & 83.93 & 88.52 & 12568 & 153 & 2050.32 & 1429 & 6 & 205.21 & 15.65 \\
\cmidrule(lr){2-12}
 & \multirow{2}{*}{Summary-based}
 & gpt-4o             & \textbf{95.00} & \textbf{97.24} & 11247 & 160 & 1902.54 & 960  & 4 & 182.16 & 14.16 \\
 &
 & mistral-large-2407 & 92.94 & 92.92 & 13229 & 154 & 2146.21 & 1440 & 6 & 204.27 & 15.54 \\
\midrule
\multirow{6}{*}{SF110}
 & \multirow{2}{*}{Tests-only}
 & gpt-4o             & 77.09 & 92.41 & 15439 & 202 & 2471.22 & 1578 & 5 & 243.07 & 18.63 \\
 &
 & mistral-large-2407 & 76.58 & 91.95 & 13185 & 148 & 2802.56 & 1589 & 6 & 255.50 & 18.85 \\
\cmidrule(lr){2-12}
 & \multirow{2}{*}{Code-centric}
 & gpt-4o             & 75.58 & 90.10 & 14829 & 209 & 2391.34 & 1580 & 5 & 239.50 & 18.32 \\
 &
 & mistral-large-2407 & 74.87 & 89.02 & 12999 & 188 & 2562.74 & 1607 & 11 & 251.40 & 18.60 \\
\cmidrule(lr){2-12}
 & \multirow{2}{*}{Summary-based}
 & gpt-4o             & \textbf{95.82} & \textbf{98.69} & 15653 & 203 & 2434.80 & 1580 & 11 & 242.08 & 18.59 \\
 &
 & mistral-large-2407 & 93.16 & 96.99 & 13253 & 189 & 2731.31 & 1641 & 11 & 253.85 & 18.77 \\
\bottomrule
\end{tabular}
}
\end{table*}

\begin{table*}[ht]
\caption{RQ4 Coverage Statistics per Dataset and Approach.}
\label{tab:rq4-coverage-stats}
\centering
\scalebox{0.72}{
\begin{tabular}{llrrrrr}
\toprule
\textbf{Dataset} & \textbf{Coverage / Source} & \textbf{Min} & \textbf{Q1} & \textbf{Mean} & \textbf{Q3} & \textbf{Max} \\
\midrule

\multirow{15}{*}{Defects4J}
& \multicolumn{6}{l}{\textbf{Instruction Coverage}} \\
& EvoSuite        & 74.93 & 91.04 & 92.83 & 97.33 & 100.00 \\
& gpt-4o          & 2.00  & 81.50 & 79.35 & 88.50 & 100.00 \\
& Tests-only      & 31.33 & 88.67 & 91.59 & 98.00 & 100.00 \\
& Code-centric    & 34.00 & 87.00 & 90.99 & 98.00 & 100.00 \\
& Summary-based   & 66.33 & 91.00 & \textbf{93.93} & 100.00 & 100.00 \\
\cmidrule(lr){2-7}

& \multicolumn{6}{l}{\textbf{Line Coverage}} \\
& EvoSuite        & 68.39 & 88.24 & 91.68 & 96.77 & 100.00 \\
& gpt-4o          & 0.94  & 76.21 & 73.76 & 85.20 & 100.00 \\
& Tests-only      & 40.71 & 83.82 & 90.03 & 97.25 & 100.00 \\
& Code-centric    & 40.71 & 82.99 & 89.54 & 97.22 & 100.00 \\
& Summary-based   & 60.78 & 86.22 & \textbf{92.29} & 100.00 & 100.00 \\
\cmidrule(lr){2-7}

& \multicolumn{6}{l}{\textbf{Method Coverage}} \\
& EvoSuite        & 78.19 & 100.00 & 99.60 & 100.00 & 100.00 \\
& gpt-4o          & 4.17  & 100.00 & 92.06 & 100.00 & 100.00 \\
& Tests-only      & 58.82 & 100.00 & 98.49 & 100.00 & 100.00 \\
& Code-centric    & 57.73 & 100.00 & 98.63 & 100.00 & 100.00 \\
& Summary-based   & 66.25 & 100.00 & \textbf{99.42} & 100.00 & 100.00 \\

\midrule

\multirow{15}{*}{SF110}
& \multicolumn{6}{l}{\textbf{Instruction Coverage}} \\
& EvoSuite        & 84.78 & 93.37 & 95.35 & 98.30 & 100.00 \\
& gpt-4o          & 10.00 & 60.00 & 56.57 & 60.00 & 100.00 \\
& Tests-only      & 30.67 & 92.00 & 94.48 & 98.00 & 100.00 \\
& Code-centric    & 76.33 & 92.00 & 94.68 & 98.00 & 100.00 \\
& Summary-based   & 78.00 & 93.00 & \textbf{95.32} & 100.00 & 100.00 \\
\cmidrule(lr){2-7}

& \multicolumn{6}{l}{\textbf{Line Coverage}} \\
& EvoSuite        & 80.77 & 90.82 & 93.89 & 98.01 & 100.00 \\
& gpt-4o          & 7.70  & 66.70 & 60.29 & 66.70 & 100.00 \\
& Tests-only      & 30.63 & 88.69 & 92.39 & 97.52 & 100.00 \\
& Code-centric    & 73.81 & 88.90 & 92.68 & 98.02 & 100.00 \\
& Summary-based   & 72.40 & 89.52 & \textbf{93.61} & 100.00 & 100.00 \\
\cmidrule(lr){2-7}

& \multicolumn{6}{l}{\textbf{Method Coverage}} \\
& EvoSuite        & 99.82 & 100.00 & 100.00 & 100.00 & 100.00 \\
& gpt-4o          & 20.00 & 25.00 & 35.23 & 25.00 & 100.00 \\
& Tests-only      & 33.33 & 100.00 & 99.88 & 100.00 & 100.00 \\
& Code-centric    & 90.95 & 100.00 & 99.95 & 100.00 & 100.00 \\
& Summary-based   & 95.22 & 100.00 & \textbf{99.98} & 100.00 & 100.00 \\

\bottomrule
\end{tabular}
}
\end{table*}

\begin{table*}[ht]
\caption{Detailed human-evaluation sample (per dataset, project, and class).}
\label{app:rq6-sample-details}
\centering
\scalebox{0.7}{
\begin{tabular}{ll l r r}
\toprule
\textbf{Dataset} & \textbf{Project} & \textbf{Class} & \textbf{LOC} & \textbf{Size bin} \\
\midrule

\multirow{15}{*}{Defects4J}
& Cli          & WriteableCommandLine         & 9   & Small  \\
& Compress     & DeflateCompressorInputStream & 24  & Small  \\
& Compress     & UnixStat                     & 67  & Small  \\
& Csv          & CSVRecord                    & 65  & Small  \\
& Mockito      & TypeBasedCandidateFilter     & 10  & Small  \\
\cmidrule(lr){2-5}
& Cli          & TypeHandler                  & 145 & Medium \\
& Codec        & Lang                         & 84  & Medium \\
& Compress     & ArchiveStreamFactory         & 118 & Medium \\
& Compress     & X7875\_NewUnix               & 164 & Medium \\
& Lang         & SerializationUtils           & 101 & Medium \\
\cmidrule(lr){2-5}
& Csv          & CSVParser                    & 212 & Large  \\
& Gson         & ISO8601Utils                 & 242 & Large  \\
& JacksonCore  & NumberInput                  & 247 & Large  \\
& JacksonDatabind & StdDateFormat              & 254 & Large  \\
& Lang         & ExtendedMessageFormat        & 245 & Large  \\

\midrule

\multirow{15}{*}{SF110}
& 21\_geo-google   & AddressToUsAddressFunctor & 54  & Small  \\
& 41\_follow       & TabbedPane                & 44  & Small  \\
& 87\_jaw-br       & Salvar                    & 16  & Small  \\
& commons-math     & AbstractSimplex           & 71  & Small  \\
& scribe           & Verifier                  & 3   & Small  \\
\cmidrule(lr){2-5}
& 60\_sugar        & FSPathExplorer            & 158 & Medium \\
& 77\_io-project   & ClientGroup               & 83  & Medium \\
& 86\_at-robots2-j & RobotRenderer             & 114 & Medium \\
& guava            & Suppliers                 & 125 & Medium \\
& javaml           & Fold                      & 174 & Medium \\
\cmidrule(lr){2-5}
& 47\_dvd-homevideo & Menu                    & 288 & Large  \\
& 69\_lhamacaw      & DisplayableListPanel    & 268 & Large  \\
& 85\_shop          & JSPredicateForm         & 218 & Large  \\
& guava             & Monitor                 & 442 & Large  \\
& guava             & Predicates              & 325 & Large  \\

\bottomrule
\end{tabular}
}
\end{table*}

\begin{table*}[ht]
\caption{Descriptive statistics of human evaluation scores (Likert 1--5 scale).}
\label{tab:rq6-human-stats}
\centering
\scalebox{0.7}{
\begin{tabular}{lllrrrrrr}
\toprule
\textbf{Metric} & \textbf{Evaluator} & \textbf{Approach} & \textbf{Min} & \textbf{Q1} & \textbf{Median} & \textbf{Mean} & \textbf{Q3} & \textbf{Max} \\
\midrule

\multirow{4}{*}{Readability}
& Advanced graduate student & EvoSuite & 2 & 2 & 2.5 & 2.71 & 3 & 4 \\
&                           & Summary  & 2 & 3 & 4   & 3.79 & 4 & 5 \\
& Senior developer          & EvoSuite & 1 & 1 & 2   & 2.14 & 3 & 4 \\
&                           & Summary  & 1 & 2 & 4   & 3.50 & 4 & 5 \\

\midrule

\multirow{4}{*}{Adoption}
& Advanced graduate student & EvoSuite & 1 & 2 & 3 & 2.57 & 3 & 4 \\
&                           & Summary  & 2 & 3 & 4 & 3.50 & 4 & 5 \\
& Senior developer          & EvoSuite & 1 & 1 & 2 & 2.14 & 3 & 4 \\
&                           & Summary  & 2 & 2 & 4 & 3.25 & 4 & 5 \\

\bottomrule
\end{tabular}
}
\end{table*}

\begin{table*}[ht]
\caption{Detailed inter-annotator agreement metrics for RQ6 (per group and approach).}
\label{tab:rq6_iaa_detailed}
\centering
\scalebox{0.6}{
\begin{tabular}{llccccc}
\toprule
\textbf{Scope} & \textbf{Approach} & $\alpha$ & QWK & \%Agree & $\rho$ & $p$ \\
\midrule
Group A & EvoSuite      & 0.79 & 0.78 & 0.71 & 0.89 & $1.75\times10^{-5}$ \\
        & TestHumanizer & 0.74 & 0.75 & 0.57 & 0.97 & $6.09\times10^{-9}$ \\
        & EvoSuite (Read.)      & 0.82 & 0.82 & 0.71 & 0.94 & $4.78\times10^{-7}$ \\
        & TestHumanizer (Read.) & 0.72 & 0.73 & 0.57 & 0.96 & $3.79\times10^{-8}$ \\
\midrule
Group B & EvoSuite      & 0.56 & 0.60 & 0.43 & 0.80 & $6.30\times10^{-4}$ \\
        & TestHumanizer & 0.94 & 0.94 & 0.93 & 0.98 & $1.10\times10^{-9}$ \\
        & EvoSuite (Read.)      & -0.15 & 0.17 & 0.21 & 0.35 & 0.216 \\
        & TestHumanizer (Read.) & 0.94 & 0.94 & 0.86 & 1.00 & $6.78\times10^{-14}$ \\
\midrule
Overall & EvoSuite (Adopt.)      & 0.80 & -- & -- & -- & -- \\
        & TestHumanizer (Adopt.) & 0.82 & -- & -- & -- & -- \\
        & EvoSuite (Read.)       & 0.71 & -- & -- & -- & -- \\
        & TestHumanizer (Read.)  & 0.85 & -- & -- & -- & -- \\
\bottomrule
\end{tabular}
}
\vspace{2mm}
\begin{minipage}{0.9\textwidth}
\scriptsize
\textit{Note:} Quadratic weighted kappa, percent agreement, and Spearman’s $\rho$ are reported only for two-rater settings (Groups A and B). For the overall four-rater setting, Krippendorff’s $\alpha$ is used as it generalizes to multiple annotators.
\end{minipage}
\end{table*}

\end{document}